\documentclass[fleqn,usenatbib]{mnras}
\usepackage{newtxtext,newtxmath}
\usepackage{comment}
\usepackage{multirow}
\usepackage{mathtools}
\usepackage[T1]{fontenc}
\usepackage{ae,aecompl}

\usepackage{graphicx}	
\usepackage{amsmath}	
\usepackage[ruled,vlined]{algorithm2e}
\usepackage{algpseudocode}
\usepackage{enumitem} 
\usepackage{xfrac} 

\setlist[itemize]{leftmargin=*}
\newcommand{\bigO}[1]{\mathcal{O}(#1)}

\newcommand{\ud}{\mathrm{d}}

\newcommand{\vect}[1]{\boldsymbol{#1}}
\newcommand{\matr}[1]{\boldsymbol{\mathrm{#1}}}

\newcommand{\derfrac}[2]{\frac{\ud #1}{\ud #2}}

\newcommand{\zsol}[1]{{#1}\:\mathrm{Z_\odot}}
\newcommand{\msol}[1]{{#1}\:\mathrm{M_\odot}}
\newcommand{\Sigmasol}[1]{{#1}\:\mathrm{M_\odot\:pc^{-2}}}
\newcommand{\rhosol}[1]{{#1}\:\mathrm{M_\odot\:pc^{-3}}}

\newcommand{\sevn}{\texttt{SEVN}}
\newcommand{\bifrost}{\texttt{BIFROST}}

\newcommand{\paperone}{{Paper I}}

\title[Star cluster assembly with binaries \& triples]{{FROST-CLUSTERS -- II. Massive stars, binaries and triples boost supermassive black hole seed formation in assembling star clusters}}
\author[A. Rantala et al.]{Antti Rantala$^{1}$\thanks{E-mail: anttiran@mpa-garching.mpg.de}, Natalia Lahén$^{1}$, Thorsten Naab$^{1}$, Gastón J. Escobar$^{2,3}$, Giuliano Iorio$^{4,5}$\\
$^{1}$Max-Planck-Institut f\"ur Astrophysik, Karl-Schwarzschild-Str. 1, 
D-85748, Garching, Germany\\
$^{2}$Instituto de Astrofísica de Canarias, E-38205 La Laguna, Tenerife, Spain\\
$^{3}$Departamento de Astrofísica, Universidad de La Laguna, E-38206 La Laguna, Tenerife, Spain\\
$^{4}$Departament de Física Quàntica i Astrofísica (FQA), Universitat de Barcelona (UB), c. Martí i Franquès, 1, 08028 Barcelona, Spain\\
$^{5}$Institut de Ciències del Cosmos (ICCUB), Universitat de Barcelona (UB), c. Martí i Franquès, 1, 08028 Barcelona, Spain\\
}
\date{Accepted XXX. Received YYY; in original form ZZZ}
\pubyear{2025}

\begin{document}
\label{firstpage}
\pagerange{\pageref{firstpage}--\pageref{lastpage}}
\maketitle

\begin{abstract}
Observations and high-resolution hydrodynamical simulations indicate that massive star clusters form through a complex hierarchical assembly. We use simulations including post-Newtonian dynamics (the \bifrost{} code) and stellar evolution (the \sevn{} module) to investigate this collisional assembly. With a full initial stellar mass function, we study the effect of initial binary, triple and massive single stars (450 $M_\odot$) on the assembly, structure, and kinematics of massive ($M_\mathrm  {cl}\sim10^6 M_\odot$, $N=1.8 \times 10^6$) star clusters. Simultaneously, intermediate mass black holes (IMBHs), potential seeds for supermassive black holes, can form and grow in our models by stellar collisions, tidal disruption events (TDEs) and black hole (BH) mergers. At a fixed cluster mass, stellar multiplicity or a high mass limit increase the numbers (up to $\sim$ 10) and masses (up to $10^4 M_\odot$) of the formed IMBHs within the first 10 Myr of cluster evolution. The TDE rates peak at $\Gamma_\mathrm  {tde}\sim 5 \times 10^{-5}$ yr$^{-1}$ after IMBH formation at $\sim 2$ Myr. In all simulations, we find gravitational wave driven mergers involving stellar BHs and IMBHs. Initial multiplicity or a high mass limit also result in IMBH-IMBH mergers. The IMBH masses correlate with the initial cluster masses, surface densities and velocity dispersions approximately as $M_\bullet \propto M_\mathrm{cl}$, $M_\bullet\propto\Sigma_\mathrm{h}^\mathrm{3/2}$ and $M_\bullet\propto\sigma^\mathrm{3}$. Our results suggest the dense $z\sim10$ star clusters recently observed by the James Webb Space Telescope host IMBHs with masses above $M_\bullet \gtrsim \msol{10^4}$.
\end{abstract}



\section{Introduction}

The origin of supermassive black holes (SMBHs; $M_\bullet \gtrsim \msol{10^6}$) remains a long-standing open question \citep{Rees1984}. Recent observations by the James Webb Space Telescope (JWST) of accreting SMBHs at high redshifts (e.g. \citealt{Maiolino2024a,Maiolino2024b,Übler2024,Scholtz2024,Juodzbalis2024,Maiolino2025}) have only deepened this puzzle. The observed SMBHs powering early active galactic nuclei (AGN) had already reached masses of $M_\bullet \gtrsim \msol{10^6}$ at $z\sim10$--$11$ \citep{Maiolino2024a} and $M_\bullet \gtrsim \msol{10^9}$ at $z\sim6$--$7$ \citep{Inayoshi2020}. Simple Eddington rate arguments place the formation redshift of these black holes (BHs) between $z\sim25$--$20$ (intermediate seeds; $M_\bullet \sim \msol{10^3}$) or $z\sim15$--$12$ (heavy seeds; $M_\bullet \sim \msol{10^5}$). Metal-free Population-III or metal-enriched Population-II light stellar BH seed models ($M_\bullet \lesssim \msol{10^2}$--$\msol{10^3}$) in general require special environments or super-Eddington accretion to explain the early AGN populations (e.g. \citealt{Volonteri2006,Inayoshi2020}).

Repeated collisions of stars in dense environments may lead to formation of SMBH seeds in the intermediate mass black hole (IMBH; \citealt{Mezcua2017,Greene2020}) range of $\msol{10^2} \lesssim M_\bullet \lesssim \msol{10^5}$ \citep{Gold1965,Spitzer1966,Spitzer1971,Peebles1972,Begelman1978}. The stellar collisional runaway\footnote{Strictly speaking a cascade of stellar collisions is 'runaway' only if $\derfrac{m}{t}\propto m^\mathrm{\beta}$ with $\beta>0$ i.e. each collision increases the collision rate.} is an especially attractive scenario for the SMBH seed origins as it relies on relatively well understood physical processes and requires no fine-tuning. Due to star cluster mass segregation and core collapse, massive stars may collide at the centres of dense and massive star clusters, forming collision products reaching masses in the range from a few times $\msol{100}$ to $\sim\msol{5\times10^4}$ \citep{PortegiesZwart2004, Vergara2025}. At low metallicities ($Z\lesssim 0.1 Z_\odot$) the stellar collision products may retain most of their mass as the stellar winds are relatively weak, and the massive collision products can directly collapse into IMBHs at the end of their lives. Still, uncertainty on the wind loss rates of massive stars remains. For certain wind loss prescriptions the mass loss rates of massive stars can be more intense even at low metallicities, which can have an impact on their later evolution (e.g. \citealt{Sabhahit2023,Shepherd2025,Simonato2025}). Furthermore, very high wind mass loss rates might inhibit the runaway stellar collisional IMBH formation channel altogether \citep{Belkus2007,Yungelson2008,Glebbeek2009,Pauldrach2012}. If they indeed form, the collisionally formed SMBH seeds are heavy enough to explain the highest redshift AGN if they grew by the Eddington rate since their formation at $z\gtrsim15$ (e.g. \citealt{Taylor2025}). However, directly observing ongoing stellar collisions especially at high redshifts remains an outstanding observational challenge, motivating the search for indirect evidence for the collisional runaway scenario. The observed nitrogen enrichment at low metallicities at high star formation densities of early galaxies and AGN (e.g. \citealt{Bunker2023,Cameron2023,Ji2024,Schaerer2024,Topping2024,Isobe2025,Ji2025,Naidu2025}) may\footnote{The enriched ejecta from stellar collisions is not the only plausible explanation for the early nitrogen enrichment (e.g. \citealt{Rizzuti2024}).} indeed point to dense low-metallicity environments in which stellar collisions can frequently occur \citep{Gieles2018,Charbonnel2023,MarquesChaves2024}.

The formation of BHs and IMBHs in and above the so-called the pulsational pair-instability (P)PISN mass gap through stellar collisions and gravitational wave (GW) driven BH mergers has been widely studied numerically in the context of idealised, isolated star clusters without an initial binary star population. Mass gap BH and IMBH formation has been explored using Pop-II single star cluster models using both Cluster Monte Carlo (CMC) codes \citep{Freitag2006a,Goswami2012,Kremer2020b} and direct \textit{N}-body simulations \citep{PortegiesZwart1999, PortegiesZwart2002, Gurkan2004, Fujii2013, Mapelli2016, Boekholt2018, Reinoso2018, Vergara2025}. Recently, studying the collisional stellar growth in molecular cloud and even dwarf galaxy scale star-by-star simulations with hydrodynamical codes coupled to accurate \textit{N}-body integrators has become feasible \citep{Fujii2024,Lahen2025a}. Overall, the studies suggest that the mass of the most massive stellar collision product in an isolated star cluster sensitively depends on the physical properties of the cluster, including its mass $M_\mathrm{cl}$, central density $\rho_\mathrm{c}$, velocity dispersion $\sigma$, virial parameter $\alpha$, tidal environment, stellar metallicity $Z$, initial mass function (IMF) and binary star fraction $f_\mathrm{bin}$. In general, star clusters with high masses, central densities, binary fractions and top-heavy IMFs at low metallicities can typically form more massive collision products.

It has been established by observations that most of the massive stars form as members of binary, triple or even higher-order multiple systems (e.g. \citealt{Duquennoy1991,Kroupa1995,Raghavan2010,Goodwin2010,Duchene2013,Tokovinin2014a,Tokovinin2014b}). Over $\gtrsim 90\%$ of stars exceeding $\msol{10}$ in mass initially reside in binaries with $\sim50\%$ of these massive stars in triples and quadruple systems \citep{Moe2017,Offner2023}, and the majority of massive stars interact with binary companions during their lifetimes \citep{Sana2012}. The high multiplicity fractions highlight the importance of binary stellar evolution processes including mass transfer, Roche lobe overflows, common envelope phases and stellar collisions for the evolution of massive stars (e.g. \citealt{Kouwenhoven2008}), potentially leading to the formation of X-ray binaries as well as gravitational wave sources. Binary stars also notably affect the evolution of their host star clusters \citep{Heggie2006,Hurley2007,Mackey2008,Wang2016}, in the extreme case halting the core collapse of the clusters \citep{Sugimoto1983,Heggie1993,Kamlah2022}.

Observationally, ancient globular clusters (GCs; \citealt{Gratton2019}) typically host total main sequence binary star fractions in the range of $0.05 \lesssim f_\mathrm{bin} \lesssim 0.2$ \citep{Rubenstein1997,Bellazzini2002,Cool2002,Sollima2007,Dalessandro2011,Milone2012} with individual clusters potentially reaching higher binary fractions. The radial GC binary fraction profiles are mostly flat or declining towards the outer parts, and the total GC binary fractions anticorrelate with GC luminosities and thus with their masses \citep{Milone2012}. Due to the extreme reddening of young massive Milky Way star clusters such as Westerlund 1 and the clusters near the Galactic centre, the best local Universe constraints for binary fractions of young massive clusters have originated (e.g. \citealt{Li2013}) from the Large Magellanic Cloud (LMC). Young massive clusters NGC 1805 and NGC 1818 have total binary fractions $f_\mathrm{bin} \sim 0.37$ and $f_\mathrm{bin} \sim 0.20$--$0.35$ \citep{Elson1998,Hu2010,Grijs2013,Li2013}, comparable to the binary population of field stars (e.g. \citealt{Duquennoy1991,Raghavan2010}). In the 30 Doradus region, the intrinsic binary fractions for massive O-type and B-type stars are high, $0.51\pm0.04$, $0.58\pm0.11$, respectively. Very recently, \cite{RamirezTannus2024} obtained a spectroscopic binary fraction of $0.27$ for the Galactic young cluster M17, indicating a high intrinsic $f_\mathrm{bin}\sim0.87$. Meanwhile, constraining the high redshift initial stellar binary fractions remains a considerable observational challenge. In general, stellar population synthesis models of high redshift galaxies point towards a fraction of massive stars in binaries to explain the spectral features of the galaxies (e.g. \citealt{Steidel2016}).

Initial\footnote{In this study we refer the metal-enriched Pop-II binary stars present at the beginning of the simulations as initial binaries to avoid confusion with initial, pristine metal-free Pop-III stars.} stellar multiplicity plays a crucial role in stellar collisions in idealised, isolated dense star clusters, and has been studied with CMC \citep{Giersz2006,Fregeau2007, Giersz2015, Askar2017, Gonzales2021, GonzalezPrieto2022,GonzalezPrieto2024} and direct \textit{N}-body codes \citep{PortegiesZwart2004, Gurkan2006, Baumgardt2011, Moeckel2011, Fujii2012, Katz2015, DiCarlo2020, DiCarlo2021, Rastello2021, Rizzuto2021, Rizzuto2022, ArcaSedda-DRAGON2a,ArcaSedda-DRAGON2b,ArcaSedda-DRAGON2c}. Initial binary populations promote collisions of massive stars via various physical processes, leading to formation of more massive BHs. First, the components of a binary may collide with each other either due to external gravitational perturbations (e.g. hardening; \citealt{Heggie1975}) or single and binary stellar evolution processes (e.g. common-envelope evolution, collisions at periastron for eccentric orbits, see \citealt{Hurley2002}). Second, due to their larger interaction cross sections compared to single stars (binary semi-major axis $a$ > stellar radius $R_\mathrm{\star}$), high binary fractions increase the probability of stellar collisions in strong single-binary and binary-binary interactions \citep{Fregeau2004,Valtonen2006}. Stellar collision cascades in dense star clusters are typically initiated by the most massive cluster members located in binary systems \citep{Gaburov2008}. Finally, massive stars in especially close to equal mass binary systems experience on average stronger mass segregation compared to isolated stars of the same mass, and are expected to sink into their host cluster centres \citep{Grijs2013}. Similarly, a top heavy IMF with a higher number of massive stars is expected to increase the masses of the stellar collision products (e.g. \citealt{GonzalezPrieto2024}) due to enhanced mass segregation and gravitational focusing as well as increased collision cross-sections. Early star formation may have produced stellar populations with a top-heavy IMF, which might potentially explain the excess of UV bright galaxies observed by the JWST at high redshifts (e.g. \citealt{Inayoshi2022,Harikane2023}). Indeed, hydrodynamical simulations suggests that the IMF might be more top-heavy in metal-poor environments (e.g. \citealt{Chon2022}, see also \citealt{Hennebelle2024}). In the local Universe, the young massive R136 star cluster in the LMC potentially hosts several very massive stars (VMSs) with masses up to a few times $\msol{100}$ \citep{Bestenlehner2020,Brands2022}.

However, isolated and spherically symmetric initial models for massive star clusters may be overly idealised for studying the very earliest phases of the cluster evolution. Both redshift $6 \lesssim z \lesssim 10$ JWST observations of star forming complexes and early bound proto-GCs (e.g. \citealt{Adamo2024, Fujimoto2024, Mowla2024}) as well as early Milky Way archaeological studies \citep{Belokurov2022} indicate that the high redshift star formation at low metallicities was dense, clumpy and clustered. In the local present-day Universe, the most intense star formation environments such as R136 in the LMC, NGC 346 in the Small Magellanic Cloud, the star formation regions of Antennae galaxies, M51 and dwarf star-bursts such as NGC 4449 show complex assemblies of young star clusters \citep{Fahrion2024,Sabbi2007,Whitmore2010,Bastian2005,Meena2025} instead of single, monolithic clusters.

Star-by-star resolution hydrodynamical simulations of star cluster formation in low metallicity star-bursts support this picture \citep{Lahen2020,Elmegreen2024}: massive star clusters form through a complex hierarchical assembly of sub-clusters (e.g. \citealt{VazquezSemadeni2017}) following a universal cluster mass function \citep{Elmegreen1996,Zhang1999,Adamo2020}. The full consequences of this hierarchical massive star cluster assembly scenario for the formation of IMBHs through repeated stellar collisions remain unexplored.

Star cluster formation out of structured non-spherical initial conditions has been studied in the literature (e.g. \citealt{Aarseth1972, Smith2011, Fujii2012, Howard2018, Grudic2018, Sills2018,  Torniamenti2022,Farias2024}) but rarely in the context of collisional IMBH formation. \cite{Fujii2013} studied the growth of massive stars via stellar collisions in ensemble star clusters, examining both spherical and filamentary setups in configurations with up to $8$ sub-clusters and found that each sub-cluster may produce a massive collision product. The dynamics of pre-existing IMBHs in star clusters or star cluster mergers has been more widely studied (e.g. \citealt{Amaro-Seoane2009,Amaro-Seoane2010,Abbas2021,Souvaitzis2025}), and it seems plausible that IMBH binaries can merge within the Hubble time, especially if more than two IMBHs are present per cluster \citep{Rantala2024b,Liu2024,Rantala2025}.

In FROST-CLUSTERS I \citep{Rantala2024b}, hereafter \paperone, we explored the hierarchical formation of three massive star clusters through a complex assembly of $\sim1000$ sub-clusters from the universal power-law cluster mass function with up to $N=2.4\times10^6$ stars ($M_\mathrm{cl}\sim \msol{10^6}$) in total. We found that hierarchical star cluster assembly boosts IMBH formation in a sense that the single star hierarchical setups produce several IMBHs reaching $M_\bullet \sim \msol{2200}$ in mass while isolated comparison clusters closely resembling the final assembled clusters do not.

In this study, FROST-CLUSTERS II, we present a sample of $12$ novel hierarchical massive star cluster simulations with models including single stars, initial binary stars, initial triple stars and finally models with massive single stars up to $\msol{450}$. Recently, we have shown that hierarchical star cluster assembly with initial binaries and triples efficiently leads to close to equal mass GW mergers in the IMBH mass range \citep{Rantala2025}. In this work we present the key features of the coupling between our \textit{N}-body code \bifrost{} and the \sevn{} fast stellar population synthesis package \citep{Iorio2023} and a detailed analysis of the hierarchical single, binary and triple simulations of \cite{Rantala2025}. Moreover, we explore the effect of a higher IMF\footnote{The IMF we use is not top-heavy but follows the standard \cite{Kroupa2001} formulation. However, we sample single star masses beyond the standard IMF upper limit of $\sim \msol{100}$--$\msol{150}$.} cut-off mass on the IMBH formation in a simulation sample with initial stellar masses up to $\msol{450}$. Throughout this study, this sample is referred to as the massive single simulation sample.

The massive single setup can result in close to equal mass GW mergers with $M_\bullet \gtrsim \msol{10^4}$ within $10$ Myr of the star cluster formation. We focus on the structure, kinematics and multiplicity content of hierarchically assembled star clusters as well as scaling relations between the collisionally formed IMBHs and their host clusters. Moreover, a very recent hydrodynamical study of \cite{Lahen2025b} provides an opportunity to decipher dissipative versus hierarchical origins of structural and kinematic properties of observed young massive clusters by comparing detailed hydrodynamical models to accurate but gas-free hierarchical \textit{N}-body simulations.

The article is structured as follows. After the introduction we present the updated \bifrost{} \textit{N}-body code and its coupling to the fast stellar population synthesis code \sevn{} as well as the initial conditions (ICs) in section \ref{section: 2}. We detail the IMBH formation and early growth in isolated and hierarchical simulation models in sections \ref{section: isolated} and \ref{section: 4}. Next, we present the structural and kinematic properties of the hierarchically assembled clusters in section \ref{section: 5} and the scaling relations between the IMBH masses and their host cluster properties in section \ref{section: 6}. Finally, after discussing the implications of our results for the next-generation GW experiments in section \ref{section: 7}, we summarize our results and conclude in section \ref{section: 8}.


\section{Numerical methods and initial conditions}\label{section: 2}

\subsection{Numerical methods: the \bifrost{} \textit{N}-body code}

The numerical simulations for this study are performed using the GPU accelerated direct summation \textit{N}-body code \bifrost{} \citep{Rantala2023}. Besides collisional IMBH and SMBH seed formation \citep{Rantala2024b,Rantala2025}, \bifrost{} has been used to examine the complex sub-pc eccentric stellar disk dynamics in Milky Way like galactic nuclei \citep{Rantala2024a}, seed black hole growth by tidal disruption events \citep{Rizzuto2023} and the production of high and hyper velocity stars in merging star clusters including IMBHs \citep{Souvaitzis2025}. The hierarchical version \citep{Rantala2021} of the fourth order forward symplectic integrator (e.g. \citealt{Chin1997,Chin2005,Chin2007,Dehnen2017a}) used in \bifrost{} is momentum conserving and especially efficient in accurately integrating the equations of motion of \textit{N}-body systems with a large dynamical range. For few-body subsystems, \bifrost{} uses post-Newtonian (PN) secular and algorithmically regularised (e.g. \citealt{Rantala2020}) techniques to treat close fly-bys, binary and multiple systems as well as small clusters around massive black holes. For a detailed description of the few-body methods used, see the Section 3.10 of \cite{Rantala2023} for secular two-body methods, and the Appendix A2 of \paperone{} for the slow-down algorithmic regularisation technique \citep{Mikkola1996,Wang2020}. Most importantly, the direct few-body PN equations of motion for BHs allow us to self-consistently model the gravitational wave emission driven inspirals of BH binaries down to $\sim 10$ Schwarzschild radii.

The criteria for stellar collisions and mergers, tidal disruption events (TDEs) of stars by black holes and GW driven black hole mergers for this study in \bifrost{} are described in Appendix B3 of \paperone. Briefly, two stars collide\footnote{A stellar collision occurs when the radii of two stars overlap. This process leads either to a stellar merger or common envelope evolution.} and can merge if their radii overlap, and the TDE criterion is based on the widely used order-of-magnitude TDE radius estimate (e.g. \citealt{Kochanek1992}). Stellar collisions in this study are mass conserving, and we accrete $50\%$ of the stellar mass disrupted in TDEs into the BHs. Two BHs merge if their innermost stable circular orbits overlap. Each BH-BH merger is accompanied by a relativistic gravitational wave recoil kick whose magnitude we obtain from fitting functions to numerical relativity \citep{Zlochower2015}.

As in \paperone, we do not assume any tidal environment for the assembling star clusters. Unbound stars with total energy $E>0$ are removed from the simulation if they reach separations of $r_\mathrm{esc}=200$ pc from the centre of the most massive central star cluster of the simulation. In order to improve the speed of the particle time-step and neighbour assignment in our code, we have implemented an octree algorithm into \bifrost{}. The octree is used only for the particle neighbour searches while the particle acceleration calculations are still performed using $\bigO{N^2}$ direct summation. In addition, for the fly-by time-steps of binary stars in subsystems, we now use binary centre-of-mass velocities in the \bifrost{} fly-by time-step criteria instead of the velocities of the binary components. In our accuracy tests, using the instantaneous binary component velocities led to overly conservative time-step estimates for particles near hard binaries. We list the key user defined \bifrost{} code parameters and their values in Table \ref{table: bifrost_params}.

\begin{table}
\begin{tabular}{l l l}
\hline
\bifrost{} parameter & symbol & value\\
\hline
integration interval /& $\epsilon_\mathrm{max}$ & $5\times10^{-3}$\\
maximum time-step &  & Myr\\
forward integrator time-step factor & $\eta_\mathrm{ff}$, $\eta_\mathrm{fb}$, $\eta_\mathrm{\nabla}$ & $0.2$\\
subsystem neighbour radius & $r_\mathrm{subsys}$ & $0.5$ mpc\\
regularisation GBS tolerance  & $\eta_\mathrm{GBS}$ & $1\times10^{-8}$\\
regularisation end-time tolerance & $\eta_\mathrm{endtime}$ & $10^{-2}$\\
regularisation highest PN order &  & PN3.5\\
\hline
\end{tabular}
\caption{The user-given \bifrost{} parameters used in the simulations of this study, closely resembling our parameter choices in \paperone.}
\label{table: bifrost_params}
\end{table}

\subsection{Numerical methods: binary stellar evolution in the \bifrost{} code}

\subsubsection{\sevn}

\bifrost{} is coupled to the fast stellar population synthesis code \sevn{}\footnote{\sevn{} is available at \url{https://gitlab.com/sevncodes/sevn}.} \citep{Iorio2023,Mapelli2020,Spera2015,Spera2017} which allows for the modelling of stellar evolution (SE) in our \textit{N}-body simulations. For this study, we adopt the \sevn{} version 2.8.0. For the evolution of single stars, \sevn{} relies on multi-dimensional (stellar mass and metallicity) on-the-fly interpolation of pre-calculated stellar tracks. These tracks include the evolution of H-rich and H-depleted stars separately. In addition, \sevn{} assumes the provided tracks already include the effect of stellar winds in the evolution of the stars. For this study we use the \texttt{PARSEC}-based \citep{Bressan2012, Chen2015, Costa2021, Nguyen2022, Costa2025} stellar tracks \texttt{SEVNtracks\_parsec\_ov04\_AGB} (with the overshooting parameter $\lambda = 0.4$) at $Z=0.0002=\zsol{0.01}$ ranging in stellar mass from $\msol{2.2} \leq m_\star \leq \msol{600}$. For stripped pure He stars we adopt the \texttt{SEVNtracks\_parsec\_pureHe36} tracks. In addition to the stellar tracks, \sevn{} incorporates a number of processes related to (core collapse) supernova (SN) physics, including SN neutrino mass loss \citep{Zevin2020,Lattimer1989}, (P)PISN processes \citep{Woosley2007,Yoshida2016,Woosley2017} and SN kicks, and includes updated prescriptions for compact remnant formation. \sevn{} also accounts for the evolution of the stellar spin through mass-loss and magnetic breaking. Binary stellar evolution prescriptions in \sevn{} are based on \cite{Hurley2002} with updates for common envelope treatments and stellar mergers \citep{Spera2019,Iorio2023}. \sevn{} models binary evolution processes such as wind mass transfer, Roche lobe overflow, common envelope phases, collisions at periastron, stellar tides, orbit decay by GW emission and stellar mergers using a set of analytic and semi-analytic formulas. In its standard set-up, \sevn{} updates the stellar and binary properties according to an adaptive time-step that guarantees the properties change in less than 5\% of their initial value before the step. This condition, i.e. the threshold for the relative change, can be modified by the user at runtime. If a single time-step produces too large change in the modelled physical quantities, the evolution step is repeated using smaller time-steps.

In the beginning of our simulations, we initialise each $\msol{2.2} \leq m_\star \leq \msol{600}$ star into the onset of its main sequence evolution phase. Stars with masses lower than $\msol{2.2}$ are not evolved in our simulations of $\lesssim10$ Myr as their evolution would be negligible. Stellar collisions or common envelope evolution may lead to stellar mergers. Also low-mass stars have radii in the models and can collide and merge with other stars in the simulations. Stellar merger products in the mass range of $\msol{2.2} \leq m_\star \leq \msol{600}$ are initialised as follows. The merger products depend on the nature of the components. If both component stars are on the main sequence or are both naked helium or core helium burning stars, the collision product will have a mass equal to the sum of its progenitors. \sevn{} then jumps to the most suitable stellar track from which the evolution of the resulting star will continue with the new stellar mass and fractional life-time of the more evolved progenitor. If the collision involves at least one evolved star (i.e., a star with a well-defined core-envelope structure), the final core mass is taken as the sum of the progenitors’ core masses, and the total mass is the sum of their total masses. In this case, \sevn{} sums their core and total masses, accounting also for the H, He, and carbon-oxygen (CO) contributions. If the core mass remains unchanged after the collision, the star is considered to continue its evolution along the interpolated tracks of the most evolved progenitor. Otherwise, a new evolutionary track is selected to match the new core mass at the same evolutionary phase as the more evolved progenitor. For additional details see section 2.4.3 in \cite{Iorio2023}. The case in which one of the components is a compact object may follow different paths; a detailed description of the merger outcomes can be found in section 2.3.7 of \cite{Iorio2023} and its corresponding appendix A5. When stars and binaries are embedded in an environment in which they can exchange members or form new ones (such as the dynamical environment provided in a stellar cluster), the formed systems may present different properties as the expected in the case of isolated evolution. For these cases \sevn{} allows to initialise customised stars that did not follow an evolution track from the beginning, but at the certain moment of initialisation they have known core and envelope properties such that they can be assigned an off-record track.

\subsubsection{\sevn-\bifrost{} binary evolution interface}

As a key improvement from \paperone{} in which we only used single stellar evolution, \bifrost{} can now fully utilise the binary stellar evolution functionalities of \sevn. This is enabled by the novel binary stellar evolution interface in our \textit{N}-body code. The binary stellar evolution interface in most parts focuses on identifying the binaries and organising them into the interface data structures on the \bifrost{} side of the coupled code. Bound binary systems with semi-major axis $a<200$ AU $\sim 1$ mpc and masses in the range of $\msol{2.2} \leq m_\star \leq \msol{600}$ are eligible for binary stellar evolution in the code. Above $a\gtrsim 200$ AU, no binary stellar evolution processes are efficient and the evolution of the two stars is equivalent to single stellar evolution. Stars exceeding $\msol{600}$ evolve as single stars using scaled high mass stellar tracks as described in \paperone{} for a straightforward comparison between the models. We note that especially massive ($\gtrsim \msol{600}$) close to equal mass binaries rarely form in the models of this study as typically only one star per sub-cluster can considerably grow by collisions \citep{Rantala2025}. As such, we expect that the assumption of single evolution for $\gtrsim \msol{600}$ stars unlikely affects the results of this study. However, we note that including binary processes such as mass transfer for $\gtrsim \msol{600}$ stars remains an important future improvement for our models if \bifrost{} and \sevn{} are to be used to model systems in which such very massive binaries form. We update the \bifrost{} time-step criterion $\epsilon = \min{(\epsilon_\mathrm{dyn},\epsilon_\mathrm{se})}$ so that the \sevn{} binary stellar evolution time-step is accounted for in the joint single and binary evolution time-step as $\epsilon_\mathrm{se}=\min{(\epsilon_\mathrm{se,single},\epsilon_\mathrm{se,binary})}$. This ensures that the dynamical and stellar evolution processes remain properly coupled. We disable the \sevn{} binary GW orbit decay and circularisation as these effects are already accounted for by the post-Newtonian few-body integrators in \bifrost.

The \sevn{} binary evolution is coupled into the \bifrost{} time evolution sequence just as the \sevn{} single stellar evolution. The single and binary evolution procedures are called after the forward symplectic integration (FSI) routine \citep{Rantala2021} in the hierarchical fourth order forward integration (HHS-FSI). In \bifrost{}, each binary system eligible for binary stellar evolution is assigned a next evolution time $t_\mathrm{next}$ based on its \sevn{} time-step $\epsilon_\mathrm{se}$. We also update the \sevn{} binary orbit data ($a$,$e$) for the binary population due to external gravitational perturbations and apply binary stellar evolution immediately if the binary semi-major axis $a$ or eccentricity $e$ has changed due to gravitational perturbations, i.e. $|\Delta a/a|>\eta_\mathrm{orbit}$ or $|\Delta e|>\eta_\mathrm{orbit}$. We use $\eta_\mathrm{orbit}=0.05$ throughout this study. After binary stellar evolution we update the position and the velocity of the binary components in \bifrost{} using the updated masses, semi-major axis and eccentricity while keeping the orbit orientation (Keplerian elements $i$, $\omega$ and $\Omega$) and mean anomaly unchanged. The maximum time-step for both single and binary stellar evolution is the \bifrost{} integration interval duration $\epsilon_\mathrm{max}$ for which we use $\epsilon_\mathrm{max}=5\times10^{-3}$ Myr in this study.

\begin{figure}
\includegraphics[width=0.8\columnwidth]{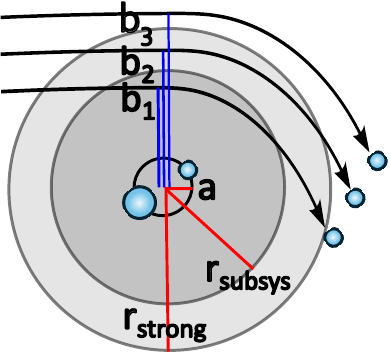}
\caption{The treatment of gravitational dynamics and binary stellar evolution for a binary star with a semi-major axis $a$ in close interactions with other stars. In weak encounters with a large impact parameter ($b_\mathrm{3}>r_\mathrm{strong}$) the dynamics of the binary is mildly perturbed and no special treatment for the binary evolution is required. If the encounter is close enough ($b_\mathrm{2}<r_\mathrm{strong}$) binary stellar evolution is temporarily changed to single evolution to avoid spurious binary evolution effects from osculating orbital elements of the binary. In addition, in very strong interactions ($b_\mathrm{1}<r_\mathrm{subsys}$) the encounter is integrated using algorithmic regularisation.}
\label{fig: binaryradii}
\end{figure}

The fact that the \cite{Hurley2002} like analytic and semi-analytic binary evolution process formulas depend on Keplerian two-body orbital elements (especially $a$ and $e$) requires careful attention when evolving binaries in very dense stellar environments. Perturbations from passing stars onto the binary as well as strong few-body interactions change the instantaneous osculating eccentricities (and thus osculating periastron separations $r_\mathrm{p}=a(1-e)$, potentially leading to onset of spurious mass transfer and common envelope phases as well as stellar mergers. Such undesirable numerical effects are possible when simulating dense cluster environments as in this study, but are likely rare in low density environments. In order to avoid such numerical effects, especially the spurious periastron collisions and mergers, we disable binary stellar evolution processes if a binary is experiencing a strong interaction with one or more external bodies. The radius for strong binary interactions $r_\mathrm{strong}$ is illustrated in Fig. \ref{fig: binaryradii}. During such a strong interaction, the two binary components evolve as single stars. Specifically, we define the strong binary interaction radius as $r_\mathrm{strong}=10\:a$. An additional time-step criterion for particles close to $r_\mathrm{strong}$ is introduced in \bifrost{} to ensure that the code captures all the strong interactions with $r_\mathrm{strong}$ between the binary components and third bodies (or other binaries). After the strong interaction, the binary is again evolved using the binary stellar evolution processes. We emphasise that stars can still collide and merge during the strong interactions if any of the \bifrost{} stellar, TDE or black hole merger criteria are fulfilled.

We note that our current code version does not, for simplicity, include potentially important specific triple stellar evolution prescriptions (e.g. \citealt{Toonen2016, Stegmann2022,Perets2025}) such as triple mass transfer or triple common envelopes \citep[e.g.][and references therein]{Hamers2021}. The dynamical processes enhancing the collision rates in triples such as the \cite{vonZeipel1910,Lidov1962,Kozai1962} oscillations are treated (without any dependence on $r_\mathrm{strong}$) using the post-Newtonian algorithmically regularised integrator.

\subsubsection{Simplified modelling of supermassive stars and stellar collision products}

Our simplified model for very massive and supermassive stars in this work relies on extrapolating the $\msol{600}$ stellar tracks provided by \sevn. Observationally, the stellar radii $R_\mathrm{\star}$ of very massive and supermassive stars (SMSs) beyond a few hundred $M_\odot$ remain unconstrained. In analytic and numerical studies it is commonly assumed that such massive stars above $m_\star \gtrsim \msol{100}$ follow $R_\mathrm{\star} \propto m_\star^\mathrm{\delta}$ with $\delta \sim0$--$1$ (e.g. \citealt{Gieles2018}) based on extrapolating the observed mass-size relation for O-type stars (e.g. \citealt{Crowther2010}). Nevertheless, this extrapolated mass radius relation may still underestimate the true radii of SMSs. Several physical processes including radiation pressure \citep{Grafener2012,Szecsi2015}, stellar collisions, or gas accretion may further enhance the SMS radii. Numerical simulations have shown that right after a collision, the collision product may be $1$--$2$ orders of magnitude larger in size compared to an unperturbed star of similar mass \citep{Suzuki2007}. In the case of accreting protostar-like massive stars \citep{Nandal2024}, the accretion rate onto the star determines the radius of the star \citep{Hosokawa2013, Schleicher2013,AlisterSeguel2020}. For high accretion rates, the stellar radii may be inflated by several orders of magnitude compared to non-accreting stars. The exact accreting SMS radius also sensitively depends on the assumptions for convection within the star \citep{RamirezGaleano2025}. In the absence of repeated collisions or gas accretion, the stars settle into their equilibrium sizes on the Kelvin-Helmholtz time scale.

In general, mass conservation may be a poorly justified approximation for stellar collisions, and more realistic mass loss  fractions may be up to $\sim10\%$ per collision in the main sequence (e.g. \citealt{Glebbeek2009}). Collisions involving supermassive stars might also lead to net mass loss as loosely bound material in the envelope of the SMS may become unbound as the incoming star spirals in and dissolves into the SMS \citep{RamirezGaleano2025}. This type of destructive mass loss may severely limit the growth of SMSs by stellar collisions.

In addition, the life-times of stellar collision products may deviate from the life-times of unperturbed stars (e.g. \citealt{Glebbeek2009}). This stellar rejuvenation process has been incorporated in \textit{N}-body simulations using a simple prescription depending on the masses of the merging stars and their current age in their evolutionary phase \citep{Hurley2002,Mapelli2016}. For SMSs colliding with low mass stars, the standard approach may lead to excessive rejuvenation, and a refined prescription for the age of the collision product is required \citep{Vergara2025}.

Our simplified collision and SMS prescriptions likely underestimate the stellar collision rates and thus the final IMBH masses as we do not model the temporarily increased radii after stellar collisions. The lack of explicit rejuvenation procedure leads to further IMBH mass underestimation. On the contrary, the assumption of no mass loss in stellar collisions likely leads to overestimation of the final number of IMBHs and their masses. Without further detailed calculations it is challenging to estimate to which degree the two competing effects balance each other, or not. We will update our stellar collision product and SMS prescriptions for the forthcoming FROST-CLUSTERS III study.

\subsection{Initial conditions}

\subsubsection{Individual star clusters}\label{section: ic-isolated}

The individual star clusters in this study (both isolated clusters and sub-clusters in the hierarchical assembly regions) are realised as spherical equilibrium \cite{Plummer1911}\footnote{We have examined in a sample of $60$ isolated simulations whether the assumed Plummer model has an effect on the masses of the collisionally grown stars. We find that for a fixed cluster half-mass radius $r_\mathrm{h}$ the Plummer setups produce very similar masses for the collisionally grown stars as King $W_\mathrm{6}$ models \citep{King1962}.} models characterised by their masses $M_\mathrm{cl}$ and half-mass radii $r_\mathrm{h}$. For the cluster sizes we assume a power-law mass-size relation (normalisation $R_\mathrm{4}$, slope $\beta$) for our star cluster population defined as
\begin{equation}\label{eq: mass-size-relation}
    \frac{r_\mathrm{h}}{\mathrm{pc}} = \frac{f_\mathrm{h} R_\mathrm{4}}{1.3} \left( \frac{M_\mathrm{\star}}{10^4M_\odot} \right)^\beta,
\end{equation}
in which $f_\mathrm{h}$ is an adjustable constant and the factor of $1.3$ originates from the properties of the Plummer model. As in Paper I, we set $\beta=0.180\pm0.028$ and $R_\mathrm{4}=2.365\pm0.106$ following \cite{Brown2021} but choose $f_\mathrm{h}=1/8$ in order to capture the small birth radii of observed embedded clusters (e.g. \citealt{Marks2012}) and star-by-star hydrodynamical simulations \citep{Lahen2020,Lahen2025b}. For the relations of \cite{Brown2021}, $f_\mathrm{h}=1$. The central densities $\rho_\mathrm{c}$ of the most massive cluster models are a few times $\rhosol{10^6}$, similar to \cite{Lahen2025b}. The half-mass surface densities $\Sigma_\mathrm{h}$ of our individual star clusters reach up to $\Sigma_\mathrm{h} \gtrsim \Sigmasol{10^5}$ and are consistent with other recent simulation studies exploring IMBH formation and the evolution of massive star clusters, as shown in Fig. 1 of \cite{ArcaSedda-DRAGON2a} and Fig. 2 of \cite{Rantala2024b}.

The masses of individual stars in the individual clusters are sampled from the standard \cite{Kroupa2001} initial mass function. The maximum initial stellar mass $m_\mathrm{max,0}$ in a cluster model depends on the cluster mass $M_\mathrm{cl}$ \citep{Weidner2006,Yan2023} as described in Paper I. For star clusters with single stars, initial binaries and triples, the maximum initial stellar mass is $m_\mathrm{max,0} \sim \msol{80}$ for $M_\mathrm{cl}=\msol{10^4}$ and $m_\mathrm{max,0} = \msol{150}$ for clusters masses above $M_\mathrm{cl} \gtrsim \msol{3\times10^4}$. For the setups involving massive single stars we have $m_\mathrm{max,0} = \msol{450}$.

\subsubsection{Hierarchical cluster assembly regions}

We setup the hierarchical star cluster assembly regions as in Paper I. We sample the masses $M_\mathrm{cl}$ of the star clusters in the region from the universal power-law cluster mass function with a slope of $\alpha=-2$ \citep{Elmegreen1996,Zhang1999,Adamo2020,Lahen2020} and $\max{(M_\mathrm{cl})}=\msol{2.5\times10^5}$. The total stellar mass in the region is $M_\mathrm{region} = \msol{10^6}$ yielding $N_\mathrm{\star}\sim1.8\times10^6$ particles in our hierarchical setups.

We sample the initial centre-of-mass positions for our star cluster population within a uniform sphere with a radius of $r_\mathrm{max} = 50$ pc. The most massive star cluster is placed at rest at the origin. The cluster region is initially collapsing with a radial velocity of $v_\mathrm{r} = -3.5$ km/s for each cluster centre-of-mass, together with an isotropic random velocity component of $3.5$ km/s. The selected star cluster velocities are motivated by the structure of the cluster formation regions in solar mass resolution hydrodynamical simulations \citep{Lahen2020}. For additional details of the topic see section 3.3 of \paperone. We finally ensure that each cluster centre-of-mass is initially bound to the region.

\subsection{Initial conditions: the initial binary and triple populations}\label{section: ic-binary-triple}
Given the uncertainties concerning the high redshift stellar multiplicity, for this study we adopt stellar multiplicity (binary and triple) fractions of the local Milky Way observations. For the initial stellar binary population properties, very similar models have been adopted for recent dynamical studies \citep{Cournoyer-Cloutier2021,Cournoyer-Cloutier2024}. In our models we do not consider initial stellar quadruples or any higher order hierarchical configurations.

\begin{table}
    \centering
    \begin{tabular}{rcc}
        \hline
        $m_\mathrm{p}$ & $f_\mathrm{bin,0}(m_\mathrm{p})$ & $f_\mathrm{trip,0}(m_\mathrm{p})$\\ 
        \hline
        $\msol{0.1}$ & 0.19 & 0.02\\
        $\msol{0.3}$ & 0.25 & 0.04\\
        $\msol{1.0}$ & 0.45 & 0.12\\
        $\msol{3.0}$ & 0.78 & 0.33\\
        $\msol{10.0}$ & 0.93 & 0.56\\
        $\msol{30.0}$ & 0.96 & 0.68\\
        $>\msol{30.0}$ & 0.96 & 0.68\\
        \hline
    \end{tabular}
    \caption{The primary mass $m_\mathrm{p}$ dependent initial binary $f_\mathrm{bin,0}(m_\mathrm{p})$ and triple star fractions $f_\mathrm{trip,0}(m_\mathrm{p})$ for $m_\mathrm{\star} \lesssim \msol{30}$ in our models as closely following the results of \citet{Offner2023}. For more massive stars we use the $f_\mathrm{bin,0}(m_\mathrm{p})$ and $f_\mathrm{trip,0}(m_\mathrm{p})$ of $\msol{30}$ stars.}
    \label{table: multifractions}
\end{table}

For the multiplicity fractions of zero age main sequence (ZAMS) stars we assume primary star mass ($m_\mathrm{p}$) dependent fractions that closely follow the observations of \cite{Offner2023}. We define the binary $f_\mathrm{bin}$ and triple fractions $f_\mathrm{trip}$ as
\begin{equation}
\begin{split}   
    f_\mathrm{bin}(m_\mathrm{p}) &= \frac{N_\mathrm{bin}(m_\mathrm{p})+N_\mathrm{trip}(m_\mathrm{p})}{N_\mathrm{sin}(m_\mathrm{p})+N_\mathrm{bin}(m_\mathrm{p})+N_\mathrm{trip}(m_\mathrm{p})}\\
    f_\mathrm{trip}(m_\mathrm{p}) &= \frac{N_\mathrm{trip}(m_\mathrm{p})}{N_\mathrm{sin}(m_\mathrm{p})+N_\mathrm{bin}(m_\mathrm{p})+N_\mathrm{trip}(m_\mathrm{p})}
\end{split}
\end{equation}
in which $N_\mathrm{sin}(m_\mathrm{p})$, $N_\mathrm{bin}(m_\mathrm{p})$ and $N_\mathrm{trip}(m_\mathrm{p})$ are the numbers of single, binary and triple systems with a primary star of mass $m_\mathrm{p}$. We note that the $f_\mathrm{bin}$ is often called the multiplicity fraction (MF) in the literature, and $f_\mathrm{trip}$ is referred to as the triple/high-order fraction (THF). The initial mass dependent binary $f_\mathrm{bin,0}(m_\mathrm{p})$ and triple $f_\mathrm{trip,0}(m_\mathrm{p})$ fractions we use are listed in Table \ref{table: multifractions}. For stars with ZAMS masses higher than $\msol{30}$ we use the $\msol{30}$ values for their binary fractions. Thus, only a small fraction ($\sim4\%$) of massive stars remain single in our initial setups that include initial binaries and triples.

For the secondary star mass $m_\mathrm{s}$, binary mass ratio $(q=m_\mathrm{s}/m_\mathrm{p}<1)$, semi-major axis $a$ and eccentricity distributions we follow the approach of \cite{Cournoyer-Cloutier2021} using the observations of \cite{Moe2017} and \cite{Winters2019}. First, given the primary star mass, we sample the orbital period $P$ of the binary. Next, the secondary star mass $m_\mathrm{s}$ is sampled using $m_\mathrm{p}$ and $P$. This yields the semi-major axis $a$ of the binary. Finally, we sample the binary eccentricity $e$ using $m_\mathrm{p}$ and $P$. The resulting IMF is still very close to the primary star \cite{Kroupa2001} IMF. We ensure that for each sampled binary the pericenter distance initially fulfils $r_\mathrm{p}=a(1-e)>R_\mathrm{p}+R_\mathrm{s}$ in which $R_\mathrm{p}$ and $R_\mathrm{s}$ are the ZAMS radii of the stars.

We sample the inner binaries of our hierarchical triple star systems as already described above. We further scale the semi-major axis $a$ of each inner binary with $m_\mathrm{p}>\msol{1}$ by $(m_\mathrm{p}/\msol{1})^{3/2}$ to take into account the fact that inner binary distributions are skewed towards smaller separations compared to isolated binaries (see Fig. 3 of \citealt{Offner2023}). The outer orbits of the triples are sampled using the non-scaled recipe. We ensure the initial stability of the hierarchical triples by enforcing the empirical relation of $P_\mathrm{out}/P_\mathrm{in}>5(1-e_\mathrm{out})^{-3}$ \citep{Tokovinin2004} as well as the criteria of \cite{Mardling2001} and \cite{Vynatheya2022}. Finally, we integrate the orbits of a subset of our triple systems for $100P_\mathrm{out}$ in isolation to ensure their dynamical stability.

In our simulations, stars retain information of their initial spin direction. Binary members in the initial conditions have their spin vectors $\vect{S}$ aligned with the binary angular momentum vector $\vect{L}$ while single stars have random spin directions. Outer stars in triples have their spins aligned with the angular momentum vector of the outer orbits. The spin direction is used for setting the spin direction of BHs formed during the simulation. Thus, binary BHs originating from initial binaries have aligned spins while dynamically formed binaries have misaligned spins. The spin magnitude of BHs is determined using the Geneva model \citep{BelczynskiKlencki2020} as in Paper I. Otherwise, the stored initial spin direction has no effect on the stellar dynamics or stellar evolution in the simulations. However, the BH spins play an important role for the relativistic GW recoil kick velocities after BH-BH mergers.

\begin{table*}
\begin{center}
\begin{tabular}{l l c c c c c c c c}
\hline
Isolated & $N_\mathrm{seed}^\mathrm{single}$ & $N_\mathrm{seed}^\mathrm{binary}$ & $N$ & $M_\mathrm{\star}$ & $\rho_\mathrm{c,init}$ & $t_\mathrm{seg}$ & $t_\mathrm{cc}$ & $m_\mathrm{max}$ with & $m_\mathrm{max}$ incl. \\
cluster & & & & $\mathrm{[M_\odot]}$ & $\rhosol{10^6}$ & Myr & Myr & singles $ \mathrm{[M_\odot]}$ & binaries $\mathrm{[M_\odot]}$\\
\hline
I1 & $10$ & $10$ & $1.7\times10^4$ & $9.4\times10^3$& 0.43 & 0.05 & 0.41 & ${197}$ & ${304}$ \\
I2 & $10$ & $10$ &  $2.6\times10^4$ & $1.5\times10^4$& 0.54 & 0.05 & 0.62 & ${290}$ & ${339}$ \\
I3 & $10$ & $10$ &  $3.9\times10^4$ & $2.3\times10^4$& 0.65 & 0.06 & 0.91 & ${419}$ & ${527}$ \\
I4 & $10$ & $10$ &  $6.3\times10^4$ & $3.7\times10^4$& 0.81 & 0.07 & 1.38 & ${932}$ & ${1000}$ \\
I5 & $10$ & $10$ &  $1.0\times10^5$ & $5.9\times10^4$& 1.00 & 0.10 & 2.09 & ${982}$ & ${1275}$ \\
I6 & $10$ & $10$ &  $1.6\times10^5$ & $9.4\times10^4$& 1.25 & 0.14 & 3.17 & ${1429}$ & ${2091}$ \\
I7 & $10$ & $10$ &  $2.5\times10^5$ & $1.5\times10^5$& 1.55 & 0.21 & 4.83 & ${1494}$ & ${2583}$ \\
I8 & $10$ & $10$ &  $4.0\times10^5$ & $2.3\times10^5$& 1.89 & 0.29 & 7.11 & ${362}$ & ${3215}$ \\
I9 & $5$ & $10$ &  $6.2\times10^5$ & $3.7\times10^5$& 2.34 & 0.43 & 10.95 & ${199}$ & ${4759}$ \\
I10 & $5$ & $10$ &  $1.0\times10^6$ & $5.9\times10^5$& 2.91 & 0.63 & 16.76 & ${150}$ & ${4392}$ \\

\hline
\end{tabular}
\caption{The isolated star cluster simulation models I1--I10 with their initial particle numbers $N$, initial masses $M_\star$ and the masses for the most massive star ($\max{m_\star}$) formed through collisions in the simulation sets with and without initial binaries. The single star models and results are adopted from \paperone. We also display the initial central stellar densities of the cluster setups $\rho_\mathrm{c,init}$ as well as estimates for the mass segregation and core collapse time-scales ($t_\mathrm{seg}$, $t_\mathrm{cc}$) of the single star cluster models.}
\label{table: isolated}
\end{center}
\end{table*}

\section{Stellar collisions in isolated clusters with and without initial binary stars}\label{section: isolated}

\subsection{Isolated cluster sample}
In \paperone, we showed that massive stellar collisions are suppressed in the high velocity dispersion environments of massive star cluster centres if only single stars are present in the isolated initial setups. We repeat the same exercise now including an initial binary star population as described in section \ref{section: ic-binary-triple}. We setup the structural properties of our isolated cluster sample as in \paperone{} and section \ref{section: ic-isolated}. The $10$ different logarithmically spaced cluster masses for the models I1--I10 range from $M_\mathrm{cl}\sim \msol{9.4 \times 10^3}$ to $M_\mathrm{cl}\sim \msol{5.9 \times 10^5}$ as in \paperone{} to allow a straightforward comparison of results without and with initial binary stars. The most massive models I10 contain $N\sim10^6$ stars. The key isolated cluster properties are collected in Table \ref{table: isolated}. We simulate $N_\mathrm{seed}^\mathrm{binary}=10$ random realisations of the cluster models with initial binaries for $t=5$ Myr, in total $100$ isolated simulations. We compare these isolated binary model results directly to our single star findings of \paperone.

\subsection{Initial binary stars are crucial for collisional IMBH formation}\label{section: role-binaries-isolated}

\begin{figure}
\includegraphics[width=\columnwidth]{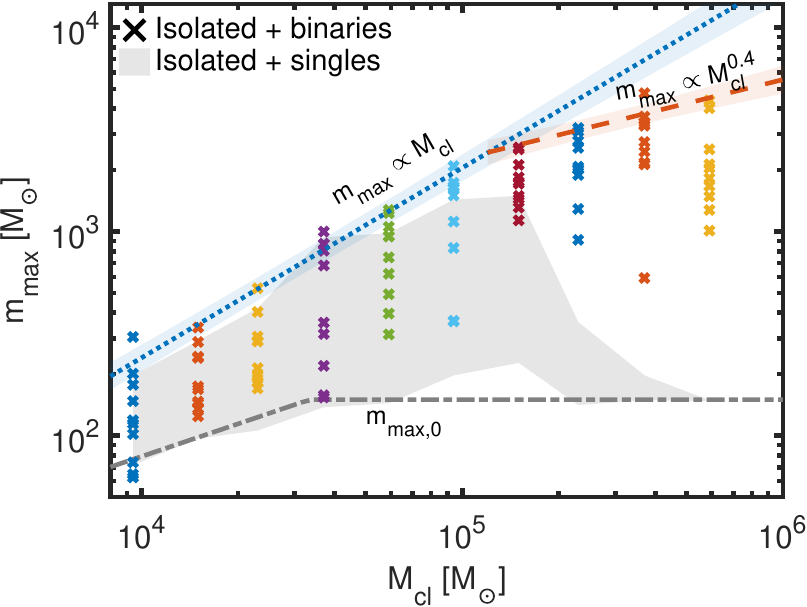}
\caption{The maximum collisionally grown stellar masses $m_\mathrm{max}$ in the isolated setups with initial binaries (small crosses) within $5$ Myr. The single star models of \paperone{} are displayed as the shaded background. The highest $m_\mathrm{max}$ for each cluster mass $M_\mathrm{cl}$ linearly increases (dotted line) until $M_\mathrm{cl}\sim \msol{10^5}$. At higher cluster masses, the stellar collisions are suppressed in the single star models. In the binary models, the collisions can proceed at $M_\mathrm{cl} > \msol{10^5}$, however the linear trend between $m_\mathrm{max}$ and $M_\mathrm{cl}$ is somewhat flattened (dashed line). The initial maximum stellar mass $m_\mathrm{max,0}$ is shown near the bottom as the dash-dotted line.}
\label{fig: isolated-mmax}
\end{figure}

Fig. \ref{fig: isolated-mmax} shows the maximum masses of collisionally formed stars $m_\mathrm{max}$ as a function of the initial isolated host cluster mass $M_\mathrm{cl}$. The different random realisations of the models I1--I10 result in different values for $m_\mathrm{max}$ with the same $M_\mathrm{cl}$. Initially, all clusters have their $m_\mathrm{max,0}$ determined by $M_\mathrm{cl}$ as detailed in section \ref{section: ic-isolated} with $m_\mathrm{max,0} \sim \msol{80}$ for $M_\mathrm{cl}=\msol{10^4}$ and $m_\mathrm{max,0} \sim \msol{150}$ for $M_\mathrm{cl} \gtrsim \msol{3\times10^4}$. In the absence of any stellar collisions, $m_\mathrm{max} = m_\mathrm{max,0}$. This is the case for multiple simulations, especially in the single star sample.

The stellar collision speeds in our simulations are high, typically in the range of $v_\mathrm{coll}\sim500$--$2300$ km/s for main sequence stars and above $v_\mathrm{coll} \gtrsim 100$ km/s for evolved stars. These collision speeds are considerably higher than our cluster central velocity dispersions at any point of their evolution, \mbox{$\sigma \lesssim 60$ km/s}, highlighting the importance of binaries, strong few-body encounters and gravitational focusing for the collision process. Due to the geometry of the binary-single star interactions the relative velocity at the moment of impact may be different than in a two-body encounter \citep{Gaburov2008}, but we do not find any evidence of a low-velocity tail in the distribution of stellar collision speeds. Most ($\gtrsim90\%$) of the collisions occur from almost parabolic orbits with eccentricities in the range of $0.95 \lesssim e \lesssim 1.05$. Stellar multiplicity has only a minor effect on the collision orbit eccentricities and collision velocities. As is shown in Fig. 11 of \paperone, the mass function of stars colliding with the growing very massive star is heavily biased towards massive stars above $\gtrsim \msol{10}$, and as such considerably deviates from an evolved Kroupa-like IMF.

We obtain the maximum collisional stellar mass for each cluster mass $M_\mathrm{cl}$ from the $10$ random realisation of each model. In practise, we fit the $M_\mathrm{cl}$--$m_\mathrm{max}$ relation for the models using the two highest $m_\mathrm{max}$ for each $M_\mathrm{cl}$. In the single star models, the maximum stellar mass scales almost linearly as $m_\mathrm{max} \propto M_\mathrm{cl}^\mathrm{0.90\pm0.07}$ until $M_\mathrm{cl} \sim \msol{10^5}$. In this regime below $M_\mathrm{cl} = \msol{10^5}$, the initial binary models produce collisionally formed stars with up to $\sim30\%$ larger masses compared to single star models. Still, the slope of the $M_\mathrm{cl}$--$m_\mathrm{max}$ relation for the binary setups $m_\mathrm{max} \propto M_\mathrm{cl}^\mathrm{0.93\pm0.06}$ is very similar to the single star case.

For a fixed $M_\mathrm{cl}$, the scatter of $m_\mathrm{max}$ among the different random realisations of the same cluster model is substantial. At $M_\mathrm{cl} \lesssim \msol{7\times10^4}$ especially in the single star models, the most massive collisionally formed stars essentially populate the entire range between $m_\mathrm{max,0}$ and the highest $m_\mathrm{max}$. Our number of different random realisations per cluster model ($10$) is not sufficient to study the distribution of $m_\mathrm{max}$ for a fixed $M_\mathrm{cl}$ in further detail and is left for future work. Furthermore, we find no obvious correlation between the highest $m_\mathrm{max}$ and the star cluster properties including $m_\mathrm{max,0}$ and the starting locations of the most massive stars in the initial conditions. This highlights the chaotic nature of the collisional \textit{N}-body problem: even models with a massive binary initially close to the centre do not necessarily result in the most massive collisionally formed stars.

For cluster masses higher than $M_\mathrm{cl} \sim \msol{10^5}$, the single and binary setups considerably differ. In the single star models the collisional stellar growth is strongly suppressed in high mass clusters. In Paper I we argued that the suppression is caused by the high velocity dispersions of the massive clusters, and the competition between the cluster core collapse timescale and the massive star lifetimes of $\lesssim 3$ Myr (e.g. \citealt{Fujii2013}). Most collisions involving massive stars proceed through a binary phase. As the dynamical binary formation rate $\bar{C}$ strongly scales as $\bar{C} \propto \sigma^{-9}$ \citep{Goodman1993,Binney2008,Atallah2024}, $\bar{C}$ decreases by a factor of $\sim40$ from model I7 to I10 as the central velocity dispersion increases from $41$ km/s to $61$ km/s. This dynamical velocity dispersion bottleneck prevents collisional stellar growth in single star models. Fig. \ref{fig: isolated-mmax} shows that this is indeed the case. When including an initial binary star population into the setups, the dynamical binary formation bottleneck vanishes and the collisional stellar growth is not suppressed at $M_\mathrm{cl} > \msol{10^5}$. We also confirm the that the velocity dispersion bottleneck for stellar collisions is avoided in models with initial triples, as expected. In addition, 2/3 of the massive single star models up to $m_\mathrm{max,0}=\msol{450}$ form $\gtrsim \msol{10^3}$ stars through collisions in $M_\mathrm{cl} > \msol{2 \times 10^5}$ clusters.

Our results thus highlight the crucial importance of initial binary stars for collisional IMBH formation in dense, massive star clusters with high velocity dispersions. First, massive stars in binaries usually drive the first collisions in star clusters \citep{Gaburov2008} in single-binary and binary-binary interactions \citep{Fregeau2004}. Second, for a fixed mass, a star with a binary companion sinks more efficiently due to dynamical friction and mass segregation compared to an isolated star of the same mass. Dynamical binary formation alone cannot produce a sufficiently hard binary population of massive stars to fuel single-binary and binary-binary interactions and to overcome the collisional velocity dispersion bottleneck.

\subsection{Maximum masses of collisionally formed stars in isolated models}

Initial binary populations allow increasingly massive stellar collision products up to $m_\mathrm{max}\sim \msol{4800}$ to form in isolated massive, dense clusters with high velocity dispersions in contrast to single star models. However, the almost linear relation between highest $m_\mathrm{max}$ and $M_\mathrm{cl}$ seems to somewhat flatten at high cluster masses. At $M_\mathrm{cl} > \msol{10^5}$, $m_\mathrm{max} \propto M_\mathrm{cl}^\mathrm{0.39\pm0.09}$ ($\propto M_\mathrm{cl}^\mathrm{0.54\pm0.03}$ without I10) for the isolated models. We attribute this trend to increasing the core collapse timescales in the most massive star clusters (see $t_\mathrm{cc}$ in Table \ref{table: isolated}).

We note that our collisional IMBH masses are very similar to the literature results $\msol{100} \lesssim M_\bullet \lesssim \msol{10^4}$ (e.g. \citealt{Volonteri2021} and references therein) with $\sim0.1\%$--$5\%$ of the initial cluster mass ending up in the formed IMBH \citep{Freitag2006a,ArcaSedda-DRAGON2a,Rantala2024b,Fujii2024,Rantala2025,Lahen2025b}. Recently, \cite{GonzalezPrieto2024} presented a large grid of Monte Carlo runaway collision simulations and found $m_\mathrm{max}^\mathrm{GP24} \propto N^\mathrm{0.26\pm0.13}$. Assuming a fixed mean stellar mass $\Tilde{m}$, then $N=M_\mathrm{cl}/\Tilde{m}$ and $m_\mathrm{max}^\mathrm{GP24} \propto M_\mathrm{cl}^\mathrm{0.26\pm0.13}$, a considerably shallower relation than our result $m_\mathrm{max} \propto M_\mathrm{cl}$ below $M_\mathrm{cl} \lesssim \msol{10^5}$. At higher cluster masses the results agree better. \cite{GonzalezPrieto2024} note that their setups are not in the extreme collisional runaway regime (as e.g. in \citealt{Freitag2006a,Freitag2006b}) in which the entire cluster core collapses (e.g. \citealt{Vergara2023}) and forms a very massive star. As our results lie closer to the extreme collisional regime, this potentially explains our steeper scaling of the collision product masses with the star cluster mass.

\begin{table*}
\begin{center}
\begin{tabular}{c c c c c c c c c}
\hline
Hierarchical & $N_\mathrm{seed}$ & $N_\mathrm{subcluster}$ & $M_\mathrm{\star}$ & $N_\mathrm{\star}$ & $m_\mathrm{max,init}$ & massive & binaries & triples\\
simulation & & (max) & $\mathrm{[M_\odot]}$ & & $\mathrm{[M_\odot]}$ & singles & (initial) & (initial)\\
\hline
HS150 & $3$ & $900$ & $1.0\times10^6$ & $1.8\times10^6$ & $150$ & $\times$ & $\times$ & $\times$\\
HS450 & $3$ & $900$ & $1.0\times10^6$ & $1.8\times10^6$ & $450$ & $\checkmark$ & $\times$ & $\times$\\
HB150 & $3$ & $900$ & $1.0\times10^6$ & $1.8\times10^6$ & $150$ & $\times$ & $\checkmark$ & $\times$\\
HT150 & $3$ & $900$ & $1.0\times10^6$ & $1.8\times10^6$ & $150$ & $\times$ & $\checkmark$ & $\checkmark$\\
\hline
\end{tabular}
\caption{The $12$ hierarchical massive star cluster assembly simulations of this study. We indicate whether each simulation initially contains massive ($>\msol{450}$) singles, initial binaries or initial triple systems.}
\label{table: hierarchical-cluster-sample}
\end{center}
\end{table*}

\section{Stellar multiplicity and collisions in hierarchically assembling star clusters}\label{section: 4}

\subsection{Overview of the hierarchical simulation setups}
\begin{table}
\begin{center}
\begin{tabular}{c c c c c c}
\hline
Hierarchical & max $M_\mathrm{\bullet}$ &  max $M_\mathrm{\bullet}$ $\mathrm{[M_\odot]}$ & $N_\bullet^\mathrm{max}$ & $N_\bullet^\mathrm{end}$ & $f_\mathrm{remain}$\\
simulation & $\mathrm{[M_\odot]}$ & ($t=10$ Myr) &  &  & \\
\hline
HS150-A & 1809 & 1809 & 4 & 3 &\\
HS150-B & 2708 & 2708 & 2 & 1 & $85\%$\\
HS150-C & 2671 & 2671 & 7 & 7 &\\
\hline
HS450-A & 5363 & 5363 & 6 & 3 & \\
HS450-B & 3601 & 3261 & 5 & 4 & $68\%$\\
HS450-C & 10052 & 894 & 8 & 6 & \\
\hline
HB150-A & 5323 & 2479 & 5 & 3 &\\
HB150-B & 3904 & 3904 & 6 & 5 & $73\%$\\
HB150-C & 6190 & 546 & 4 & 3 & \\
\hline
HT150-A & 3374 & 3374 & 9 & 6 &\\
HT150-B & 4676 & 4676 & 9 & 5 & $73\%$\\
HT150-C & 2479 & 1091 & 11 & 8 &\\
\hline
\end{tabular}
\caption{The IMBH content of the hierarchically assembled star cluster models. The fraction $f_\mathrm{remain}$ describes the percentage of formed IMBHs that still remain in the clusters at $t=10$ Myr, though not necessarily at their centres.}
\label{table: hierarchical-cluster-1}
\end{center}
\end{table}

We construct in total $12$ hierarchical massive star cluster assembly setups for this study. The simulations differ by their stellar multiplicity properties (four different options) and by their random seeds (three different random seeds per multiplicity option). We consider models with single stars only, as in \paperone, with $m_\mathrm{max,0}=\msol{150}$ for massive clusters. Next, we examine models with an initial binary population and setups with an initial triple star population as descried section \ref{section: ic-binary-triple}. In the initial binary and triple setups we also have $m_\mathrm{max,0}=\msol{150}$. Finally, we construct three hierarchical single star models with a higher IMF cut-off $m_\mathrm{max,0}=\msol{450}$. We note that the massive $m_\mathrm{max,0}=\msol{450} = 3\times\msol{150}$ single setups correspond to the most optimistic case for stellar collisions in the triple setups. The key properties of the $12$ hierarchical assembly setups are listed in Table \ref{table: hierarchical-cluster-sample}.

We run the $12$ hierarchical cluster assembly simulations for $t=10$ Myr after which most of the individual sub-clusters have merged into the final massive star cluster. In \paperone, we ran the three hierarchical setups for a longer time until $t=50$ Myr. From $t=10$ Myr to $t=50$ Myr, the central stellar and stellar BH densities decreased by a factor of $\lesssim10$ while $0$--$2$ IMBHs were ejected from the final clusters through strong Newtonian interactions and relativistic gravitational wave recoil kicks. In this study, all the assembled cluster contain at least a single IMBH at $t=10$ Myr, though not necessarily at their centres.

\subsection{Merger histories of star clusters, their VMSs and IMBHs}\label{section: mergertrees}

\subsubsection{The general picture}

Stellar multiplicity has little effect on the orbits of the infalling sub-clusters and how they disrupt and merge into the growing massive central star cluster. While the hierarchical cluster assembly process is unaffected by stellar multiplicity, initial binary and triple stars play a crucial role in the collisional formation of the massive collision products and IMBHs in the sub-clusters\footnote{We confirm by running the individual IMBH forming sub-cluster models in isolation that the cluster in-fall itself does not considerably affect the stellar collision process or IMBH masses.} as discussed in section \ref{section: role-binaries-isolated} in the context of isolated clusters. In this section, we describe in detail the build-up of multiple IMBHs in the mass range of $\sim \msol{1000} \lesssim M_\bullet \lesssim \msol{5400}$ in the hierarchical assembly simulation HB150-A which contains an initial binary population. Later, the massive star and IMBH growth histories in the other $11$ hierarchical models are discussed in further detail.

\subsubsection{The hierarchical cluster assembly simulation HB150-A with initial binaries}

At $t=0$ Myr, the cluster assembly region HB150-A consists of $711$ individual star clusters. Out of these clusters eight exceed a threshold mass of $M_\mathrm{cl}=\msol{10^4}$. In \paperone{} and in the isolated simulations of section \ref{section: isolated}, clusters with lower masses than this mass rarely form IMBHs through stellar collisions. Three initial sub-clusters are more massive than $M_\mathrm{cl}=\msol{10^5}$ while the central cluster initially has a mass of $M_\mathrm{cl} = \msol{2.3\times10^5}$.

The merger tree of the sub-clusters and the stars leading to the formation of the IMBHs is illustrated in Fig. 3 of \cite{Rantala2025}. Five sub-clusters form IMBHs above $M_\bullet \gtrsim \msol{1000}$ through stellar collisions. Three sub-clusters above the threshold mass of $M_\mathrm{cl}=\msol{10^4}$ do not form any IMBHs. Two of these three sub-clusters have a star growing by stellar collisions, but the collision products are tidally disrupted by BHs before the massive stars end their lives. The first stellar collisions in the clusters occur rapidly, before $t\lesssim0.5$ Myr, and in most cases involve the members of an initial binary system, either colliding with each other or with a third star. As opposed to single star setups, several massive stars per cluster may initially grow through collisions, but they typically rapidly merge with each other producing only a single massive object per sub-cluster. 

In the setup HB150-A, the fastest growing stars reach $m_\star \sim \msol{1000}$ by $t\sim 1.6$ Myr and $m_\star \sim \msol{2000}$ by $t\sim 2.0$ Myr. By $t=2.5$ Myr, the five sub-clusters form IMBHs in the range of $\sim \msol{1000} \lesssim M_\bullet \lesssim \msol{2550}$. As opposed to the single star setups of \paperone, the most massive formed seed black hole originates from a collision cascade in the central most massive cluster. A massive sub-cluster merges with the central cluster at $t \sim 3.0$ Myr. The central IMBHs of the two massive clusters rapidly form an IMBH binary with component masses of $M_\mathbf{\bullet,1} = \msol{2881}$ and $M_\mathbf{\bullet,2} = \msol{2665}$. The IMBH binary efficiently hardens by interacting with the stars at the cluster core, reaching a semi-major axis of $a=1$ mpc by $t=3.26$ Myr and $a=0.2$ mpc by $t=4.49$ Myr. Around $t\sim4.7$ Myr, a third massive sub-cluster merges with the central growing cluster. At this point the infalling cluster contains an IMBH with $M_\mathrm{\bullet} = \msol{1968}$. By $t\sim4.80$ Myr the third IMBH has sunk into separations of $\sim 10$ mpc from the inner binary, forming a hierarchical triple IMBH system. The eccentric outer binary $0.85 \lesssim e_\mathrm{out} \lesssim 0.9$ with its low pericentre $r_\mathrm{p,out}=a_\mathrm{out}(1-e_\mathrm{out}) \sim 1$ mpc begins to excite the orbital eccentricity of the inner binary $e_\mathrm{in}$. At this point the hierarchical triple system is only marginally stable with $r_\mathrm{p,out} / a_\mathrm{in} \sim 3.5$--$5$. At $t=4.93$ Myr the inner binary eccentricity $e_\mathrm{in}$ reaches $\sim 1$. This leads to a GW merger of the inner binary at $t=5.05$ Myr resulting in a formation of an IMBH with $M_\bullet=\msol{5323}$. However, the formed IMBH receives a relativistic GW recoil kick of $v_\mathrm{kick} \sim 310$ km s$^\mathrm{-1}$ and becomes unbound of the cluster assembly region. The tertiary IMBH subsequently sinks into the centre of the assembling central star cluster.

At $t \sim 6$ Myr, two subsequent massive sub-clusters merge with the main cluster, bringing the two final seed IMBHs of $M_\bullet = \msol{2479}$ and $M_\bullet = \msol{991}$ into the central cluster. By the end of the simulation at $t=10$ Myr, the IMBH from the more massive sub-cluster has sunk to the centre of the main cluster, forming a binary system with the central IMBH ($M_\bullet=\msol{1985}$). Meanwhile, the lowest mass seed IMBH orbits in the cluster outskirts, slowly sinking towards its central regions. This IMBH will eventually reach the centre of the cluster, initiating another triple IMBH interaction. Whether the final assembled cluster of the HB150-A simulation would retain any BHs in the IMBH mass range depends on the outcome of this interaction. In case of a very strong Newtonian interaction, the most likely outcome is the ejection of the lowest mass IMBH with $M_\bullet = \msol{991}$. This might result in the merger of the two remaining IMBHs with $M_\bullet = \msol{2479}$ and $M_\bullet=\msol{1985}$ with their merger product potentially ejected from the cluster via GW recoil. The potential final IMBH content of the cluster is therefore either an IMBH with $M_\bullet \sim \msol{4000}$, $M_\bullet \sim \msol{1000}$, or no IMBHs at all. 

As in \paperone, different random realisations of the same hierarchical cluster assembly setup yield qualitatively different outcomes at the end of the simulations. The final clusters of the other two hierarchical assembly region simulations with initial binaries, HB150-B, HB150-C, contain either a central massive IMBH binary or three IMBHs sinking in the outskirts of the cluster. In the model HB150-B, one IMBH with $M_\bullet=\msol{542}$ is ejected from the final cluster in a strong Newtonian few-body interaction with a central IMBH binary. This IMBH binary has a moderate mass ratio of $q \sim 0.25$ with IMBH masses of $M_\bullet=\msol{935}$ and $M_\bullet=\msol{3904}$ at the end of the simulation. In addition, there is a third IMBH with $M_\bullet=\msol{638}$ from a cluster merger slowly sinking at the outskirts of the cluster. The setup HB150-C produced a close to equal mass GW event with IMBH masses of $M_\bullet=\msol{3516}$ and $M_\bullet=\msol{2980}$ with the merger remnant with $M_\bullet=\msol{6190}$ being ejected out of its host cluster by a relativistic GW recoil kick. At $t=10$ Myr, three lower mass IMBHs are sinking towards the cluster centre with masses in the range of $\msol{314} \lesssim M_\bullet \lesssim \msol{546}$. These three IMBHs originate from three sub-cluster mergers, but have not yet had enough time to reach the centre of the cluster. 

\subsubsection{Hierarchical cluster assembly simulations with single stars (HS150)}

At $t=10$ Myr, the centre of the model HS150-A with an initial single star population contains an IMBH triple with BHs of $M_\bullet=\msol{1809}$ and $M_\bullet=\msol{1461 }$ forming an inner binary with a third BH with $M_\bullet=\msol{1059}$ that orbits the inner binary. In the simulation one IMBH with $M_\bullet=\msol{1235}$ becomes marginally unbound and escapes the cluster after a Newtonian few-body interaction. The model HS150-B closely resembles the simulation H2 of \paperone. The most massive cluster does not form an IMBH, an infalling cluster provides an IMBH of $M_\bullet=\msol{2708}$ for the main assembling cluster, and a lower mass IMBH ($M_\bullet=\msol{582}$) is never accreted by the main cluster due to cluster interactions. Finally, the assembling cluster of the setup HS150-C hosts a binary IMBH of $M_\bullet=\msol{1804}$ and $M_\bullet=\msol{2671}$. Several lower mass seeds reside in the outer parts of the assembled cluster. An IMBH of $M_\bullet=\msol{618}$ has a complex interaction history with the main binary, resembling the triple interactions of the simulation H1 in \paperone{}. BHs with $M_\bullet=\msol{684}$, $M_\bullet=\msol{644}$ and $M_\bullet=\msol{619}$ originating from cluster mergers have not had yet enough time to sink into the cluster centre. Finally, an IMBH of $M_\bullet=\msol{934}$ is ejected from the cluster in a strong Newtonian few-body interaction.\\

\subsubsection{Hierarchical cluster assembly simulations with massive singles (HS450) up to $\msol{450}$}

The main difference between the single star models (HS150) and setups with massive singles (HS450) is the fact that the massive single models may form BHs in the IMBH mass range above the pulsational pair-instability mass gap ($M_\bullet \gtrsim \msol{300}$) from their stars even in the absence of collisions. Thus, in the massive single models IMBHs in the mass range of $\msol{300} \lesssim M_\bullet \lesssim \msol{450}$ are somewhat more numerous (the number of IMBHs above $\msol{300}$; $N_\bullet^\mathrm{max}=5$-$8$) than in the $m_\mathrm{max,0} = \msol{450}$ single star models ($N_\bullet^\mathrm{max}=2$-$7$). Due to their higher maximum initial stellar masses, the massive star models produce the most massive IMBHs of this study.

The massive single star model HS450-A produces a massive IMBH binary with component masses $M_\bullet=\msol{5363}$ and $M_\bullet=\msol{4102}$, the two IMBHs originating from different sub-clusters. The massive binary ejects two lower-mass IMBHs out of the cluster via Newtonian interactions ($M_\bullet=\msol{1525}$ and $M_\bullet=\msol{2507}$), and is itself ejected into the cluster outskirts in the latter strong interaction. At $t=10$ Myr, the centre of the model HS450-A hosts no IMBHs. However, a low mass IMBH of $M_\bullet=\msol{301}$ and the previously kicked massive binary are sinking at the cluster outer parts. In the model HS450-B, the main cluster collisionally forms a star of $\msol{3999}$, which is tidally disrupted by an IMBH with $M_\bullet=\msol{1265}$. The resulting BH of $M_\bullet=\msol{3261}$ is retained at the core of the cluster until the end of the simulation at $t=10$ Myr. The IMBHs is accompanied by two lower mass seeds of $M_\bullet=\msol{318}$ and $M_\bullet=\msol{353}$ while another BH of $M_\bullet=\msol{305}$ sinks at the outer parts of the cluster.

The simulation HS450-C involves the most massive GW driven IMBH merger of this study. IMBHs with masses of $M_\bullet=\msol{6776}$ and $M_\bullet=\msol{3753}$ merge forming a $M_\bullet=\msol{10052}$ BH. The remnant receives a GW recoil kick of $v_\mathrm{kick}=120$ km s$^{-1}$ which marginally unbinds the massive IMBH from the cluster. We note that a recoil kick of such magnitude would not be enough to unbind the remnant from a dense, massive nuclear star cluster or a proto-bulge with a higher escape velocity. The cluster core contains no IMBHs at $t=10$ Myr, however six lower mass IMBHs from $M_\bullet=\msol{299}$ up to $M_\bullet=\msol{894}$ orbit at the outer parts of the cluster.

In summary, including massive singles up to $\msol{450}$ into the single star models has two main effects. First, there are somewhat more IMBHs forming in the IMBH mass range, either directly from stellar evolution above the mass gap, or via stellar collisions. More massive stars can sink into the centre of the cluster more rapidly and have higher collision cross sections due to their increased radii and gravitational focusing. In addition, they frequently form massive binaries with other high mass stars. This leads to the second effect: simulation models with initially more massive singles can produce $3$--$4$ times higher IMBH masses compared to the lower mass singles, up to $M_\bullet \gtrsim \msol{10^4}$ in star clusters with similar central densities and velocity dispersions. In general, more top-heavy IMF slopes have been suggested to result in increasingly massive collisionally formed IMBHs (e.g. \citealt{GonzalezPrieto2024}). Our results highlight that even ordinary Kroupa-like IMFs can lead to increasingly massive IMBHs if initial massive stars beyond $> \msol{150}$ are allowed to be sampled from the IMF. Such initially massive stars may exist in massive star formation regions as observed in the R136 cluster in the LMC \citep{Bestenlehner2020,Brands2022}.

\subsubsection{Hierarchical cluster assembly simulations with initial triple stars}

Perturbed by the star cluster potential and weak distant fly-by interactions and stellar evolution, the initial hierarchical triple systems may eventually become unstable, potentially leading to one or more mergers in the disrupting system (e.g. \citealt{Hamers2021}). In addition, the outer body may drive the inner binary to a merger through von Zeipel-Lidov-Kozai inclination-eccentricity oscillation cycles \citep{vonZeipel1910, Lidov1962, Kozai1962}. Furthermore, triple-single, triple-binary and triple-triple encounters promote chaotic Newtonian few-body interactions that boost the number of mergers in the models with initial triples.

The simulation model HT150-A forms and hosts a central IMBH binary with masses of $M_\bullet=\msol{3374}$ up to $M_\bullet=\msol{2256}$ at $t=10$ Myr. Two lower mass IMBHs with $M_\bullet=\msol{290}$ and $M_\bullet=\msol{395}$ orbit in the cluster close to the core region while another BH with $M_\bullet=\msol{671}$ sinks at the cluster outskirts. Two massive seeds with $M_\bullet=\msol{2686}$ and $M_\bullet=\msol{1530}$ are ejected from the final cluster, the former after a GW merger with an IMBH just above the mass gap and the latter after a strong Newtonian interaction with the central binary.

The core of the setup HT150-B contains in total three IMBHs with masses $M_\bullet=\msol{529}$, $M_\bullet=\msol{652}$ and $M_\bullet=\msol{4676}$ in a wide configuration. In addition, two IMBHs with $M_\bullet=\msol{413}$ and $M_\bullet=\msol{447}$ orbit in the outer parts of the cluster. Two IMBHs are ejected from the assembling cluster: a $M_\bullet=\msol{585}$ BH and its cluster are not accreted by the main system due to cluster interactions, and a $M_\bullet=\msol{1048}$ BH is ejected after a strong few-body interaction with the central IMBH system.

In the model HT150-C at $t=9.3$ Myr, the central massive IMBH binary of the cluster with component masses of $M_\bullet = \msol{1117}$ and $M_\bullet = \msol{2669}$ merge, producing an IMBH of $M_\bullet = \msol{3657}$. This merger remnant received a recoil kick of $v_\mathrm{kick}=172$ km s$^{-1}$, unbinding it from the host cluster. At $t=10$ Myr the setup HT150-C has a three IMBHs with masses of $M_\bullet=\msol{933}$, $M_\bullet=\msol{1091}$ and $M_\bullet=\msol{454}$ in the final cluster, though outside its core region. In addition, several IMBHs in the mass range of $\msol{330} \lesssim M_\bullet \lesssim \msol{999}$ are marginally bound to the cluster at large separations. In summary, hierarchical models including initial triples form IMBHs with similar masses as in the initial binary models. However, the enhanced collision rates in the triple models increase the number of formed IMBHs as shown in Table \ref{table: hierarchical-cluster-1}.

\subsection{The collisional growth histories of very massive stars and IMBHs}

\subsubsection{Stellar collision histories, masses and orbits}

\begin{figure*}
\includegraphics[width=0.9\textwidth]{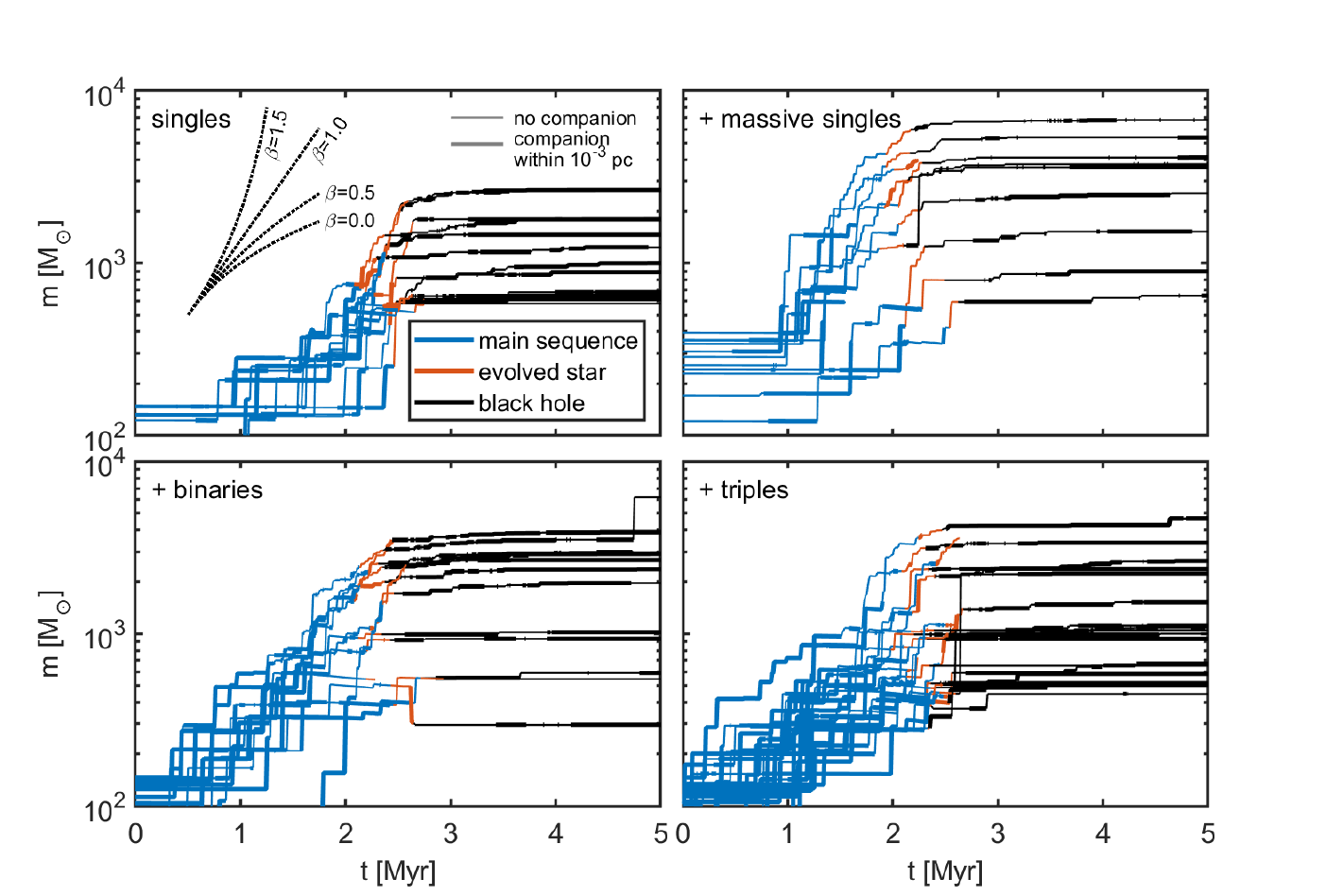}
\caption{The mass assembly histories of selected IMBHs and their VMS progenitors that reached masses above $\gtrsim \msol{500}$ during the first $5$ Myr of their evolution. The collisional growth is rapid and usually truly runaway ($\beta>0$) between $0.5$ Myr $\lesssim t \lesssim 2.5$ Myr. First stellar collisions occur earlier in models with initial stellar multiplicity (binaries and triples). Most collisions occur in the main sequence phase of the stars and are associated with stars having close companions within $\lesssim 10^{-3}$ pc as indicated by the thick line-style. After $t\gtrsim5$ Myr the IMBHs mainly grow by IMBH-IMBH mergers.}
\label{fig: vms-growth}
\end{figure*}

We present the detailed mass growth histories the most massive IMBHs and their stellar progenitors in Fig. \ref{fig: vms-growth}. Not all collisionally grown stars and massive BHs are shown: we focus on objects that reached $\gtrsim \msol{500}$ in mass through one or more collisions. In our simulations, runaway growth by stellar collisions occurs, i.e.
\begin{equation}\label{eq: mdot}
    \derfrac{m}{t} = \dot{m}_\mathrm{0} \left( \frac{m}{m_\mathrm{0}} \right)^\beta
\end{equation}
with $\beta>0$, i.e. each collision increases the collision rate. The solution of Eq. \eqref{eq: mdot} for $t\gg0$ Myr is $m(t) \propto t$ for $\beta=0$, $m(t) \propto t^\mathrm{2}$ for $\beta=0.5$ and $m(t) \propto \exp{(t)}$ for $\beta=1.0$. We show mass growth curves with $m_\mathrm{0} = \msol{500}$ and $\dot{m}_\mathrm{0} = 10^\mathrm{-3} M_\mathrm{\odot}$ yr$^{-1}$ for $0 \lesssim \beta \lesssim 1.50$ in the top-left panel of Fig. \ref{fig: vms-growth}. This corresponds to repeated collisions of $\msol{10}$ stars with a rate of $\Gamma_\mathrm{coll} = 10^{-4}$ yr$^{-1}$, similar to the peak stellar collision rates in our runs. The comparison shows that the our runaway collision cascades can reach growth phases with $\beta \sim 1$, e.g. exponential runaway growth. However, the phases for the most rapid growth are brief and only last for $\lesssim 0.5$ Myr.

The first collisions in the runaway cascades leading to IMBHs occur earlier if initial binaries and especially triples are included in the models. The increased interaction cross sections of multiple systems (binary semi-major axis $a$ vs stellar radius $R_\star$) allow for more frequent early interactions before the mass segregation and core collapse of the host cluster is complete. Both the single star models and the setups with massive singles require $\sim 1$ Myr for the first massive star collisions to occur. As discussed in \paperone, at this point the clusters are already mass segregated but not necessarily core collapsed. In the binary models, the first massive star collisions take place at $t\sim0.3$ Myr and in the triple star models already within the first $0.1$ Myr of the simulation. The times of the first collisions are typically associated with the stars having binary companions within $a<10^{-3}$ pc. This highlights the crucial role of stellar multiplicity in collisional IMBH formation. The runaway growth index $\beta$ of Eq. \eqref{eq: mdot} is initially low in the triple models as the first early collisions occur in interacting triple systems. Exponential growth phases ($\beta=1.0$) are reached later ($t>1.5$ Myr) when the clusters mass segregate and core collapse.

The progenitors for the IMBHs rapidly grow during the first $2$--$3$ Myr of the simulations, and the IMBHs can reach masses up to $M_\bullet \sim \msol{6700}$ in $\sim 3$ Myr mostly via stellar collisions and TDEs. Further IMBH growth at $t>5$ Myr up to masses of $M_\bullet \gtrsim \msol{10^4}$ is fuelled by GW driven IMBH-IMBH mergers. For seeds above $\msol{1000}$, the mass of the IMBH and its massive progenitor star is built up in total $\sim 10$--$600$ collisions. Most of these are star-star collisions. As in \paperone, the mass function of the stars that collide with the main IMBH progenitor strongly deviates from a typical IMF. While following a Kroupa-like mass function at $m_\mathrm{\star} \lesssim \msol{5}$, high mass stars with $m_\star \gtrsim \msol{50}$ are strongly overrepresented in the colliding star mass function due to mass segregation of their host clusters. Most stellar mass is contributed to the IMBH progenitor star in these high mass collisions. In the models with single stars, $50\%$ of the IMBH mass originates from collisions with stars more massive than $m_\mathrm{50\%} = \msol{109}$. In models with massive singles, binaries and triples, even more mass is acquired in a relatively small number of massive collisions. For massive single models $m_\mathrm{50\%} = \msol{225}$ while for the binary and triple models $m_\mathrm{50\%} = \msol{128}$ and $m_\mathrm{50\%} = \msol{139}$, respectively.

As in \paperone, most stellar collisions in our models occur from almost parabolic orbits. Typical collision velocities are over an order of magnitude higher than the star cluster velocity dispersions, highlighting the importance of mergers from wide, bound and extremely eccentric orbits for the IMBH progenitor growth. Even though the geometry of single-binary encounters alters the velocity at stellar collisions compared to single only models \citep{Gaburov2008}, we do not find evidence for an excess of low velocity collisions in our models with initial binary and triple stars.

\begin{figure*}
\includegraphics[width=0.9\textwidth]{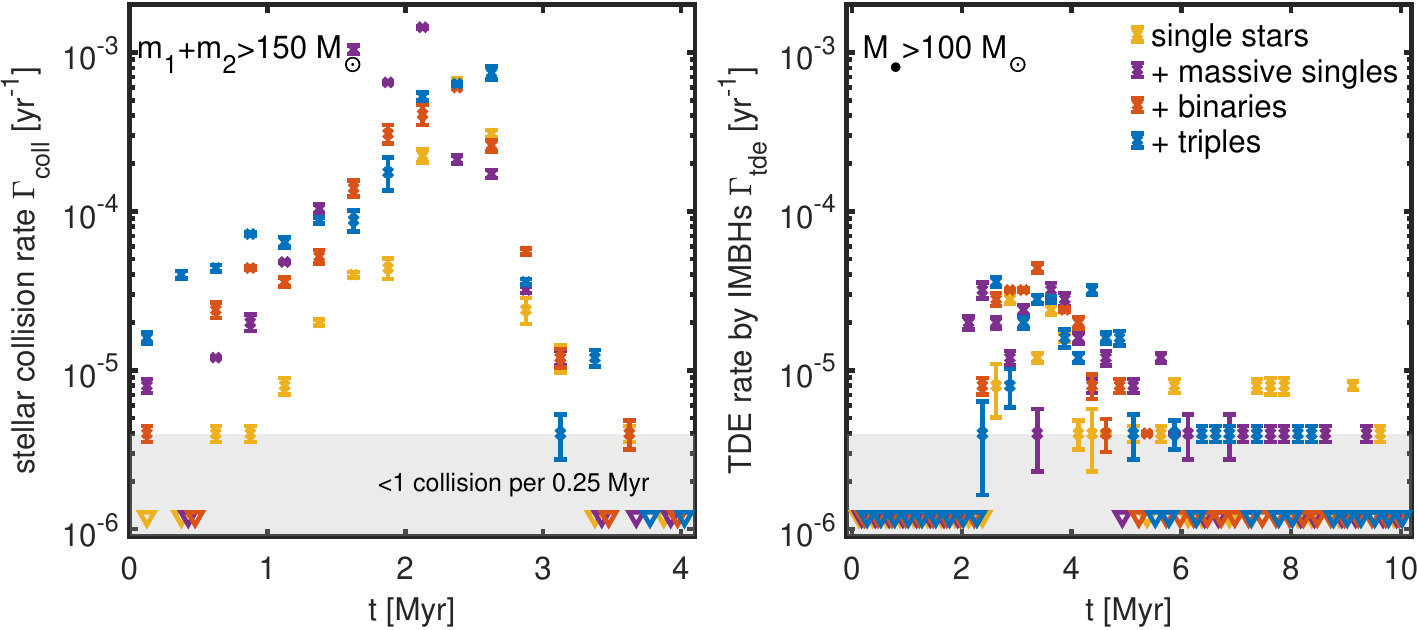}
\caption{Left panel: the total VMS stellar collision rates ($m_\mathrm{1}+m_\mathrm{2}>\msol{150}$) in the hierarchical simulations in time bins of $0.25$ Myr. The peak VMS collision rates $\Gamma_\mathrm{coll} \sim 6\times10^{-4}$--$1.4\times10^{-3}$ yr$^{-1}$ are reached near the end of the lifetimes of the collisionally grown stars at $2$ Myr $\lesssim t \lesssim 2.5$ Myr. Stellar multiplicity and massive singles lead to higher collision rates especially at early times ($t\lesssim 1.5$ Myr) compared to the single star setups. Right panel: the tidal disruption event rates by IMBHs ($M_\bullet>\msol{100}$) in the hierarchical simulations in time bins of $0.25$ Myr. The peak TDE rates of $\Gamma_\mathrm{tde} \sim3$--$4.5\times10^{-5}$ yr$^{-1}$ occur right after the formation of the IMBHs. Overall, stellar multiplicity has only a small effect on the TDE rates.}
\label{fig: collrate}
\end{figure*}

\subsubsection{Stellar collision rates}

The stellar collision rates $\Gamma_\mathrm{coll}$ in the hierarchical setups are presented in the left panel of Fig. \ref{fig: collrate} as a function of time. The peak stellar collision rates involving the growing massive stars ($m_\mathrm{1}+m_\mathrm{2}>\msol{150}$) of $\Gamma_\mathrm{coll} \sim 6\times10^{-4}$ yr$^{-1}$ -- $1.4 \times 10^{-3}$ yr$^{-1}$ occur between $2.0$ Myr $\lesssim t \lesssim 2.5$ Myr in the simulations. The highest instantaneous collision rates are comparable in the models with singles ($\Gamma_\mathrm{coll} \sim 6.4\times 10^{-4}$ yr$^{-1}$) and binary stars ($\Gamma_\mathrm{coll} \sim 6.0\times 10^{-4}$ yr$^{-1}$) but somewhat higher in the models with triples ($\Gamma_\mathrm{coll} \sim 7.4\times 10^{-4}$ yr$^{-1}$) and massive singles ($\Gamma_\mathrm{coll} \sim 1.4\times 10^{-3}$ yr$^{-1}$). These high collision rates are not sustained over long timescales. The rates rapidly decline after $\sim 3$ Myr as the massive stars end their lives and the central densities $\rho_\mathrm{c}$ of their host clusters decrease.

In the simulations with singles, the collisionally growing objects gain $43\%$ of their mass while being in the main sequence (see Fig. \ref{fig: vms-growth}), $42\%$ in the evolved star phase and the remaining $15\%$ as a BH during the first $5$ Myr. In the setups with massive singles and binaries the corresponding fractions of the mass growth are $58$--$59\%$ (in the main sequence), $21$--$22\%$ (as an evolved star) and $19$--$20\%$ (as a BH). In the models with triple stars, somewhat more mass is acquired by the growing stars in the main sequence ($68\%$) compared to the single and binary models. In the evolved star and BH phases the mass growth fractions are $18\%$ and $14\%$ for the triple models, respectively.

\subsubsection{TDEs and GW driven BH mergers}

After their formation, the IMBHs may subsequently grow by tidally disrupting main sequence stars as well as evolved stars (see e.g. \citealt{Stone2017}). Most mass of the most massive collisionally growing objects in acquired through stellar collisions (up to $85\%$). Tidal disruption events account for at least $\sim 15\%$ of the total mass budget. This result depends on our assumed TDE accretion fraction of $50\%$. We estimate that for accretion fractions of $10\%$ and $100\%$ the corresponding TDE portions of the total IMBH mass budget would be $\sim3\%$ and $\sim26\%$, respectively. GW driven mergers with stellar BHs play a minor role ($<1\%$) for the growth during the first $10$ Myr. Most of the TDEs occur soon ($t\sim2.25$--$3.75$ Myr) after the formation of the IMBHs with peak TDE rates $\Gamma_\mathrm{tde} \lesssim$ $3$--$5 \times 10^{-5}$ per assembled cluster. This is illustrated in the right panel of Fig. \ref{fig: collrate}. The peak TDE rates are by a factor of $\sim35$ lower than peak stellar collision rates in our hierarchical models. At later times ($t\gtrsim 4$--$5$ Myr), the central density of the formed cluster decreases and the TDE rate accordingly decreases approximately by an order of magnitude. Overall, the TDE rates by IMBHs seem to be little affected by stellar multiplicity in the models. During the $10$ Myr of the simulations, IMBH growth by GW mergers with stellar mass BHs is not very important with a single IMBH-BH merger occurring in the run HB150-A. Individual IMBHs may grow by factor of up to $\sim 2$ in close to equal mass GW mergers in the IMBH mass range. As briefly discussed in \cite{Rantala2025}, these IMBH-IMBH mass GW mergers may provide an important growth channel for SMBH seeds without any electromagnetic (AGN or TDE) accretion signature, if the merger remnants are retained in their dense merger environments.

\subsection{Forming and retaining IMBHs in early star clusters}

In general, the infalling star clusters with $M_\mathrm{cl}\lesssim \msol{10^4}$ do not form collisional IMBHs in the hierarchical setting. Even though they have shorter mass segregation and core collapse time-scales, their initial densities are lower and the initially most massive stars less massive compared to $M_\mathrm{cl}>\msol{10^4}$ clusters. The result also holds in the simulation sets with massive singles, binaries and triples. The probability $f_\bullet$ for a star cluster to form a collisional IMBH rapidly increases with increasing cluster mass. Approximately $\sim50\%$ of the infalling sub-clusters with masses of $M_\mathrm{cl}\sim2$--$3\times \msol{10^4}$ can form a BH with $M_\bullet>\msol{100}$. For clusters with $M_\mathrm{cl}\gtrsim \msol{10^5}$ the probability to form a $M_\bullet>\msol{100}$ BH is already $f_\bullet \sim 80\%$--$100\%$. As each of our hierarchical setup contains only of the order of $\sim10$ star clusters with $M_\mathrm{cl}\gtrsim \msol{10^4}$, we refrain from performing a more detailed analysis of the exact functional form of $f_\bullet(M_\mathrm{cl})$ at $\msol{10^4} \lesssim M_\mathrm{cl}\lesssim \msol{10^5}$.

We count the maximum number of IMBHs exceeding $\msol{300}$ in mass, $N_\bullet^\mathrm{max}$, at any time in the simulations and count the number of BHs above this mass threshold at $t=10$ Myr, $N_\bullet^\mathrm{end}$. The ratio $f_\mathrm{remain} = N_\bullet^\mathrm{end} / N_\bullet^\mathrm{max}$ is the fraction of formed per remaining IMBHs at the end of the simulation. We have collected $N_\bullet^\mathrm{max}$ and $N_\bullet^\mathrm{end}$ from each of the $12$ hierarchical assembly simulations in Table \ref{table: hierarchical-cluster-1}. We show the value of $f_\mathrm{remain}$ collectively for our four different stellar multiplicity types as the number of individual formed, retained and ejected IMBHs is relatively small in a single simulation $(\lesssim10)$. 

The hierarchical single star models HS150 form $N_\mathrm{max}=2$-$7$ IMBHs per simulation, and retain on average $f_\mathrm{remain}=11/13 \approx85\%$ of them by $t=10$ Myr. The models HS450 with massive singles already include a number of stars with masses of $m_\star>\msol{300}$ at $t=0$ Myr, leading to a somewhat larger number of formed IMBHs, $N_\bullet^\mathrm{max}=5$-$8$. As the most massive stars in the simulations HS450 are more massive than their counterparts in HS150, they reach the cluster centres faster and their mutual few-body interactions are stronger. These interactions lead to more efficient mergers and ejections of massive stars from the cluster models. As such the fraction $f_\mathrm{remain}=13/19\approx68\%$ is lower with massive singles compared to the HS150 single star models.

The hierarchical models HB150 with initial binary stars form $N_\bullet^\mathrm{max}=4$-$6$ IMBHs above $\msol{300}$, and on average retain $f_\mathrm{remain}=11/15 \approx 73\%$ of them by $t=10$ Myr. For the models with initial triples the corresponding values are $N_\bullet^\mathrm{max}=9$-$11$ and $f_\mathrm{remain}=19/26 \approx 73\%$, a similar fraction as in the binary setups. Compared to the $2$--$7$ IMBHs in the $m_\mathrm{max,0}=\msol{150}$ single star models, the triple setups produce $9$--$11$ IMBHs. This further highlights the importance of the initial stellar multiplicity to collisional IMBH formation. Overall, our hierarchical models produce $\sim0.2$--$1.1$ IMBHs per $\msol{10^5}$ of clustered stellar mass. In comparison, isolated models are expected to form $\sim0.1$ IMBHs per $\msol{10^5}$ of clustered star formation. Finally, we note that all IMBHs present in the assembled clusters at $t=10$ Myr are not likely be retained over Gyr timescales. Especially $M_\bullet \lesssim \msol{1000}$ IMBHs can be easily ejected from their host clusters through GW recoil kicks after mergers with stellar BHs (e.g. \citealt{ArcaSedda-DRAGON2a, Rantala2024b}).

\section{Structure, kinematics and stellar content of young, massive star clusters}\label{section: 5}

\subsection{Density profiles of the assembled clusters}

\renewcommand{\arraystretch}{1.2}
\begin{table*}
\begin{center}
\begin{tabular}{c c c c c c c c c c c c}
\hline
Hierarchical & $M_\mathrm{cl}^\mathrm{orig}$ & $M_\mathrm{cl}^\mathrm{cutoff}$ & $r_\mathrm{h}$ & $\rho_\mathrm{c,\star}$ & $r_\mathrm{b,\star}$ & $\beta_\mathrm{\star}$ & $\gamma_\star$ & $\rho_\mathrm{c,\bullet}$ & $r_\mathrm{b,\bullet}$ & $\beta_\mathrm{\bullet}$ & $\gamma_\bullet$\\
simulation & $[\msol{10^5}]$ & $[\msol{10^5}]$ & [pc] & $[\rhosol{10^5}]$ & [pc] & & &  $[\rhosol{10^5}]$ & [pc] & &\\
\hline
HS150-A & 7.62 & 5.77 & 3.52 & 3.51 & 0.58 & 3.23 & 0.33 & 11.94 & 0.22 & 3.14 & 0.89\\
HS150-B & 7.75 & 5.29 & 4.02 & 12.44 & 0.36 & 3.07 & 0.47 & 89.25 & 0.10 & 3.03 & 1.32\\
HS150-C & 7.70 & 5.26 & 2.89 & 5.60 & 0.41 & 3.16 & 0.29 & 9.15 & 0.10 & 3.05 & 0.31\\
\hline
HS450-A & 7.42 & 5.49 & 4.39 & 0.34 & 0.83 & 3.28 & 0.00 & 0.21 & 0.35 & 3.17 & 0.00\\
HS450-B & 7.63 & 5.16 & 4.41 & 5.45 & 0.51 & 3.11 & 0.46 & 17.74 & 0.18 & 3.04 & 1.04\\
HS450-C & 7.50 & 5.04 & 3.66 & 0.55 & 0.64 & 3.17 & 0.00 & 0.49 & 0.31 & 3.10 & 0.22\\
\hline
HB150-A & 7.48 & 5.35 & 3.59 & 2.09 & 0.59 & 3.21 & 0.25 & 4.61 & 0.22 & 3.13 & 0.64\\
HB150-B & 7.19 & 5.07 & 4.48 & 10.45 & 0.37 & 2.89 & 0.57 & 15.61 & 0.10 & 2.95 & 0.82\\
HB150-C & 6.92 & 5.06 & 2.77 & 6.00 & 0.29 & 2.99 & 0.17 & 15.74 & 0.10 & 3.12 & 0.48\\
\hline
HT150-A & 7.39 & 5.40 & 4.14 & 2.39 & 0.79 & 3.23 & 0.44 & 1.04 & 0.26 & 3.11 & 0.27\\
HT150-B & 7.52 & 4.91 & 4.71 & 4.27 & 0.54 & 3.10 & 0.45 & 4.06 & 0.16 & 3.01 & 0.49\\
HT150-C & 7.50 & 5.06 & 3.35 & 1.90 & 0.40 & 3.11 & 0.00 & 3.17 & 0.13 & 3.04 & 0.12\\
\hline
\end{tabular}
\caption{The double power-law density profiles from Eq. \eqref{eq: nuker} of the final assembled hierarchical clusters at $t=10$ Myr. The columns list the cluster masses without ($M_\mathrm{cl}^\mathrm{orig}$) and with ($M_\mathrm{cl}^\mathrm{cutoff}$) the tidal cut-off (see the main text for details), the half-mass radii $r_\mathrm{h}$, central densities $\rho_\mathrm{c}$ as well as outer and inner power-law slopes $\beta$ and $\gamma$ for stars and stellar BHs.}
\label{table: hierarchical-cluster-2}
\end{center}
\end{table*}
\renewcommand{\arraystretch}{1.0}

\begin{figure*}
\includegraphics[width=0.9\textwidth]{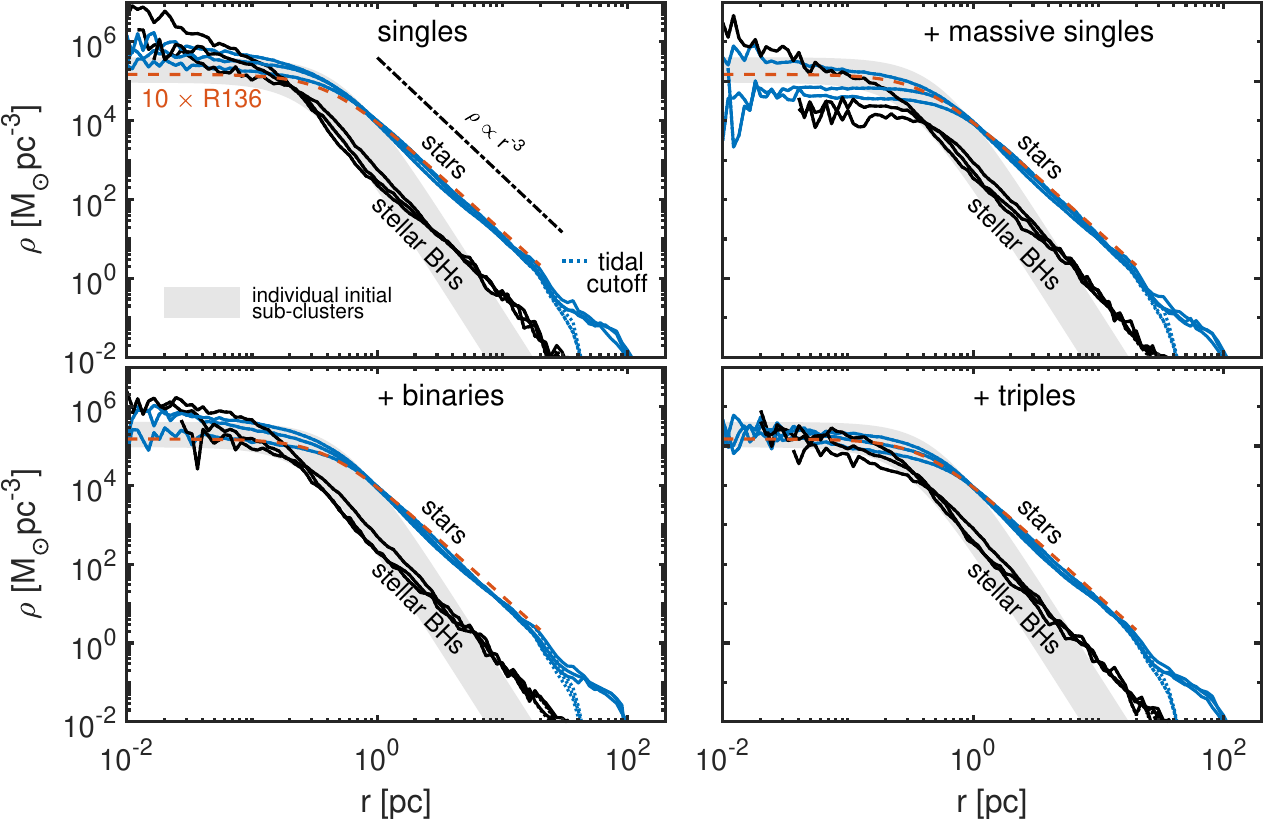}
\caption{The 3D mass density profiles for stars and stellar BHs of the assembled hierarchical clusters at $t=10$ Myr. BHs in the IMBH mass range are excluded from the analysis. Double power-law fits of the profiles are provided in Table \ref{table: hierarchical-cluster-2}. The 3D density profiles follow a slope of $-3$ in the outer parts, consistent with the R136 profile and literature simulation results. The clusters can reach $\rhosol{10^5}$--a few times $\rhosol{10^6}$ in their central density and show more diversity in their central structure compared to their outer parts, especially for the BH density profiles. We connect these central profile variations to the cluster assembly and IMBH interaction histories during the first $10$ Myr. The density profile range of the initial individual sub-clusters is displayed as the shaded background. The density profile of the young massive LMC cluster R136 multiplied by a factor of $10$ is shown as the dashed line. The tidal cut-off above $r>37$ pc is shown as the dotted line in the stellar profiles. See the main text for further details.}
\label{fig: densityprofiles}
\end{figure*}

\subsubsection{Double power law fits}\label{section: densityprofiles}

We present the final 3D density profiles $\rho(r)$ of the assembled clusters at $t=10$ Myr in Fig. \ref{fig: densityprofiles} for the single star setups and models with massive singles, initial binaries and triples. We distinguish the stellar component and stellar mass BHs in the analysis. BHs in the IMBH mass range with $M_\bullet>\msol{100}$ are excluded from the BH density profile. Other compact remnants, neutron stars and white dwarfs, have not had time to form by $t=10$ Myr in numbers.

At $t=10$ Myr the hierarchically assembled star clusters have density profiles $\rho(r)$ relatively well described by a double power law model (e.g. \citealt{Saha1992, Zhao1996}). Such models are commonly used to characterise surface density profiles of both star clusters (Elson-Fall-Freeman EFF profile; \citealt{Elson1987}) and galactic nuclei (the Nuker model; \citealt{Lauer1995}). For our double power-law fits we use a 3D Nuker-like density profile defined as
\begin{equation}\label{eq: nuker}
    \rho(r) = 2^\mathrm{\frac{\beta-\gamma}{\alpha}} \rho_\mathrm{b} \left(\frac{r}{r_\mathrm{b}} \right)^{-\gamma} \left[ 1 + \left(\frac{r}{r_\mathrm{b}} \right)^\alpha \right]^\mathrm{\frac{\gamma-\beta}{\alpha}}
\end{equation}
in which $r_\mathrm{b}$ is the break radius between the inner ($\gamma$) and outer ($\beta$) power-law slopes and $\rho_\mathrm{b}$ is the density at break radius. Compared to the EFF profile, the double power law Nuker density profile allows for a non-zero central power-law instead of being restricted to a flat $\gamma=0$ core. The parameter $\alpha$ controls how sharp the transition between the two slopes is, and we fix its value to $\alpha=3$, resulting in a sharp break between the inner and outer power-laws. Instead of the density at break radius $\rho_\mathrm{b}$ we express the overall profile normalisation using the central density $\rho_\mathrm{c}$ which we define at $r=0.01$ pc. The density profile fits are performed in the radial range of $0.01$ pc $\lesssim r \lesssim 200$ pc. The density profile fit parameters for the stars ($\rho_\mathrm{c,\star}$, $r_\mathrm{b,\star}$, $\beta_\star$, $\gamma_\star$) and stellar BHs ($\rho_\mathrm{c,\bullet}$, $r_\mathrm{b,\bullet}$, $\beta_\bullet$, $\gamma_\bullet$) are displayed in Table \ref{table: hierarchical-cluster-2} for all the $12$ hierarchical models.

The outer power-law slopes $\beta$ of stellar and BH density profiles are in the range of $2.89 \lesssim \beta_\star \lesssim 3.28$ (stars) and $2.95 \lesssim \beta_\bullet \lesssim 3.17$ (stellar BHs). The $\beta\sim3$ profiles continue well beyond $r=100$ pc until the simulation escaper removal radius of $r_\mathrm{esc}=200$ pc. The outer parts of the cluster beyond a few times $10$ pc exhibit a diffuse low-density ($\lesssim \rhosol{1}$) envelope. In reality, such loosely bound stellar material would be stripped away by the tidal field of the host galaxy. In our models, we did not include a tidal background for simplicity, however, this simplified assumption leads to overestimation of the final cluster masses $M_\mathrm{cl}$ and especially half-mass radii $r_\mathrm{h}$ due to the diffuse extended cluster outer parts. For this reason we introduce a tidal cut-off radius of $r_\mathrm{t} \sim37$ pc. This value for $r_\mathrm{t}$ is the expected tidal radius of a $M_\mathrm{cl}=\msol{5\times10^5}$ cluster orbiting at $r=1$ kpc in a proto-galaxy of $M=\msol{10^{10}}$. Supporting this estimate for $r_\mathrm{t}$, the recent low metallicity dwarf star-burst star cluster formation simulations of \cite{Lahen2025b} show a cut-off like feature above $r\gtrsim10$ pc in their density profile of a young massive proto-cluster of $M_\mathrm{cl}\sim \msol{2\times10^5}$. Including the tidal cut-off, the masses and half-mass radii differ from the original values by $r_\mathrm{h}^\mathrm{orig}/r_\mathrm{h}^\mathrm{cutoff} \sim 2.5\pm0.4$ and $M_\mathrm{cl}^\mathrm{orig}/M_\mathrm{cl}^\mathrm{cutoff} \sim 1.4\pm0.1$. Unless otherwise stated we hereafter use the $M_\mathrm{cl}^\mathrm{cutoff}$ and $r_\mathrm{h}^\mathrm{cutoff}$ as the star cluster mass $M_\mathrm{cl}$ and half-mass radius $r_\mathrm{h}$ in the further analysis of this study. Both the total and tidal cut-off masses for the hierarchically assembled clusters are listed in Table \ref{table: hierarchical-cluster-2}.

\subsubsection{Central structure of assembled star clusters}

In Fig. \ref{fig: densityprofiles}, the centres of the star clusters show more structural differences (both flat cores and mild cusps) than their outer parts. Still, a number of general remarks can be made about the central structure. More centrally cuspy (larger $\gamma$) clusters have larger central densities $\rho_\mathrm{c}$ and smaller break radii $r_\mathrm{b}$. Clusters with larger half-mass radii $r_\mathrm{h}$ also have larger break radii $r_\mathrm{b}$. Clusters with a central stellar cusp and high stellar density also host central BH cusps and have higher central BH densities. The central density profile slopes $\gamma$ range from flat cores ($\gamma=0$) to mild and moderate stellar and BH cusps $\max{(\gamma_\star)}=0.57$ and $\max{(\gamma_\bullet)}=1.32$. We note that the final assembled clusters, many of which have central IMBHs at $t=10$ Myr, have not yet had enough time to relax into a steep Bahcall-Wolf cusp \citep{Bahcall1976,Bahcall1977,Preto2004}. For such single-mass and multi-mass cusps central power-law slopes in the range of $1.50 \lesssim \gamma \lesssim 2.00$ would be expected depending on the dominating mass component. Overall, the fact that most clusters at least briefly host multiple IMBHs at their centres likely complicates the Bahcall-Wolf cusp formation on $<10$ Myr time-scales.

The central stellar BH profiles slopes are more cuspy than the stellar slopes with all but one cluster satisfying $1 \lesssim \gamma_\bullet / \gamma_\star \lesssim 3$. The stellar break radii are in the range of $0.29$ pc $\lesssim r_\mathrm{b,\star} \lesssim 0.83$ pc while for the stellar BHs $0.10$ pc $\lesssim r_\mathrm{b,\bullet} \lesssim 0.35$. The stellar break radii are larger than the corresponding BH break radii with $0.24$ pc $\lesssim r_\mathrm{b,\bullet}/r_\mathrm{b,\star} \lesssim 0.48$. For flat non-cuspy models this implies that the BH cores of young star clusters are smaller than their stellar cores. The central densities of the star clusters span a wide range of $\rhosol{3.4\times10^5} \lesssim \rho_\mathrm{c,\star} \lesssim \rhosol{1.2 \times 10^6}$ for the stars and $\rhosol{2.1\times10^5} \lesssim \rho_\mathrm{c,\bullet} \lesssim \rhosol{ 6.4\times 10^6}$ for the stellar BHs. In general, the central parts of the clusters can be BH dominated due to mass segregation. This is seen in especially in the single star models.

We now wish to connect the structural properties of especially the inner cluster regions to their stellar multiplicity and IMBH content as well as their formation histories. From models with single stars to binaries and triples, the stellar and BH densities moderately decrease with increasing stellar multiplicity. The central stellar BH density slope also decreases but the central stellar profile slopes do not show such a trend. The half-mass size of the assembled clusters are not affected by their stellar multiplicity contents. We attribute these central structural trends to the (temporary) presence of IMBH binaries and multiples at the cluster centres. As shown in Table \ref{table: hierarchical-cluster-1}, our models with higher stellar multiplicity form on average a larger number of IMBHs, and they are on average more massive. As the massive IMBH binaries and triples interact with the central cluster BH and stellar components in an IMBH mass range variant of the stellar core scouring process in massive galaxies (e.g. \citealt{Hills1980,Quinlan1997,Milosavljevic2001,Merritt2006,Rantala2018}). Stellar BHs mass segregated into the cluster centres on radial orbits are subsequently ejected from their host clusters by the $\gtrsim \msol{2000}$ IMBH binaries and triples. As shown in binary star cluster merger simulations, scouring effects by $\lesssim \msol{1000}$ IMBHs are not generally very strong (see e.g. \citealt{Souvaitzis2025}).

\subsubsection{Density profiles: comparison to observed clusters and hydrodynamical simulations}

We proceed to briefly compare our initial and final star cluster density profiles to observed massive star clusters. Star clusters of high ($\rho_\mathrm{c} \gtrsim \msol{10^6}$) central densities seem to be rare in the local Universe except in extreme star-burst environments (e.g. \citealt{McCrady2007,Levy2024}). For comparison purposes we adopt the density profile \citep{Selman2013} for young massive star cluster R136 in the 30 Doradus complex in the LMC and scale it up by a factor of $10$, $\rho(r) = 10\times\rho_\mathrm{R136}(r)$, as our assembled clusters are $5$--$10$ times more massive than R136 \citep{Selman2013}. This allows us to compare the profile shapes of our hierarchically assembled models against the observed young cluster. Fig. \ref{fig: densityprofiles} shows that the agreement between our hierarchical models and the $10\times$ R136 profile is very good, especially in the outer parts of the cluster where the R136 profile has a slope of $\beta_\mathrm{R136} = 2.85$.

This outer density profile slope of $\sim -3$ appears to be a robust outcome from repeated mergers of star clusters. Our result supports previous results in the literature. Studying nuclear star cluster formation via globular cluster infall up to $12$ consecutive cluster mergers, \cite{Antonini2012a} found that the final NSC profile shape is well described by the modified Hubble profile \citep{Rood1972} which also has an outer density profile slope of $-3$. Moreover, \cite{Guszejnov2018} and \cite{Grudic2018} suggest using theoretical arguments and numerical simulations that $\sim -3$ will be reached after a large number of mergers even starting from Plummer-like or steeper outer profiles. An outer power-law profile slightly shallower than $-3$ was also recently reported in a star-by-star hydrodynamical simulation of the formation of a massive $M_\mathrm{cl} \sim \msol{2 \times 10^5}$ in its galactic environment \citep{Lahen2025b}. The outer density profile slope of $-3$ seems to be essentially independent of the stellar multiplicity properties and details of the cluster merger histories. However, our results potentially indicate that there is no similar universal central density structure for hierarchically assembled young massive star clusters, i.e. for a given $M_\mathrm{cl}$ the inner slopes $\gamma$ and the central densities $\rho_\mathrm{c}$ for the stellar and BH components can have a range of values. Nevertheless, the central stellar profiles are in general shallow with $0 \lesssim \gamma_\mathrm{\star} \lesssim 0.57$. This is consistent with the latest hydrodynamical results of \cite{Lahen2025b} that also include collisional dynamics around massive stars, finding a shallow $\gamma_\mathrm{\star}\sim0.5$ for their central number density profiles. Crucially, the central cluster properties at early times seem to somewhat depend on the details of the cluster formation process. IMBH interactions can shape the central structure of their host clusters, including the central stellar and BH density and the inner density profile slope. Increasing stellar multiplicity increases the number and mean masses of the formed IMBHs, which form binaries and triple systems scouring the central stellar and especially stellar BH cusps. Finally, our results indicate that stellar BH cusps can co-exist in young massive star cluster centres with one or multiple IMBHs, at least on timescales of the order of $10$ Myr. 

\begin{table*}
\begin{center}
\begin{tabular}{c c c c c c c c c c}
\hline
Hierarchical & $R_\mathrm{e}$ & $\sigma_\mathrm{1D}$ & $\max(|V_\mathrm{los}|)$ & $\frac{V}{\sigma}$ & $\Sigma_\mathrm{c} [10^4$ & $\Sigma_\mathrm{h} [10^4$ & $\frac{b}{a}$ & $\frac{c}{a}$ & $T$\\
simulation & [pc] & [km s$^{-1}$] & [km s$^{-1}$] & & $\rhosol{}]$ & $[\rhosol{}]$ & & &\\
\hline
HS150-A & 2.79 & 15.23 & $3.68$ & $0.24$ & 4.05 & 1.44 & $0.94$ & $0.83$ & $0.36$\\ 
HS150-B & 3.11 & 15.49 & $2.95$ & $0.19$ & 6.51 & 1.04 & $0.99$ & $0.86$ & $0.04$\\
HS150-C & 2.29 & 15.90 & $2.85$ & $0.18$ & 5.75 & 1.88 & $0.95$ & $0.85$ & $0.34$\\
\hline
HS450-A & 3.40 & 12.44 & $2.99$ & $0.24$ & 1.63 & 0.91 & $0.96$ & $0.81$ & $0.26$\\
HS450-B & 3.49 & 13.92 & $2.09$ & $0.15$ & 3.92 & 0.81 & $0.99$ & $0.87$ & $0.04$\\
HS450-C & 2.90 & 13.11 & $3.00$ & $0.23$ & 2.54 & 1.13 & $0.99$ & $0.85$ & $0.05$\\
\hline
HB150-A & 2.97 & 16.18 & $2.87$ & $0.18$ & 3.28 & 1.19 & $0.93$ & $0.84$ & $0.44$\\
HB150-B & 3.54 & 16.28 & $2.71$ & $0.17$ & 4.55 & 0.76 & $0.99$ & $0.88$ & $0.08$\\
HB150-C & 2.27 & 17.62 & $4.20$ & $0.24$ & 6.33 & 1.84 & $0.99$ & $0.86$ & $0.06$\\
\hline
HT150-A & 3.34 & 14.81 & $3.69$ & $0.25$ & 2.41 & 0.94 & $0.97$ & $0.85$ & $0.18$\\
HT150-B & 3.71 & 15.13 & $2.52$ & $0.17$ & 3.30 & 0.68 & $0.99$ & $0.86$ & $0.11$\\
HT150-C & 2.69 & 15.91 & $2.55$ & $0.16$ & 4.19 & 1.32 & $0.98$ & $0.86$ & $0.18$\\
\hline
\end{tabular}
\caption{The structure and kinematic properties of the final assembled star clusters at $t=10$ Myr. The 1D velocity dispersions $\sigma_\mathrm{1D}$ and the mean half-mass surface densities $\Sigma_\mathrm{h}$ are calculated within $R_\mathrm{e}$.}
\label{table: hierarchical-cluster-3}
\end{center}
\end{table*}

\subsection{Kinematics and cluster rotation}

\begin{figure}
\includegraphics[width=\columnwidth]{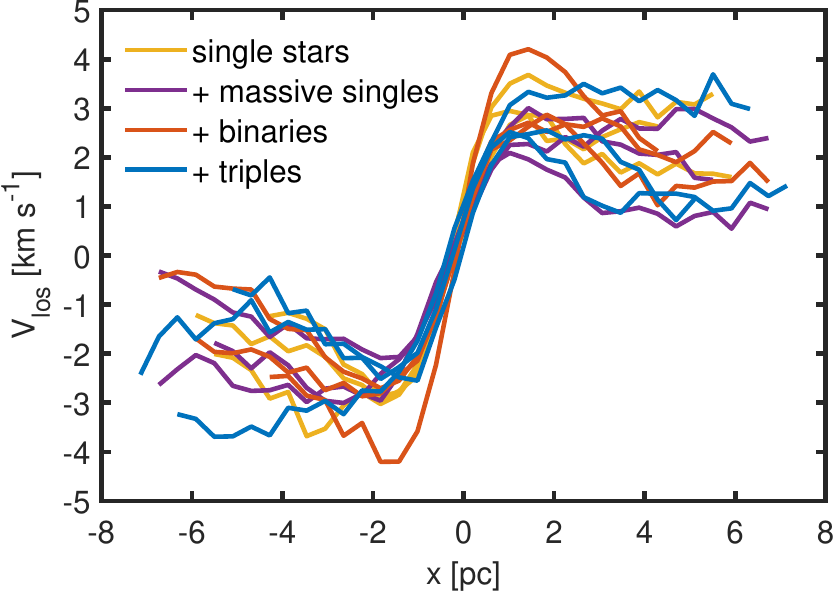}
\caption{The line-of-sight (LOS) velocity profiles $V_\mathrm{los}$ of the hierarchically assembled clusters at $t=10$ Myr within $<2 R_\mathrm{e}$ measured in a narrow slit of $|z|<1$ pc. For the analysis the cluster is oriented along its angular momentum vector within $R_\mathrm{e}$.}
\label{fig: vlos}
\end{figure}

\subsubsection{Cluster rotation}

By construction, our hierarchical cluster initial conditions have only a little angular momentum. The individual infalling sub-clusters are initially non-rotating, and the initial sub-cluster centre-of-mass positions and random initial velocities (in addition to the radial inflow velocity) are isotropic. Any angular momentum in the initial setup results from the fact that the sub-clusters on their initial orbits have different masses. Consequently, different star cluster orbits have different angular momenta which does not exactly sum up to zero total angular momentum. Tidal torques and especially cluster mergers transfer this angular momentum of the cluster orbits into the growing central massive cluster. As most of the angular momentum is brought into the main cluster by relatively low number ($\lesssim10$) of mergers with $M_\mathrm{cl} \gtrsim \msol{10^4}$, these few individual cluster mergers are responsible for the net rotation of the final assembled cluster.

We measure the line-of-sight 1D velocity profiles ($V_\mathrm{los}$) of the clusters within narrow slits of $|z|<1$ pc at $<2 R_\mathrm{e}$. For the $|V_\mathrm{los}|$ measurement the clusters are aligned with their angular momentum vector calculated within their $R_\mathrm{e}$. The $V_\mathrm{los}$ profiles of the hierarchical clusters at $t=10$ Myr are illustrated in Fig. \ref{fig: vlos}, and the peak values for $|V_\mathrm{los}|$ are listed in Table \ref{table: hierarchical-cluster-3}. The peak $V_\mathrm{los}$ velocities reach $|V_\mathrm{los}| \sim2.1$--$4.2$ km s$^{-1}$ in our models despite the non-rotating initial conditions. It seems that $2$--$4$ km s$^{-1}$ of rotation is inevitable for a massive hierarchically assembled cluster (see also e.g. \citealt{Ballone2021}). Any additional angular momentum in the initial setup, either sub-cluster rotation or global cluster assembly region rotation inherited from the parent molecular cloud \citep{Mapelli2017, Lahen2020b,Lahen2025b} would only increase the final cluster rotation. The cluster rotation inherited from the hierarchical cluster assembly process has major consequences for the subsequent dynamical evolution of the cluster on $>10$ Myr timescales (e.g. \citealt{Kamlah2022} and references therein).

\subsubsection{Cluster velocity anisotropy}

\begin{figure}
\includegraphics[width=0.9\columnwidth]{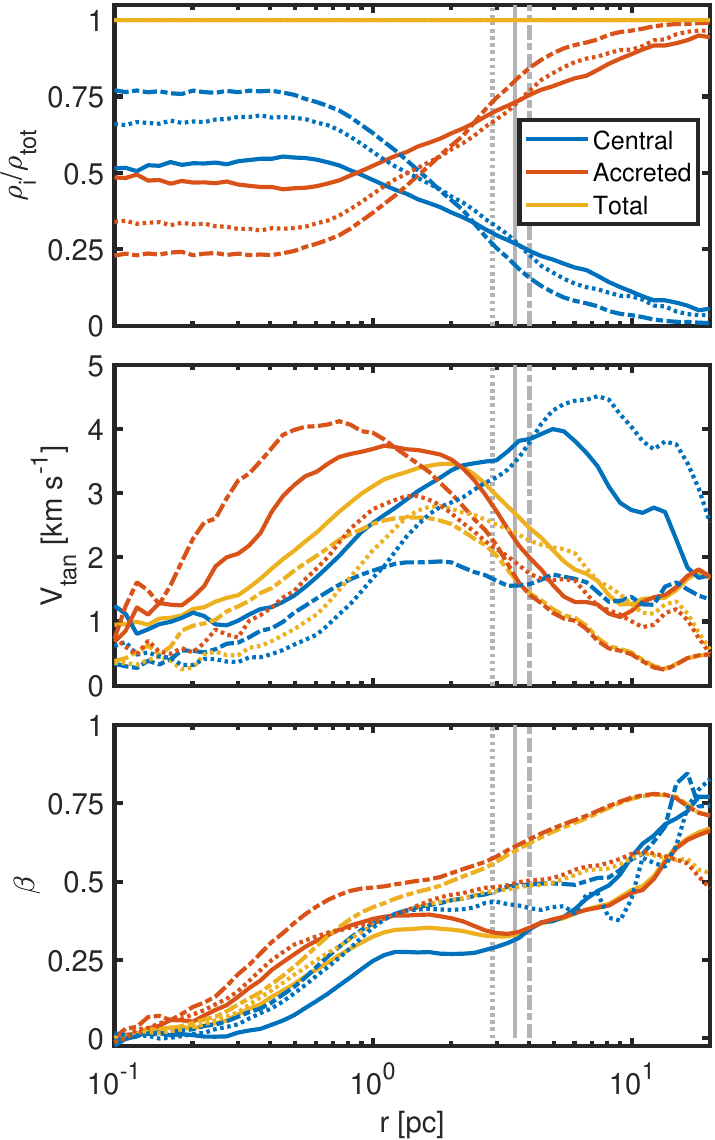}
\caption{The three hierarchical single star clusters in different line-styles separated into their initially central, accreted and total components in their density $\rho$ (top panel), rotation velocity $V_\mathrm{tan}$ (middle panel) and velocity anisotropy $\beta$ (bottom panel). The vertical lines indicate the 3D half-mass radii $r_\mathrm{h}$ of the clusters. The initially central population is more prominent within the central $r\lesssim 1$ pc, its rotation peaks at larger radii and it is less radially biased than the accreted population.}
\label{fig: tag-profiles}
\end{figure}

The assembled clusters are not isotropic in their velocity dispersions as shown in the bottom panel of Fig. \ref{fig: tag-profiles}. We define the velocity anisotropy parameter as $\beta = 1 - \sigma_\mathrm{t}^\mathrm{2}/\sigma_\mathrm{r}^\mathrm{2}$ \citep{Binney2008} in which $\sigma_\mathrm{r}$ and $\sigma_\mathrm{t}$ are the radial and tangential velocity dispersions in spherical coordinates. In our simulations at $t=10$ Myr $\beta \sim 0$ within the central $0.2$--$0.4$ pc (or $\sim10\%$ of $r_\mathrm{h}$) of the main cluster. Outside this central region, assembled clusters are radially biased with $\beta>0$. This is expected: stars accreted from the infalling clusters retain their predominantly radial orbits until $t=10$ Myr. Stars ejected from the main cluster centre into marginally bound orbits also contribute to $\beta>0$ at $r>r_\mathrm{h}$. Within $10 r_\mathrm{h}$ the maximum anisotropy parameter values are in the range of $0.5 \lesssim \beta \lesssim 0.7$. The exact value for $\beta$ in the outskirts of the final cluster depends on the detailed merger and stellar ejection histories of the cluster. Our cluster velocity anisotropy profiles built in a large number of cluster mergers are very similar to the anisotropy profiles in \cite{Souvaitzis2025} who recently studied mergers of two equal-mass star clusters with IMBHs. Recent hydrodynamical simulations of cluster formation by \citet{Karam2025} and \citet{Lahen2025b} find qualitatively similar results, wherein $\beta$ transitions from isotropic to radially biased toward outer radii.

Several massive simulated star clusters host IMBH binaries exceeding the typical $M_\bullet$--$M_\star$ ratios (e.g. \citealt{Häring2004}) in stellar bulges and massive early type galaxies (ETGs). Still, we do not find evidence of tangentially biased ($\beta<0$) central stellar orbits which for ETGs is a dynamical fingerprint of SMBH core scouring \citep{Thomas2014, Rantala2018,Rantala2019}. This is most probably due to the considerably shorter relaxation timescales at the star cluster cores compared to the nuclei of massive galaxies.

\subsubsection{Cluster shapes}

\begin{figure}
\includegraphics[width=\columnwidth]{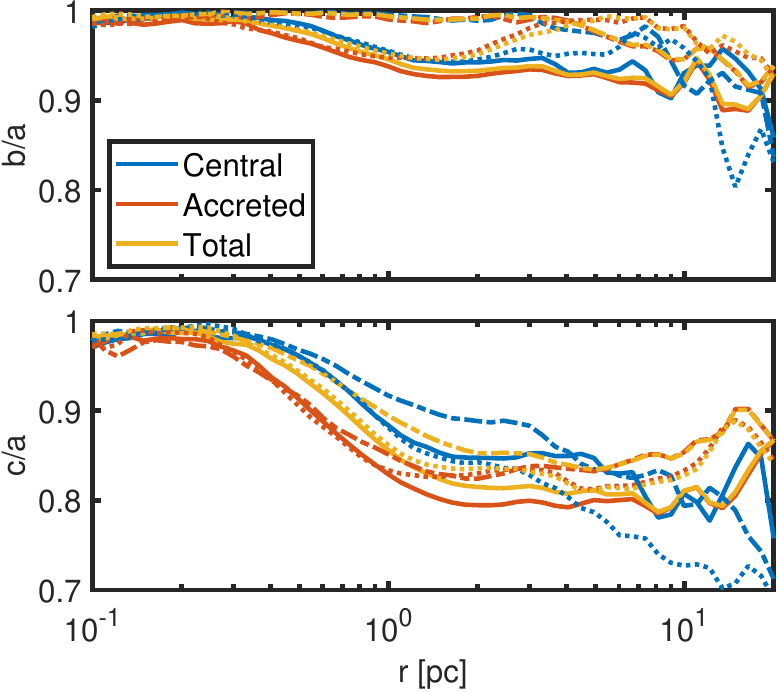}
\caption{The axis ratio profiles of the three single star models (different line-styles) divided in to the initially central and accreted stars as well as the total profiles. Overall, both $b/a$ and $c/a$ reach near unity at the cluster centres while the cluster shapes become more axisymmetric or even mildly triaxial in the cluster outskirts.}
\label{fig: shape}
\end{figure}

We determine the 3D shapes of the final assembled clusters at $t=10$ Myr, both within $R_\mathrm{e}$ and in radial bins, using the eigenvalues ($c \leq b \leq a$) of the inertia tensor $\matr{S}$ of the clusters (e.g. \citealt{Jesseit2005,Zemp2007,Binney2008}). We calculate the axis ratios $b/a$, $c/a$ and derive the cluster triaxiality parameters $T$ as
\begin{equation}
T = \frac{1-\left( \frac{b}{a} \right)^2}{1-\left( \frac{c}{a} \right)^2}.
\end{equation}
The axis ratios $b/a$, $c/a$ as well as triaxiality parameters $T$ of the final hierarchically assembled clusters within their $R_\mathrm{e}$ are listed in Table \ref{table: hierarchical-cluster-3}. The clusters are relatively axisymmetric and mildly flattened by rotation within their effective radii with $0.93 \lesssim b/a \lesssim 0.99$ and $0.81 \lesssim c/a \lesssim 0.88$. Their shapes range from oblate spheroids ($T=0.04$) to moderately triaxial systems $T=0.44$. As with the cluster rotation, the initial stellar multiplicity seems to have no large effect on cluster shapes during the first $10$ Myr of their evolution.

We present the final cluster shape profiles $b/a$ and $c/a$ of the hierarchical simulations with initial binaries in radial bins in Fig. \ref{fig: shape}. All the three cluster models are almost spherical or mildly axisymmetric at their centres ($r \lesssim 0.3$ pc) with $b/a \gtrsim 0.98$ and $0.95 \lesssim c/a \lesssim 0.98$. One of the clusters remains axisymmetric $b/a \sim 1$ until $r=4$--$5$ pc while the two other models have either a relative flat shape profiles with $b/a \sim 0.93$ above $r_\mathrm{h} \gtrsim 1$ pc or a mild minimum at $r=1$ pc. At the outer parts ($r>10$ pc) the clusters have more non-spherical shapes with $0.80 \lesssim b/a \lesssim 0.97$. In the hydrodynamical simulation of \citet{Lahen2025b}, $b/a$ and $c/a$ for a similar mass cluster show similar values, except in the inner \mbox{$<0.1$ pc} where central disk-like star formation results in a more elliptic stellar shape.

\subsubsection{Rotation and shape: comparison to simulations and observations}

Observed GCs rotate \citep{Bianchini2018,Vasiliev2021,Leitinger2025}, and in dynamically young GCs their angular momentum may originate from their formation process. These dynamically young GCs are observed with $(V/\sigma)_\mathrm{tot}=0.1$--$0.6$ while for the young massive LMC cluster R136, $\sigma_\mathrm{1D}\sim5 \pm 1$ km s$^{-1}$ and $V_\mathrm{los} \sim 3 \pm 1$ km s$^{-1}$ \citep{HenaultBrunet2020}, yielding $(V/\sigma)_\mathrm{tot}=0.6\pm0.3$ in projection. These values are overall consistent with hydrodynamical simulations (e.g. \citealt{Lahen2020b, Lahen2025b}) that indicate $(V/\sigma)_\mathrm{tot}=0.3$--$0.6$ for young, massive low-metallicity clusters. We find $(V/\sigma)_\mathrm{tot}=0.15$--$0.25$ in our hierarchical star cluster models, consistent with the dynamically young GCs with the least ordered rotation. Overall the maximum rotation velocities up to $\sim 4$ km s$^{-1}$ of our hierarchical models are consistent with the \textit{N}-body models in the literature (e.g. \citealt{Gavagnin2016, Ballone2021}). Star clusters in hydrodynamical simulations show higher $(V/\sigma)_\mathrm{tot}$ than our \textit{N}-body models, and e.g. \cite{Lahen2025b} presents a central star forming disk within their cluster responsible for the higher cluster rotation. Such a dissipative origin for the central cluster rotation cannot be modelled in pure gas-free direct \textit{N}-body simulations. Nevertheless, our results indicate that dissipationless hierarchical assembly is a plausible explanation for the rotation of slowly rotating young massive clusters and dynamically young GCs, and comparisons between hierarchical simulations with and without hydrodynamics can still be beneficial in order to distinguish between the effects of collisional and dissipative physical processes in massive star cluster formation.

Recently, \cite{Souvaitzis2025} has studied the structure of star cluster merger remnants after a single equal-mass cluster merger. After a single merger on an almost parabolic orbit, $0.96 \lesssim b/a \lesssim 0.99$ and $0.93 \lesssim c/a \lesssim 0.95$, consistent with our hierarchical models. With moderately eccentric or almost circular cluster merger orbits, the merger remnants of \cite{Souvaitzis2025} are more triaxial compared to our models, approaching $T=1$ with $0.92 \lesssim b/a \lesssim 0.95$ and $0.91 \lesssim c/a \lesssim 0.94$. This supports the scenario that the rotation and shape properties of hierarchically assembled star clusters largely originate from a single almost parabolic cluster merger with a mass ratio close to unity. For massive ($M_\mathrm{cl}>\msol{10^5}$) clusters, the maximum cluster ellipticities $\epsilon = 1-c/a$ within $R_\mathrm{e}$ are in the range of $\epsilon\sim 0.12$--$0.19$. These values for $\epsilon$ are similar or somewhat smaller compared to the maximum cluster ellipticities of $\epsilon \sim 0.2$ in hydrodynamical cluster formation simulations \citep{Lahen2020,Lahen2025b}. The reason for our somewhat less elliptic cluster shapes is the lesser amount of rotation in our clusters. Observations of GCs have projected values of $\epsilon\sim0.0$--$0.3$ \citep{Harris1996}.

\subsection{Initially central and accreted stars in the final cluster}\label{section: tag-profiles}

Star cluster mergers might at least partially explain the age and metallicities spreads in the observed GCs, NSCs and dwarf nuclei, even though they are unlikely to explain the long-standing multiple populations problem of GCs in its full complexity \citep{Gavagnin2016,Bastian2018,Gratton2019}. Even though our initial models use zero age and metallicity spreads, we can study how the final hierarchically forming star cluster is assembled from the sub-clusters of the cluster region, and whether the different cluster components can still be distinguished after $10$ Myr in their structural or kinematic properties.
 
We separate the density profiles of the three hierarchically assembled star clusters with initial binaries at $t=10$ Myr into the stars from the initially central component and the accreted stars from the in-falling clusters in the top panel of Fig. \ref{fig: tag-profiles}. The separation into the two components is very convenient using \bifrost{} as the simulation particles have initial cluster IDs in addition to their particle IDs. The central stars comprise up to $25\%$ of the total stellar mass in the simulations. After $10$ Myr, the initially central population is still the dominant component within the central $<1$ pc with $50\%$--$75\%$ of the central stars belonging to the initial population. The accreted population dominates the density profiles outside $r \gtrsim 1$--$2$ pc. For comparison, the effective radii $R_\mathrm{e}$ (azimuthally averaged 2D half-mass radii) of the assembled clusters are in the range of $R_\mathrm{e} \sim 2.1$--$3.7$ pc (see Table \ref{table: hierarchical-cluster-3}).

As with the total stellar density profiles, we distinguish the rotation $V_\mathrm{tan} = |\vect{v}-\vect{v}_\mathrm{rad}|$ of the initially central and later accreted stars in the three hierarchical models with binaries in the middle panel of Fig. \ref{fig: tag-profiles}. In the definition, $\vect{v}_\mathrm{rad}$ is the radial velocity vector. The two components have very similar rotation amplitudes of $V_\mathrm{tan}\sim1.9$--$4.5$ km s$^{-1}$ (initially central stars) and $V_\mathrm{tan}\sim3.0$--$4.1$ km s$^{-1}$ (accreted stars). The rotation $V_\mathrm{tan}$ peaks within $R_\mathrm{e}$ at $r=0.7$--$1.5$ pc for the accreted stellar population while for the initially central stars the rotation reaches its maximum in the outer parts of the cluster at $r=1.8$--$7.4$ pc. We attribute this difference to moment of inertia of the initially non-rotating central population and the interactions between its outer parts and the infalling sub-clusters. The maximum $|V_\mathrm{tan}|$ of all stars reaches $\sim3.5$ km s$^{-1}$ at $r=2$ pc which is within the effective radii of the clusters. Both the initially central and accreted stellar populations are nearly isotropic at their centres and anisotropic ($\beta>0$) above $r \gtrsim 0.2$--$0.3$ pc. The accreted population is somewhat more radially biased within $r_\mathrm{h}$ than the initially central population, as expected from the assembly histories of the clusters.

As shown in Fig. \ref{fig: shape}, the initially central and accreted stars in the final clusters have very similar $b/a$ profiles while the initially central population is less axisymmetric (higher $c/a$ for the same $b/a$) than the accreted population at $0.3$ pc $\lesssim r \lesssim 2$--$3$ pc due to its somewhat slower rotation at these separations. Above $r=1$ pc the accreted population has $0.8 \lesssim c/a \lesssim 0.9$ while for the initially central population $0.7 \lesssim c/a \lesssim 0.9$, lower values reached in the cluster outskirts. This is consistent with the rotation properties of the two populations shown in Fig. \ref{fig: tag-profiles}.

\subsection{The evolution of stellar multiplicity fractions}

\begin{figure*}
\includegraphics[width=\textwidth]{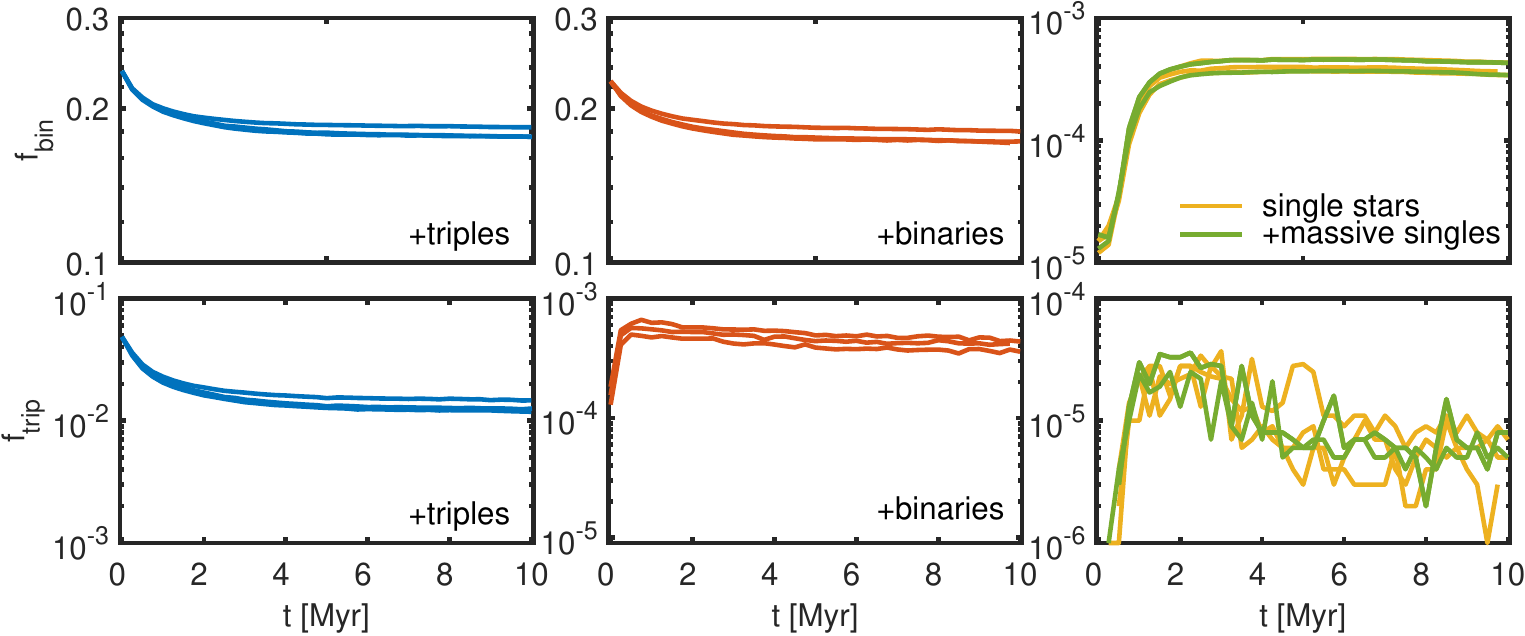}
\caption{The time evolution of binary fractions $f_\mathrm{bin}$ (top panels) and triple fractions $f_\mathrm{trip}$ (bottom panels) in the hierarchical cluster assembly models with singles, massive singles, binaries and triples. In the models without initial stellar multiplicity the binary and triple fractions remain low. In the models including stellar multiplicity, binaries and triple stars survive until $t=10$ Myr although their fractions somewhat decline due to internal dynamics, stellar evolution and strong interactions in dense stellar environments.}
\label{fig: multiplicity}
\end{figure*}

\begin{figure}
\includegraphics[width=\columnwidth]{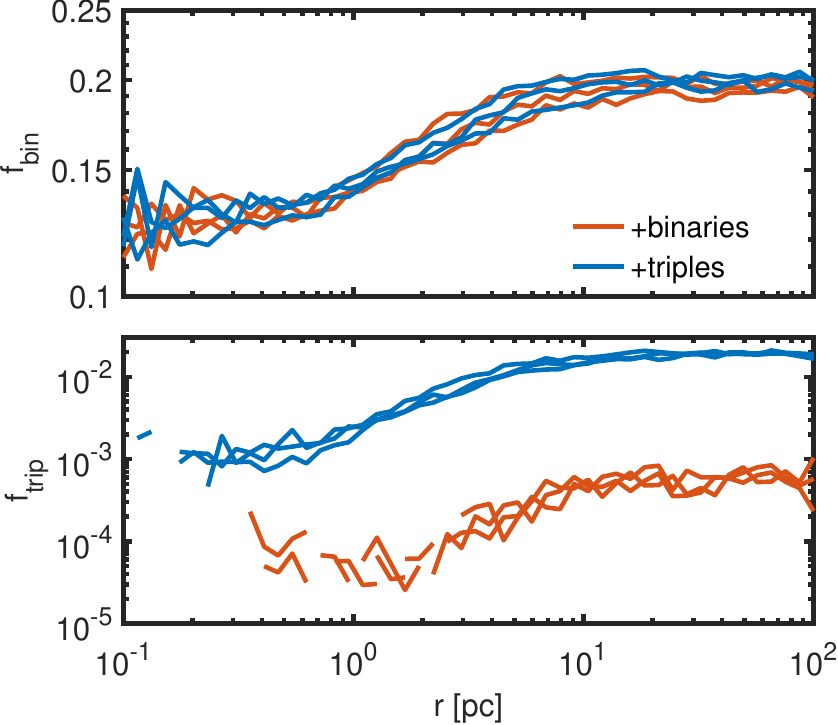}
\caption{The final $t=10$ Myr radial profiles for total binary $f_\mathrm{bin}$ and triple fractions $f_\mathrm{trip}$ in the assembled clusters of setups with initial binary and triple systems.}
\label{fig: multiplicity-spatial}
\end{figure}

\subsubsection{Identifying binaries and triples}

We identify binary and hierarchical triple systems from the \bifrost{} snapshot outputs using a k-nearest neighbours method. We focus on close binary systems (both individual binaries and inner binaries of hierarchical triples) with semi-major axes $a_\mathrm{in}<10^{-3}$ pc and triples with outer orbits $a_\mathrm{out}<5 \times 10^{-3}$ pc. With these definitions, the initial binary fraction of the binary and triple models is $f_\mathrm{bin}\sim0.23$--$0.24$ while the initial triple fraction of the triple model is $f_\mathrm{trip}\sim0.048$. These fractions, especially $f_\mathrm{bin}$,  somewhat depend on the choices for the threshold semi-major axis values of $a_\mathrm{in}=10^{-3}$ pc and  $a_\mathrm{out}=5 \times 10^{-3}$ pc. Without any semi-major axis cuts, the initial conditions generation algorithm described in section \ref{section: ic-binary-triple} results in $f_\mathrm{bin}\sim 0.29$ and $f_\mathrm{trip}\sim0.055$. However, increasing $a_\mathrm{in}$ e.g. by an order of magnitude results in a substantial increase of wide, weakly bound, short-lived binary systems in the crowded centres of our simulated clusters. The chosen values for $a_\mathrm{in}$ and $a_\mathrm{out}$ thus represent a compromise between omitting the widest initial binaries and including the least bound dynamically formed binaries.

The generation of star cluster initial conditions results in a small binary fraction of the order of $f_\mathrm{bin}\sim 10^{-5}$ even in the single and massive single star models. The initial triple fraction is very low in the single models, $f_\mathrm{trip} \lesssim 10^{-6}$, corresponding of the order of $\sim1$ hierarchical triple system from the initial conditions generation procedure. We note that these numerical fractions are $3$-$4$ orders of magnitude smaller than the binary and triple fractions in the models with initial binary and triple systems. The models with initial binaries have initially $f_\mathrm{trip} \lesssim 10^{-4}$, again $2$--$3$ orders of magnitude lower than in the simulations with an initial triple population. We conclude that even though the initial conditions generation procedure results in chance binaries and triples, their numbers are very low and thus do not affect the results of our analysis. See the Appendix A5 of \cite{Rantala2023} for more details on the subject of chance binaries.

\subsubsection{The time evolution of binary and triple fractions}

Even in the absence of the cluster environment, the binary, triple and high-order stellar multiplicity fractions of stellar populations strongly evolve on time-scales of $10^6$--$10^8$ Myr \citep{Preece2024}. The dense cluster environment can further enhance this complex evolutionary process. We present the time evolution of global binary and triple star fractions $f_\mathrm{bin}$ and $f_\mathrm{trip}$ in Fig. \ref{fig: multiplicity} in the hierarchical cluster assembly simulations. During the simulations, various physical processes cause the $f_\mathrm{bin}$ and $f_\mathrm{trip}$ to evolve. First, strong dynamical single-binary and binary-binary interactions can destroy initial binary and triple systems. However, new binary and triple systems form dynamically from initially unbound single stars (binaries) and in single-binary interactions (triples). Next, besides dynamical interactions, stellar evolution processes can also lead to stellar mergers or disruption of binary and triple systems. Finally, single stars and multiple systems that reach a separation of $200$ pc from the centre of the main assembling cluster are removed from the simulation.

In the models with initial triples, the triple fractions decrease during the first $1$-$2$ Myr of the simulations as the star clusters mass segregate and clusters with $M_\mathrm{cl} \lesssim \msol{10^5}$ undergo core collapse. The final triple fractions decline by a factor of $3$--$4$ to $f_\mathrm{trip} \sim 0.012$--$0.015$ from the initial value of $f_\mathrm{trip}\sim0.048$. The initial binary fractions of the triple models also diminish from $f_\mathrm{bin}\sim0.24$ to $f_\mathrm{bin} \sim 0.18$ over the course of the $10$ Myr simulations. In the initial binary star simulations, the binary fractions evolve similarly, declining from their initial value of $f_\mathrm{bin}\sim0.23$ to $f_\mathrm{bin} \sim 0.17$--$0.18$. The binary models dynamically form a triple population of $f_\mathrm{trip} \sim 6\times10^{-4}$ only in $\sim0.5$ Myr. Later the dynamical triple fraction slowly declines, reaching $f_\mathrm{trip} \sim 4\times10^{-4}$ at $t=10$ Myr. We note that the dynamically formed triple fractions in the initial binary models are by a factor of $>30$ lower than in the simulations with an initial triple population. This outcome is consistent with the binary star population results of \cite{Cournoyer-Cloutier2021} generalised to higher stellar multiples. Star cluster models without an initial binary population do not dynamically form a binary population that would resemble the observed stellar binary population properties. The same holds true for the triple populations.

\subsubsection{The spatial variations of binary and triple fractions in assembled clusters}

Next, we study the spatial variations of the binary and triple fractions in the final assembled clusters at $t=10$ Myr. We show the radial $f_\mathrm{bin}$ and $f_\mathrm{trip}$ profiles of the models with initial binaries and triples in Fig. \ref{fig: multiplicity-spatial}. In the outer parts of the assembled cluster, $r \gtrsim 2$--$3 R_\mathrm{e}$ the stellar multiplicity fractions are $f_\mathrm{bin} \sim 0.2$, $f_\mathrm{trip} \sim 0.02$ (with initial triples) and $f_\mathrm{trip} \sim 5\times10^{-4}$ (without initial triples). The multiplicity fractions at the outer parts of the clusters are somewhat higher than the total multiplicity fractions due to the lower stellar densities in the cluster outer parts and the consequently less frequent strong interactions with singles and other binaries and triples destroying the multiples. From $r=10$ pc to $r=1$ pc, both the binary and triple fractions decrease due to the inner denser stellar environments. Within the central $\sim 1$ pc, the radial stellar multiplicity profiles are almost constant with $f_\mathrm{bin} \sim 0.12$--$0.14$, $f_\mathrm{trip} \sim 10^{-3}$ (with initial triples) and $f_\mathrm{trip} \sim$ a few times $10^{-5}$ (without initial triples). While a considerable number of stars reside in binary systems in the dense centres of the assembled clusters up to $\Sigma_\mathrm{c} \sim \Sigmasol{10^5}$, central triples exist but are rare especially in the models without an initial triple population.

Binary stars can survive over Gyr timescales in GCs given high enough initial binary fractions (e.g. \citealt{Ivanova2005}). Our results demonstrate that initial binary and triple populations can survive the hierarchical assembly of a dense, massive low-metallicity star cluster, and that binaries and even triples can be found at the dense centres of $10$ Myr old clusters. The binary fractions of our assembled clusters are widely consistent with the main sequence binary fractions of $0.05 \lesssim f_\mathrm{bin} \lesssim 0.2$ in ancient GCs (e.g. \citealt{Rubenstein1997,Bellazzini2002,Cool2002,Sollima2007,Dalessandro2011,Milone2012}. In GCs, the radial binary fraction profiles are typically flat or declining towards the outer parts \citep{Milone2012}. This is in contrast to the $f_\mathrm{bin}$ profiles of our $10$ Myr old clusters. As our clusters are considerably younger than ancient GCs and our models do not include an external galactic tidal field for simplicity, our binary fractions at the outer parts of the cluster are likely overestimated. In Gyr long integrations with a galactic tidal field, outer binaries would be either stripped and lost, or sink into the cluster centre due to enhanced dynamical friction and mass segregation experienced by the binary member stars (e.g. \citealt{Grijs2013}). On the other hand, LMC young massive clusters NGC 1805 and NGC 1818 host somewhat higher total binary fractions $f_\mathrm{bin} \sim 0.37$ and $f_\mathrm{bin} \sim 0.20$--$0.35$ \citep{Elson1998,Hu2010,Grijs2013,Li2013}. For NGC 1805 the observed binary fraction decreases towards the outer parts \citep{Li2013} while both radially increasing and decreasing binary fractions have been suggested for NGC 1818. We note that our cluster models are several orders of magnitude denser than the young massive LMC clusters, potentially explaining why our binary and triple fractions show inward declining radial stellar multiplicity profiles. Models with lower densities more similar to LMC young massive clusters should lead to more constant radial profiles of the multiplicity fractions.

\section{IMBH -- host star cluster scaling relations}\label{section: 6}

\begin{table*}
    \begin{center}
    \begin{tabular}{lcccccc}
    \hline
        relation & $M_\mathrm{cl}$--$M_\bullet$ & & $\Sigma_\mathrm{h}$--$M_\bullet$ & & $\sigma$--$M_\bullet$\\
        fit parameter & $A_\mathrm{M}$ & $B_\mathrm{M}$ & $A_\mathrm{\Sigma}$ & $B_\mathrm{\Sigma}$ & $A_\mathrm{\sigma}$ & $B_\mathrm{\sigma}$\\
         \hline
single stars & $-0.19 \pm 0.36$ & $0.66 \pm 0.08$ & $-1.79 \pm 0.50$ & $1.00 \pm 0.11$ & $0.69 \pm 0.26$ & $2.03 \pm 0.25$\\
+ massive singles & $-2.70 \pm 0.64$ & $1.21 \pm 0.13$ & $-5.47 \pm 1.40$ & $1.79 \pm 0.30$ & $-1.17\pm0.30$ & $3.77\pm0.31$\\
+ binaries & $-1.76 \pm 0.35$ & $1.02 \pm 0.08$ & $-4.26 \pm 0.96$ & $1.55 \pm 0.21$ & $-0.43 \pm 0.26$ & $3.13 \pm 0.29$\\
+ triples & $-0.61 \pm 0.21$ & $0.77 \pm 0.05$ & $-2.61 \pm 0.44$ & $1.19 \pm 0.10$ & $0.40 \pm 0.20$ & $2.36 \pm 0.22$\\
\hline
    \end{tabular}
    \caption{The fit parameters for the IMBH--host star cluster scaling relations of Eq. \eqref{eq: Mbh-Mcl}, Eq. \eqref{eq: Mbh-Sigmacl} and Eq. \eqref{eq: Mbh-sigmacl}.}
    \label{table: scalingrelations}
    \end{center}
\end{table*}

\subsection{IMBH masses versus star cluster masses}

\begin{figure*}  
\includegraphics[width=0.92\textwidth]{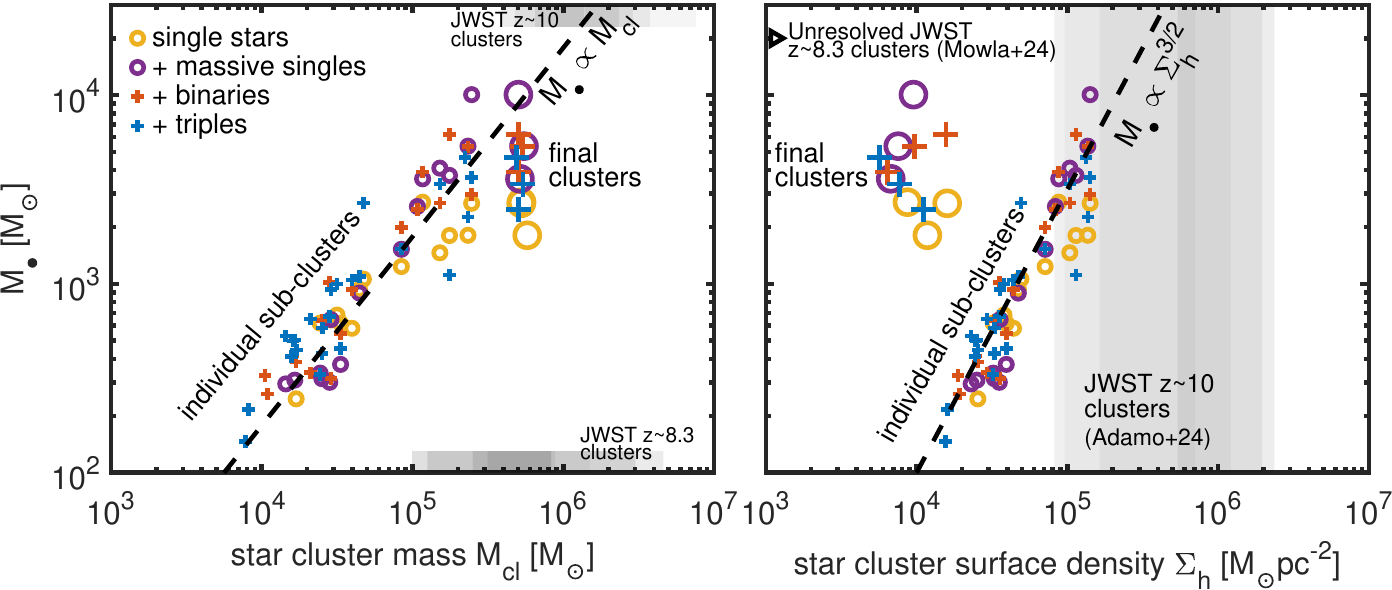}
\caption{Left panel: the star cluster mass $M_\mathrm{cl}$ vs the IMBH mass $M_\bullet$ (above $M_\bullet \gtrsim \msol{150}$) for our individual sub-clusters (small symbols) that formed at least a single IMBH in the hierarchical setups as well as the final assembled clusters (large symbols). Most massive IMBHs form in the massive single models up to $M_\bullet \gtrsim \msol{10^4}$ followed by the initial binary and triple models. Right panel: the half-mass surface densities $\Sigma_\mathrm{h}$ (within $R_\mathrm{e}$) vs $M_\bullet$. The IMBH masses in the sub-clusters approximately follow $M_\bullet \propto \Sigma_\mathrm{h}^\mathrm{3/2}$ while the final assembled sub-clusters have surface densities by a factor of $\sim 10$ lower compared to the most dense sub-clusters. For the early $z\sim10$ JWST Cosmic Gems Arc star clusters \citep{Adamo2024} our results imply $M_\bullet \gtrsim$ a few times $\msol{10^4}$. The unresolved $z\sim8.3$ Firefly Sparkle clusters \citep{Mowla2024} might host $M_\bullet \gtrsim \msol{10^3}$ IMBHs based on their masses alone.}
\label{fig: scalingrelations-1}
\end{figure*}

Correlations and simple power-law scaling relations exist between SMBH masses and the properties of their nuclear star clusters and the bulges of their host galaxies, potentially characterising their co-evolution (e.g. \citealt{Kormendy2013}). Next, we examine whether similar scaling relations exist between the IMBH masses $M_\bullet$ and their host star cluster properties.

For typical massive galaxies $M_\bullet / M_\mathrm{bulge} \sim 10^{-3}$ or slightly above (e.g. \citealt{Häring2004,Rein2015}), and for typical low-mass ($M_\mathrm{NSC} \lesssim \msol{10^6}$) NSCs $M_\bullet/M_\mathrm{NSC}\sim 0.1$ \citep{Neumayer2020}. We have shown in Fig. \ref{fig: isolated-mmax} that for isolated clusters with binaries the ratio $M_\bullet / M_\mathrm{cl}$ can reach $\sim$ a few times $10^{-2}$. We show the individual sub-cluster masses $M_\mathrm{cl}$ with their IMBH masses $M_\bullet$ in the hierarchical models, as well as for the final assembled star clusters in the left panel of Fig. \ref{fig: scalingrelations-1}. For low mass nuclear star clusters, $-3.5 \lesssim \log_\mathrm{10}{(M_\mathrm{NSC}/M_\bullet)} \lesssim 0$ \citep{Neumayer2020}. Our final assembled clusters and their IMBHs are well within this range with our IMBHs being by a factor of $\sim 5$ lower mass than typical SMBHs in low-mass NSCs. It has been recently demonstrated that NSCs aid the IMBH growth by gas accretion (\citealt{Partmann2025, Shin2025}, see also \citealt{Petersson2025}), providing a growth channel for IMBHs with low $M_\bullet / M_\mathrm{NSC}$ towards more typical values of $M_\bullet / M_\mathrm{NSC}$. Overall, simple SMBH gas accretion models (e.g. \citealt{Bondi1952}) typically have $\dot{M}_\bullet \propto M_\bullet^\mathrm{2}$, resulting in the preferred steepening of the initial scaling relations as more massive BHs can grow more efficiently via gas accretion.

The IMBH masses in the individual sub-clusters approximately follow $M_\mathrm{\bullet} \propto M_\mathrm{cl}$ as in the case of isolated star cluster models. We consider a power-law $M_\mathrm{cl}$--$M_\bullet$ relation of the logarithmic form
\begin{equation}\label{eq: Mbh-Mcl}
    \log_\mathrm{10}\left( \frac{M_\bullet}{M_\mathrm{\odot}} \right) = A_\mathrm{M} + B_\mathrm{M} \log_\mathrm{10}\left( \frac{M_\mathrm{cl}}{\msol{}} \right).
\end{equation}
The fit results for single stars, massive singles, initial binaries and triples are listed in Table \ref{table: scalingrelations}. For single star models the power-law slope is relatively shallow, $B_\mathrm{M} = 0.66 \pm 0.08$, because stellar collisions are suppressed in high-mass clusters as explained in \paperone{} and section \ref{section: isolated}. We note that very recently a qualitatively similar power-law slope was obtained in the numerical simulations of \cite{Vergara2025b}. For massive single star and initial binary setups the relation is nearly linear with $B_\mathrm{M} = 1.21 \pm 0.13$ and $B_\mathrm{M} = 1.02 \pm 0.08$, respectively. The initial triple models have a somewhat shallower slope $B_\mathrm{M} = 0.77 \pm 0.05$, however, the shallower slope is caused by more massive (compared to e.g. binary models) IMBHs forming in lower-mass sub-clusters with $M_\mathrm{cl} \sim 2$--$4 \times \msol{10^4}$ and not by suppressed collisions at higher cluster masses above $M_\mathrm{cl} \gtrsim \msol{10^5}$. 

For massive galaxies and their SMBHs, observations indicate values for $B_\mathrm{M}$ close to unity both in the local Universe and at high redshifts. For example, \cite{Häring2004} find $B_\mathrm{M} = 1.12 \pm 0.06$ while $B_\mathrm{M} = 1.05 \pm 0.11$ in \cite{Reines2015} (low redshifts). At $4 \lesssim z \lesssim 7$ \cite{Pacucci2023} finds $B_\mathrm{M} = 1.06 \pm 0.09$ using the JWST. This power-law slope of $B_\mathrm{M}\sim1$ (however, see \citealt{Graham2015}) is similar to our results. Even so, the overall observed normalisation of the $M_\bullet$--$M_\mathrm{bulge}$ relation is considerably lower compared to our $M_\mathrm{cl}$--$M_\bullet$ results. Compared to extrapolated low-redshift scaling relations, our IMBHs in their star clusters are by a factor of $\sim 60$--$110$ over-massive, and even exceed the extrapolated high-redshift relation of \cite{Pacucci2023} by a factor of $\sim2.5$. Our final clusters lie approximately on the extrapolated \cite{Pacucci2023} relation. However, we emphasise that our scaling relations between IMBH masses and star cluster masses cannot be interpreted in the context of galaxy--SMBH scaling relations in a straightforward manner. The key reason for this is the fact that most of the stellar mass in galaxies is not concentrated into a single massive nuclear star cluster. At low redshifts typical massive (nuclear) star clusters of $M_\mathrm{cl} \sim \msol{10^6}$ are hosted by galaxies with stellar masses of $\sim \msol{10^8}$ with a relatively large scatter \citep{Neumayer2020}. Given a stellar bulge of $\msol{10^8}$ and our most massive IMBHs of $M_\mathrm{\bullet} \sim \msol{10^4}$, the resulting mass ratio of $10^\mathrm{-4}$ lies below the present-day SMBH--bulge ratio of  $M_\bullet / M_\mathrm{bulge} \sim 10^{-3}$.

\subsection{IMBH masses versus star cluster surface densities}

Next, we consider the relation between the IMBH masses and the mean half-mass surface densities $\Sigma_\mathrm{h}$ calculated within $R_\mathrm{e}$ of the clusters. As $M_\bullet \propto M_\mathrm{cl}$, we expect $M_\bullet \propto M_\mathrm{cl} \propto \Sigma_\mathrm{h}^\mathrm{1/(1-2\beta)}$ in which $\beta$ is the power-law slope of the star cluster mass size relation. For our shallow $\beta=0.18$, the simple consideration yields $M_\bullet \propto \Sigma_\mathrm{h}^\mathrm{1.56}$. The simulation results are presented in the right panel of Fig. \ref{fig: scalingrelations-1}. The IMBH masses in the individual sub-clusters approximately follow $M_\bullet \propto \Sigma_\mathrm{h}^\mathrm{3/2}$ which is close to the simple estimate. We consider power-law fits of the form
\begin{equation}\label{eq: Mbh-Sigmacl}
    \log_\mathrm{10}\left( \frac{M_\bullet}{M_\mathrm{\odot}} \right) = A_\mathrm{\Sigma} + B_\mathrm{\Sigma} \log_\mathrm{10}\left( \frac{\Sigma_\mathrm{h}}{\Sigmasol{}} \right).
\end{equation}
The fit results for $A_\mathrm{\Sigma}$ and $B_\mathrm{\Sigma}$ can be found in Table \ref{table: scalingrelations}. As with the $M_\bullet$--$M_\mathrm{cl}$ relation, the single star models have the least steep relation with $B_\mathrm{\Sigma}=1.00\pm0.11$ followed by the triple modes which have $B_\mathrm{\Sigma}=1.19\pm0.10$. The binary and massive single models result in relations with $B_\mathrm{\Sigma}=1.55\pm0.21$ and $B_\mathrm{\Sigma}=1.79\pm0.30$, respectively.

We note that the surface densities of the final assembled massive star clusters are by a factor of $\sim 10$ lower compared to the highest $\Sigma_\mathrm{h}$ of the sub-clusters. The effective radius $R_\mathrm{e}$ of the assembling cluster increases (while $\Sigma_\mathrm{h}$ decreases) due to stellar evolution mass loss, internal dynamical processes and especially star cluster mergers. The mass of the central assembling star clusters in our models grows by a factor of $\sim 3$ due to cluster mergers. For our $-2$ power-law cluster mass function, each infalling $M_\mathrm{cl}=\msol{10^5}$ cluster is accompanied by $10$ $M_\mathrm{cl}=\msol{10^4}$ clusters, $100$ $M_\mathrm{cl}=\msol{10^3}$ clusters and so on. These minor star cluster mergers with highly unequal mass ratios are very efficient in increasing the $R_\mathrm{e}$ of the assembling cluster according to virial arguments (e.g. \citealt{Naab2009}). As noted by \cite{Lahen2025b}, central star formation will decrease the cluster half-mass radii and increase their mean surface densities, so our gas-free models likely underestimate the final densities and surface densities of the assembled clusters. The low final cluster surface densities raise the possibility that massive star clusters might host considerably more massive IMBHs than what could be inferred from their current surface densities. This is simply because the early sub-clusters that formed the IMBHs were denser than the final star clusters they built up.

Dense, bound early ($z>8$) star clusters have been recently discovered by the JWST \citep{Adamo2024, Mowla2024}. We show the masses of these early JWST star clusters as well as the resolved half-mass surface densities of the \cite{Adamo2024} clusters in Fig. \ref{fig: scalingrelations-1} along our scaling relations. The lensed Cosmic Gems Arc star clusters of \cite{Adamo2024} are bound and have sub-pc sizes at $z\sim10$. The cluster stellar masses in the range of $\sim$ a few times $\msol{10^6}$ demonstrate the high surface densities of early star formation in the range of $\Sigmasol{10^5} \lesssim \Sigma_\mathrm{h} \lesssim \Sigmasol{2\times10^6}$. Assuming that the five proto-GCs of \cite{Adamo2024} are the brightest and the most massive members of an assembling region resembling our initial setup, we can use our $M_\mathrm{cl}$--$M_\mathrm{\bullet}$ and $\Sigma_\mathrm{h}$--$M_\mathrm{\bullet}$ scaling relations to estimate the potential IMBH masses in the observed cluster system. For typical \cite{Adamo2024} $z\sim10$ clusters, the fiducial relations in the panels of Fig. \ref{fig: scalingrelations-1} yield IMBH masses of $M_\bullet \sim \msol{3.6\times10^4}$ for $M_\mathrm{cl}=\msol{2\times10^6}$ and $M_\bullet \sim \msol{4.6\times10^4}$ for $\Sigma_\mathrm{h}=\Sigmasol{6\times10^5}$. JWST observations have also revealed early massive star clusters in the lensed Firefly Sparkle galaxy at $z\sim8.3$ \cite{Mowla2024}. These early star clusters remain unresolved. With cluster masses in the range of $\msol{10^5} \lesssim M_\mathrm{cl} \lesssim \msol{10^6}$, the limits for the cluster effective radii below $3.9$--$6.8$ pc yield lower limits of  $\Sigma_\mathrm{h} \gtrsim \Sigmasol{10^3}$ for the effective stellar surface densities of the clusters. Using our $M_\mathrm{cl}$--$M_\bullet$ relation, the IMBH masses corresponding to the Firefly Sparkle cluster mass range are $\msol{10^3} \lesssim M_\bullet \lesssim \msol{10^4}$, if the clusters are dense enough ($\Sigma_\mathrm{h} \gtrsim \Sigmasol{10^4}$). As the effective surface densities are only lower limits, it is not possible to directly estimate the potential cluster IMBH mass range using our $\Sigma_\mathrm{h}$--$M_\bullet$ relation. Assuming that the early clusters of \cite{Mowla2024} follow a shallow mass-size relation $R_\mathrm{e} \propto M_\mathrm{cl}^\mathrm{0.2}$ as local star clusters, then $\Sigma_\mathrm{h} \propto  M_\mathrm{cl}^\mathrm{-0.4}$, which indicates half-mass surface densities above $\Sigma_\mathrm{h} \gtrsim \Sigmasol{10^5}$. For these inferred surface densities our scaling relations yield IMBH masses in excess of $M_\bullet \gtrsim \msol{2000}$. In summary, our estimates suggest that the early clustered star formation environments observed by the JWST at $z \gtrsim 8$--$10$ were dense enough to form IMBHs well above $M_\mathrm{\bullet}\gtrsim \msol{10^4}$. In this mass range, the IMBHs are plausible candidates for SMBH seeds if they can further grow via gas accretion, TDEs or GW driven BH mergers in their dense environments.

\subsection{IMBH masses versus star cluster velocity dispersions}

\begin{figure}
\includegraphics[width=\columnwidth]{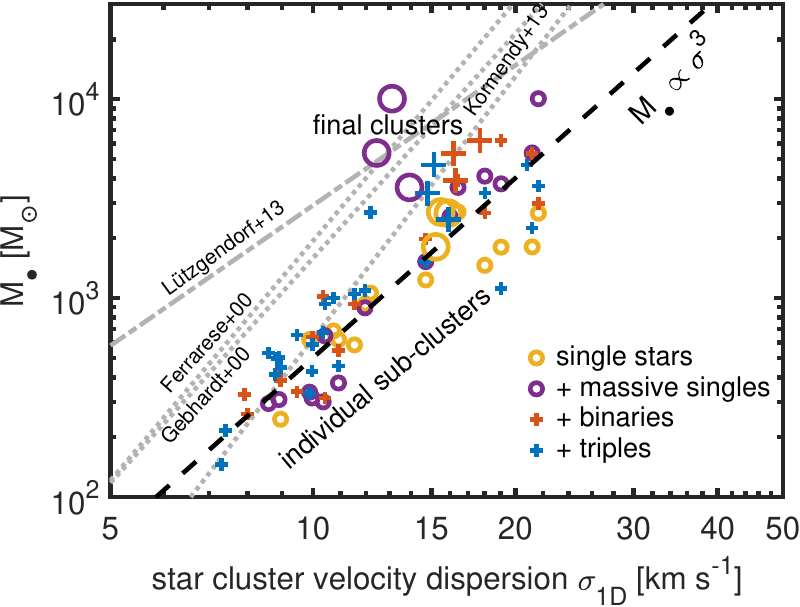}
\caption{The 1D star cluster velocity dispersions $\sigma_\mathrm{1D}$ measured within $R_\mathrm{e}$ for the individual sub-clusters (small symbols) and the final assembled clusters (large symbols). The power-law $M_\bullet$--$\sigma$ scaling relation is less steep, approximately $M_\bullet \propto \sigma^\mathrm{3}$ for our cluster models compared to the relations in the SMBH regime \citep{Ferrarese2000,Gebhardt2000,Kormendy2013}.}
\label{fig: mbh-sigma}
\end{figure}

The potential existence of IMBHs in massive star clusters and GCs is commonly motivated by the extrapolation of the $M_\bullet$--$\sigma$ relation for SMBHs and their host bulges into low velocity dispersions of $\sim$ a few times $10$ km s$^{-1}$. As $M_\bullet \propto M_\mathrm{cl}$, we expect $M_\bullet \propto M_\mathrm{cl} \propto \sigma^\mathrm{2/(1-\beta)}$ which yields $M_\bullet \propto \sigma^\mathrm{2.44}$ for our shallow cluster mass size relation. We present the IMBH masses in the sub-clusters in our hierarchical simulations as a function of their initial 1D velocity dispersions in Fig. \ref{fig: mbh-sigma}, as well as the most massive IMBHs in the final assembled clusters. We compare our results to a number of SMBH-host galaxy $M_\bullet$--$\sigma$ relations which have their power-law slopes in the range of $\sim 3.75$--$4.8$ \citep{Ferrarese2000,Gebhardt2000,Kormendy2013}. For the individual sub-clusters and their IMBHs, we fit scaling relations of the form
\begin{equation}\label{eq: Mbh-sigmacl}
    \log_\mathrm{10}\left( \frac{M_\bullet}{M_\mathrm{\odot}} \right) = A_\mathrm{\sigma} + B_\mathrm{\sigma} \log_\mathrm{10}\left( \frac{\sigma}{\text{km s$^{-1}$}} \right).
\end{equation}

The fit results for parameters $A_\mathrm{\sigma}$ and $B_\mathrm{\sigma}$ are listed in Table \ref{table: scalingrelations} for single stars, massive singles, initial binaries and triples. The fitted relations approximately follow $M_\bullet \propto \sigma^\mathrm{\alpha}$ with $2.0 \lesssim \alpha \lesssim 3.8$ and are thus typically shallower than $M_\bullet$--$\sigma$ relations for massive galaxies. The simple estimate of $M_\bullet \propto \sigma^\mathrm{2.44}$ is within the range of the fitted relations. As with the cluster masses $M_\mathrm{cl}$ and half-mass surface densities $\Sigma_\mathrm{h}$, the single star models yield the least steep scaling relations with $B_\mathrm{\sigma} = 2.03 \pm 0.25$. For the triple models, the scaling relation is again shallower than for binaries ($B_\mathrm{\sigma} = 2.36 \pm 0.22$ vs $B_\mathrm{\sigma} = 3.13 \pm 0.29$) as low velocity dispersion clusters with initial triples produce slightly more massive IMBHs than the initial binary models. The massive single models have the steepest IMBH-$\sigma$ relation with $B_\mathrm{\sigma} = 3.77 \pm 0.31$, only somewhat shallower than typical fits for the $M_\bullet$--$\sigma$ relation for massive galaxies and their SMBHs. As such, we conclude that in our simulations the IMBHs and their host star clusters do not follow the galactic $M_\bullet$--$\sigma$ relations extrapolated down into low velocity dispersions. If SMBH seeds indeed formed collisionally with masses in the range of $\msol{10^3} \lesssim M_\bullet \lesssim \msol{10^4}$ at redshifts beyond $z>10$, the tight low redshift scaling relations between massive black holes and their host galaxies must be established at later times when the SMBH seeds substantially grow by gas accretion and mergers during the hierarchical cosmological structure formation.

\subsection{Discussion: IMBHs in GCs and dwarf nuclei}
\subsubsection{Potential IMBHs in the Milky Way GCs}
Dynamical mass measurements of Milky Way GCs indicate that several GCs host a non-luminous component at their centres, in many cases consistent with both an IMBH and a centrally concentrated population of stellar BHs (e.g. \citealt{Gebhardt2002,Ibata2009,Noyola2010,vanderMarel2010,Lutzgendorf2011,Jalali2012,Kamann2016,Baumgardt2017}). In general, the issue remains debated. \cite{Lutzgendorf2013} analysed $14$ claimed dynamical mass measurements or upper limits for IMBH masses in GCs in the Local Group and compared these IMBH masses to the velocity dispersions of the GCs. Depending on the fitting method, the resulting power-law slope is between $B_\mathrm{\sigma} = 2.21 \pm 0.69$ and $B_\mathrm{\sigma} = 3.32 \pm 0.92$. A $M_\bullet$--$\sigma$ relation of \cite{Lutzgendorf2013} is displayed in Fig. \ref{fig: mbh-sigma} together with our simulation results. Interestingly, we find a similar shallow $M_\bullet$--$\sigma$ power-law relation, although with a factor of $\sim4$ times lower normalisation. In addition, the relation of \cite{Lutzgendorf2013} (see also \citealt{Kruijssen2013}) very closely matches the final hierarchically assembled clusters in $M_\bullet$ and $\sigma$ in the massive single setups.

\subsubsection{The IMBH of $\omega$ Centauri}
To this date, the most robust IMBH detection besides the $M_\bullet \sim \msol{150}$ GW merger GW190521 \citep{Abbott2020} remains the IMBH in the massive Milky Way GC $\omega$ Centauri ($\omega$ Cen), based on proper motions of fast-moving stars in its nucleus \citep{Häberle2024}. Even though $\omega$ Cen is most probably a stripped nucleus of a dwarf galaxy \citep{Lee1999,Bekki2003} instead of an ancient GC, we still explore whether the $\omega$ Cen IMBHs is consistent with any of our scaling relations between $M_\bullet$ and the host star cluster properties. \cite{Häberle2024} provides a lower limit of $M_\bullet \gtrsim \msol{8200}$ for the IMBH mass based on escape velocities alone. Further dynamical models of the overall velocity distribution and the fraction of fast-moving stars favour IMBH masses of the order $\sim3.9$--$\msol{4.7\times10^4}$.

The $\omega$ Cen has a mass of $M_\mathrm{cl}= \msol{3.55\times10^6}$ as well as central and half-light velocity dispersions of $21.1$ km s$^{-1}$ and $13.6$ km s$^{-1}$, respectively. The cluster has a half-light radius of 287 arcseconds, corresponding to $7.64$ pc given its distance of $d=5494 \pm 61$ pc \citep{Häberle2025}. This corresponds to the mean half-light surface density of $\Sigma_\mathrm{h} \sim \Sigmasol{9.7\times10^3}$. 

A $\msol{8200}$ IMBH in an environment with $\sigma=20$ km s$^{-1}$ is consistent with our results (Fig. \ref{fig: mbh-sigma}), between the individual sub-clusters with the most massive IMBHs and the final assembled clusters. The lower limit of the $\omega$ Cen IMBH mass and its half-mass surface density $\Sigma_\mathrm{h} \sim \Sigmasol{9.7\times10^3}$ very closely correspond to the surface densities and IMBH masses of our final assembled clusters (the right panel of Fig. \ref{fig: scalingrelations-1}). However, a $\msol{4.7\times10^4}$ IMBH suggested by dynamical models of the $\omega$ Cen is by a factor of $\gtrsim5$ more massive than our most massive IMBHs. We note that our star cluster models are by a factor of $\sim5$--$8$ less massive than $\omega$ Cen. Our example $M_\mathrm{cl}$--$M_\bullet$ relation in Fig. \ref{fig: scalingrelations-1} would predict $M_\mathrm{\bullet} \sim \msol{6.3\times10^4}$ for a cluster with the mass of $\omega$ Cen. This is by a factor of $7$--$8$ higher than the lower limit of $M_\bullet \gtrsim \msol{8200}$ and somewhat higher than $\sim3.9$--$\msol{4.7\times10^4}$ inferred from velocity distribution and fast moving star fraction mass estimates. Nevertheless, the mass estimate of $M_\mathrm{\bullet} \sim \msol{6.3\times10^4}$ is still consistent with the current stellar acceleration constraints on the $\omega$ Cen IMBH mass within 1 standard deviation (see Extended Data Fig. 6 of \citealt{Häberle2024}).

\section{Gravitational waves}\label{section: 7}

\begin{figure}
\includegraphics[width=\columnwidth]{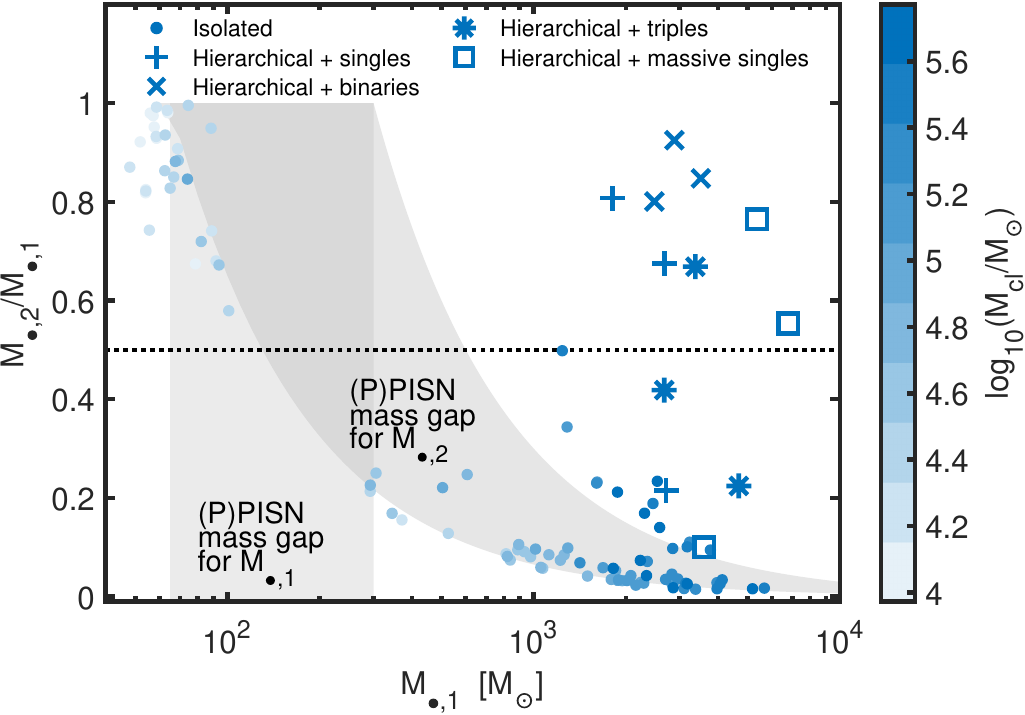}
\caption{The mass ratios of two most massive BHs formed through stellar collisions in the isolated (small symbols) and hierarchical models (large symbols). The massive single models (large squares) form the most massive IMBHs in this study, and $2$ out of $3$ hierarchical models result in $M_\mathrm{\bullet,2} / M_\mathrm{\bullet,1}>0.5$ (horizontal dotted line). Two of the massive IMBHs merge during the simulations, producing a GW merger event with a total IMBH mass of $\gtrsim \msol{10^4}$.}
\label{fig: q}
\end{figure}

\begin{figure*}
\includegraphics[width=0.85\textwidth]{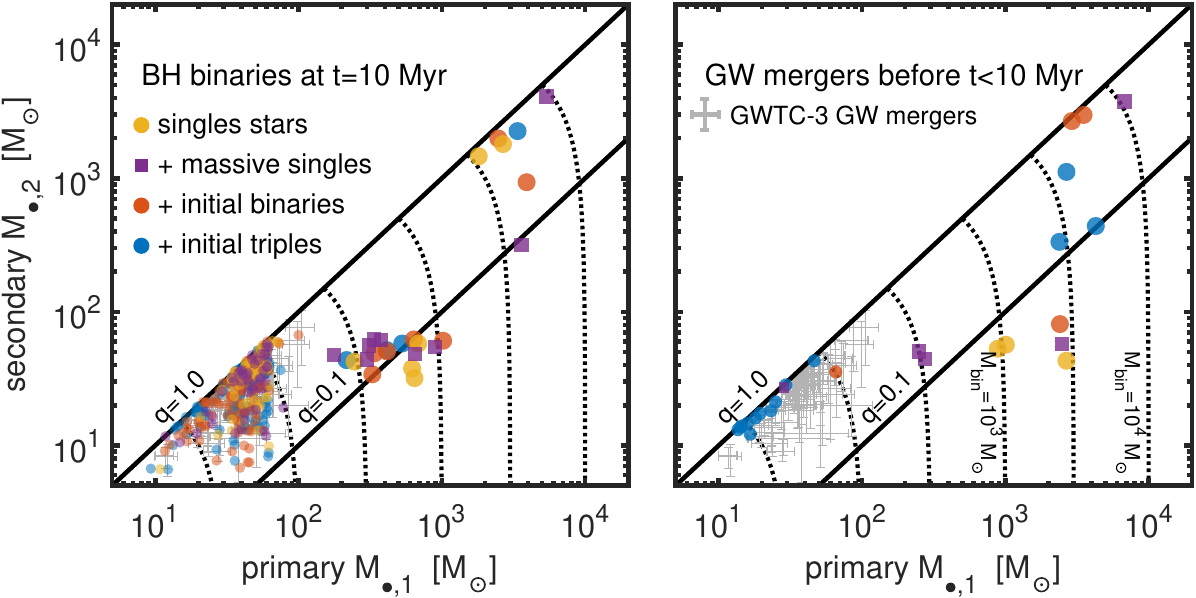}
\caption{The primary and secondary masses of (IM)BH binaries at $t=10$ Myr (left panel) and the GW mergers in the simulations at $t<10$ Myr (right panel) in the hierarchical cluster assembly models. The Advanced LIGO / Advanced Virgo GW mergers in the GWTC-3 catalogue are shown as the cross symbols \citep{Abbott2023}. Besides the Advanced LIGO like stellar BH population, the hierarchical models result in IMBH-BH binaries and IMBH-BH GW mergers with $\msol{300} \lesssim M_\mathrm{\bullet,1} \lesssim \msol{3000}$ and $0.03 \lesssim q \lesssim 0.1$. The massive single models produce the most massive IMBH-IMBH binaries and GW driven mergers. The hierarchical cluster assembly drives the formation of close to equal mass IMBH binaries and GW mergers up to $M_\mathrm{bin} = M_\mathrm{\bullet,1} + M_\mathrm{\bullet,2} \gtrsim \msol{10^4}$ in less than $t=10$ Myr for million solar mass star cluster formation regions.}
\label{fig: gw}
\end{figure*}

\subsection{Multiple IMBHs per hierarchically assembled cluster}

In \cite{Rantala2025} we showed that a close to equal mass GW merger of IMBHs in the mass range above $M_\bullet \gtrsim \msol{10^3}$ within $10$ Myr is a characteristic GW fingerprint of the hierarchical star cluster assembly. In our isolated models, especially with stellar multiplicity, several stars may grow collisionally, but they almost always merge with each other before forming IMBHs. Thus, the isolated star cluster channel rarely produces close to equal mass IMBH-IMBH binaries and GW mergers. On the other hand, in our hierarchical models several massive sub-clusters can contribute an IMBH to the final assembled cluster, more frequently leading to the formation of IMBH binaries with mass ratios $q=M_\mathrm{\bullet,2} / M_\mathrm{\bullet,1}$ near unity. Next, we show that this is the case for the hierarchical simulations with massive singles up to $m_\mathrm{max,0}=\msol{450}$ as well.

We show the masses of two most massive (IM)BHs formed in the isolated and hierarchical simulations in Fig. \ref{fig: q}. The results are presented using the largest to the second largest mass ratio $q=M_\mathrm{\bullet,2} / M_\mathrm{\bullet,1}<1$. Note that the two most massive (IM)BHs do not necessarily form a binary or merge in the simulations. The isolated cluster simulation results as well as the result of the hierarchical single, binary and triple models were briefly presented in \cite{Rantala2025}. In summary, for our isolated models almost always $q\lesssim0.25$ with fewer than $1/20$ of the models with $M_\mathrm{\bullet} \gtrsim \msol{1000}$ have $0.3 \lesssim q \lesssim 0.5$. On the other hand, $6/9$ of the hierarchical single, binary and triple models in the same mass range have $q \gtrsim 0.65$. The massive single models form the most massive IMBHs of this study, and continue the trend of having $2/3$ of the models above $q>0.5$. In the hierarchical massive single model HS450-B $q\sim0.1$ as only one sub-cluster forms a massive IMBH above $M_\bullet > \msol{3000}$. The two other massive single setups form massive IMBHs with $M_\bullet > \msol{5000}$ and $q\sim0.55$ and $q=0.76$, respectively. Two of these IMBHs will merge in the simulations, producing a GW merger with $M_\bullet > \msol{10^4}$. 

We note that other astrophysical channels may also lead to IMBH-IMBH mass GW mergers. If only a single IMBH forms per star cluster, IMBHs can still pair and merge in later star cluster mergers or in galactic nuclei \citep{Elmegreen2008, ArcaSedda2019, Dekel2025, Souvaitzis2025}. Central IMBHs of dwarf galaxies may merge during hierarchical galaxy assembly \citep{Tamfal2018,Bellovary2019,Khan2021}, and AGN disks of SMBHs potentially allow for growth, pairing and mergers of IMBHs \citep{McKernan2012,McKernan2014}. Metal-free Pop-III star clusters may also be a source of GWs from IMBH mergers (\citealt{Wang2022} and \citealt{Liu2024, Wang_Hanzhang2025}). However, the hierarchical star cluster assembly channel enables particularly rapid IMBH-IMBH mergers in less than $10$ Myr of the formation of the IMBHs, intrinsically connecting the extreme star formation environments at high redshifts to potential SMBH seed formation and gravitational wave observations.

\subsection{IMBH binaries and gravitational waves in the massive single star models}
We present the (IM)BH binaries in the hierarchical simulations at $t=10$ Myr as well as GW driven (IM)BH mergers at $t<10$ Myr. The results from simulations with single stars, initial binaries and triples have been briefly presented in \cite{Rantala2025}. 

The most massive GW merger in this study occurs in a model with massive singles. A massive IMBH binary with component masses $M_\mathrm{\bullet,1}=\msol{6776}$ and  $M_\mathrm{\bullet2}=\msol{3753}$ (mass ratio $q=0.55$) merges, resulting in a formation of an IMBH with $M_\bullet=\msol{10052}$. As discussed in \cite{Rantala2025}, the IMBH-IMBH GW mergers in the single star, initial binary and triple samples would be only marginally detectable by LISA at high redshifts as the $z\gtrsim10$ sensitivity for LISA rapidly increases between $M_\bullet\sim \msol{10^3}$ and $M_\bullet\sim \msol{10^4}$ \citep{Amaro-Seoane2012}. Our $M_\bullet=\msol{10052}$ GW merger in the massive single sample would most probably be observable by LISA at $z\gtrsim10$ with SNR $\sim10$. Furthermore, more massive (or more dense) assembling star clusters compared to our current models beyond $M_\mathrm{cl} > \msol{10^6}$ would likely produce merging IMBHs of higher masses. Lower-mass IMBHs remain promising targets for future space-based GW observatories with shorter interferometer arm lengths such as TianQuin \citep{Luo2016} or DECIGO \citep{Kawamura2021}. On the ground, third-generation ground-based GW observatories such as the Einstein Telescope and the Cosmic Explorer will be sensitive to IMBH-IMBH mergers in the mass range of $M_\mathrm{bin} = M_\mathrm{\bullet,1}+M_\mathrm{\bullet,2} \lesssim \msol{10^3}$ at $z<10$ \citep{Reali2024,EinsteinTelescopeCollaboration2025}.

\section{Summary and conclusions}\label{section: 8}

We have studied the effect of initial stellar binaries and triples as well as massive single stars up to $\msol{450}$ drawn from a standard IMF on the formation of IMBHs in hierarchically assembling massive star clusters. Our simulations with structured initial conditions with up to $N=1.8\times10^6$ stars are performed using the GPU accelerated post-Newtonian \textit{N}-body code \bifrost{} now fully coupled to the fast stellar population synthesis code \sevn{} \citep{Iorio2023}. Stellar multiplicity promotes strong interactions and stellar collisions \citep{Fregeau2004,Gaburov2008}, and runaway collision cascades ($dm/dt \propto m^\beta$ with $\beta>0$) can proceed also in massive clusters as opposed to single star models \citep{Rantala2024b}. At low metallicities ($Z=\zsol{0.01}$), our final assembled star clusters with masses of $M_\mathrm{cl} \gtrsim \msol{5\times10^5}$ rapidly form IMBHs up to $M_\bullet \sim \msol{10^4}$ in $<10$ Myr via stellar collisions, TDEs and gravitational wave driven black hole mergers. At a fixed cluster mass, initial stellar multiplicity and the massive single stars above $\msol{150}$ boost both the masses and numbers of the formed IMBHs up to $0.2$--$1.1$ IMBHs per $\msol{10^5}$ of dense, clustered star formation.

As opposed to simplified isolated setups, the hierarchical cluster assembly scenario frequently leads to the formation of close to equal mass IMBH-IMBH binaries. In models with initial stellar multiplicity and massive single stars, these IMBH-IMBH binaries with mass ratios near unity can be driven to GW mergers within $10$ Myr. This characteristic GW fingerprint of the hierarchical assembly scenario intrinsically connects early extreme star formation environments to GW observations. The most massive GW merger of our study reaches $M_\bullet \sim \msol{10^4}$ and would be observable at $z>10$ by LISA. Lower-mass IMBH-IMBH and IMBH-BH mergers are suitable candidates for next-generation space and ground based GW observatories such as TianQuin, DECIGO, the Einstein Telescope and the Cosmic Explorer. Besides GW transients, the IMBH formation is accompanied by high TDE rates up to $\Gamma_\mathrm{tde} \sim 3$--$5\times10^{-5}$ per assembling cluster until $\sim 2$ Myr after the IMBH formation.

The effective radii $R_\mathrm{e} \sim 2.3$--$3.7$ pc and mean half-mass surface densities of $\Sigma_\mathrm{h}\sim0.7$--$1.9\times\Sigmasol{10^4}$ of the final ($t=10$ Myr) hierarchically assembled star clusters are consistent with the sizes of low mass ($\lesssim \msol{10^6}$) nuclear star clusters \citep{Neumayer2020}. The final 3D density profiles of the hierarchically assembled clusters at $t=10$ Myr are well characterised double power law profiles both for stars and stellar BHs. The outer stellar power law of $\rho \propto r^\mathrm{-\beta}$ with $\beta \sim 3$ seems a robust prediction of the hierarchical assembly model with little variation between the models, supported by previous studies (e.g. \citealt{Antonini2012a,Guszejnov2018,Grudic2018,Lahen2025b}). At $t=10$ Myr, the cluster inner parts show more variation between the models compared to their outer parts. Only $1/4$ of our $12$ hierarchical models show truly flat EFF like central stellar cores ($\gamma_\mathrm{\star}=0$). Most clusters host a very shallow central stellar cusp with $0.17 \lesssim \gamma_\mathrm{\star} \lesssim 0.57$. The central profiles are expected to flatten towards $\gamma_\mathrm{\star} \sim 0$ over time (see Fig. 5 of \paperone). The core (break) radii for stellar BHs are smaller than for the stars, and central BH profile shapes span a wide range of $0.0 \lesssim \gamma_\mathrm{\bullet} \lesssim 1.32$. We find that central parts of the final clusters are still dominated by the stars and BHs that originate from the initially most massive central star cluster, and that IMBHs and cusps of stellar BHs can co-exist at least in young massive clusters at least on a $10$ Myr timescale.

The individual sub-clusters of the hierarchical assembly regions are initially non-rotating, but the assembling massive star cluster acquire non-zero rotation velocities due to star cluster mergers. For the final cluster at $t=10$ Myr both the peak 1D line-of-sight velocity $V_\mathrm{los}$ and the maximum 3D rotation velocity $V_\mathrm{tan}$ are in the range of $\sim2$--$4$ km s$^{-1}$. Thus, a few km s$^{-1}$ of rotation seems inevitable for massive hierarchically assembled star clusters. Our models with $(V/\sigma)_\mathrm{tot}\sim0.15$--$0.25$ are consistent with the observed slowly rotating young massive clusters and dynamically young GCs \citep{Bianchini2018,Leitinger2025}. The final clusters are isotropic at their centres while being radially biased ($\beta>0$) outside their cores due to their assembly histories through mergers. Consequently, the final clusters are almost spherical at their centres while having axisymmetric or even mildly triaxial shapes outside the central \mbox{$\sim 1$ pc}. The structural and kinematic properties of our assembled clusters can be used to construct improved initial conditions for long term integrations and Monte Carlo simulations. Such setups for the cluster initial conditions would avoid the early computationally expensive phase when the sub-clusters are very dense and the structure of the system is still far from spherical symmetry, and facilitate exploring the late-time evolution of hierarchically assembled massive star clusters and their IMBHs.

Models without initial stellar multiples do not dynamically form binary and triple star populations comparable to the models with initial binaries and triples, consistent with the binary population results of \cite{Cournoyer-Cloutier2021, Cournoyer-Cloutier2024b}. We find that initial binaries and even triple systems can survive the hierarchical cluster assembly, even though the central multiplicity fractions in the final assembled clusters are lower up to a factor of $\sim 2$ than in their outer parts. Our final binary fractions are in line with observed cluster binary fractions. The final central binary fractions are in the range of $f_\mathrm{bin}$--$0.12$--$0.14$ for the binary and triple models while the central triple fraction is $f_\mathrm{trip}\sim 10^{-3}$. The multiplicity fractions are higher in the low-density outer parts of the clusters, consistent with a subset of young massive clusters such as NCC 1818 in the LMC for which such a spatial binary fraction profile has been suggested.

The masses of the IMBHs correlate with the physical properties of the star clusters in which they formed, including the cluster mass $M_\mathrm{cl}$, the mean cluster half-mass surface density $\Sigma_\mathrm{h}$ and the cluster velocity dispersion $\sigma$. We find that the IMBHs and the sub-clusters follow the approximately relations of $M_\bullet \propto M_\mathrm{cl}$, $M_\bullet \propto \Sigma_\mathrm{h}^\mathrm{3/2}$ and $M_\bullet \propto \sigma^\mathrm{3}$. We emphasise that our scaling relations between IMBHs and their host clusters differ from the tight scaling relations between SMBHs and their host galaxies. This suggest that the tight observed scaling relations originate from later growth phases by gas accretion and mergers from the IMBH mass range into the SMBH range of $M_\bullet \gtrsim \msol{10^5}$. In addition, models with single stars, massive singles, initial binaries and triples follow somewhat different relations. Especially single star models show shallower relation slopes due to suppressed IMBH formation at high $M_\mathrm{cl}$. Initial triple models show shallower relation slopes than the binary models as the triple models produce more massive IMBHs in low-mass clusters. The final assembled star clusters with their IMBHs do not lie on the scaling relations between the IMBHs and their birth sub-clusters, especially in the case of the $M_\bullet$--$\Sigma_\mathbf{h}$ relation. Compared to sub-clusters with similar IMBH masses, the final assembled clusters have higher masses but lower velocity dispersions and lower surface densities due to the hierarchical merging process. Thus, GC progenitors observed at high redshifts and fully assembled young massive star clusters might host IMBHs of higher mass than what would be expected from their observed surface densities.

Finally, JWST observations have recently begun to reveal the early, dense and clustered star formation environments at redshifts $z\gtrsim8$--$10$ \citep{Adamo2024, Mowla2024}. Our results suggest that the Cosmic Gems Arc and the Firefly Sparkle star clusters and similar early clustered environments, if sufficiently dense, can form IMBHs with masses well above $M_\mathrm{\bullet}\gtrsim \msol{10^4}$. In this mass range, the IMBHs are plausible candidates for supermassive black hole seeds if they can further grow via gas accretion, TDEs or GW driven BH mergers.

\section*{Data availability statement}
The data relevant to this article will be shared on reasonable request to the corresponding author.

\section*{Acknowledgments}
TN acknowledges the support of the Deutsche Forschungsgemeinschaft (DFG, German Research Foundation) under Germany’s Excellence Strategy - EXC-2094 - 390783311 of the DFG Cluster of Excellence ''ORIGINS''. The numerical simulations were performed using facilities in Germany hosted by the Max Planck Computing and Data Facility (MPCDF) and the JUWELS Booster of the Jülich supercomputing centre (GCS project 59949 frost-smbh-origins). GI was supported by a fellowship grant from
the la Caixa Foundation (ID 100010434). The fellowship code is
LCF/BQ/PI24/12040020. GJE acknowledges support by the Spanish Ministry of Science via the Plan de Generación de Conocimiento through grant PID2022-143331NB-100. GJE and GI acknowledge financial support from the European Research Council for the ERC Consolidator grant DEMOBLACK, under contract no. 770017.


\bibliographystyle{mnras}
\interlinepenalty=10000
\bibliography{manuscript}


\bsp	
\label{lastpage}
\end{document}